\documentclass{cernrep}
\usepackage{xcolor,colortbl}
\usepackage{amsmath}
\begin{document}
\title{A Short Guide to Flavour Physics and CP Violation}
\author{Seung J. Lee and Hugo Ser\^odio
%\thanks{On leave from another institute somewhere.}
}
\institute{Department of Physics, Korea Advanced Institute of Science and Technology, \\
335 Gwahak-ro, Yuseong-gu, Daejeon 305-701, Korea}
\maketitle

\begin{abstract}
We present the invited lectures given at the second Asia-Europe-Pacific School of High-Energy Physics (AEPSHEP),
which took place in Puri, India in November 2014. The series of lectures aimed at graduate students in particle experiment/theory, covering the very basics of flavor physics and CP violation, some useful theoretical methods such as OPE and effective field theories, and some selected topics of flavour physics in the era of LHC.
\end{abstract}

\section{Short introduction}\label{sec:0}

We present the invited lectures given at the second Asia-Europe-Pacific School of High-Energy Physics (AEPSHEP), which took place in Puri, India in November 2014. The physics background of students attending the school are diverse as some of them were doing their PhD studies in experimental particle physics, others in theoretical particle physics. The lectures were planned and organized, such that students from different background can still get benefit from basic topics of broad interest in a modern way, trying to explain otherwise complicated concepts necessary to know for understanding the current ongoing researches in the field, in a relatively simple language from first principles.

These notes present a small compilation of several results that over the years has become standard in particle physics, and more concretely in the area of flavour physics. These are by no means a complete and self-contained course in flavour physics, but rather a brief introduction to several topics that should be explored in more detail by additional references for the interested readers. For the topics addressed in these notes there are several textbooks and review articles that have become standard references; here we compile an incomplete list:
\begin{itemize}
\item[$\bullet$] For aspects concerning the building blocks of gauge theories and the standard model see, for example,~\cite{Pokorski:1987ed}
\item[$\bullet$] For $CP$ and flavour aspects in particle physics the books~\cite{Branco:1999fs,Bigi:2000yz,Leader:1996hm} are two excellent sources, as well as the more specific reviews~\cite{Buchalla:1995vs,Buras:1998raa,Buras:2001pn,Buchalla:2001ux,Silva:2004gz,Nir:2005js,Hocker:2006xb,
Antonelli:2009ws,Grossman:2010gw,Nir:2010jr,Gedalia:2010rj,Isidori:2013ez,Grinstein:2015nya,Ligeti:2015kwa}
\item[$\bullet$] For topics related with effective field theories we refer the reader to~\cite{Georgi:1990um,Neubert:1993mb,Mannel:1997ky,Manohar:2000dt,Neubert:2005mu}
\end{itemize} 

\section{The building blocks in particle physics}\label{sec:1}

\subsection{What is flavour and why do we care?}\label{subsec:1.1}

In Particle Physics one attributes quantum numbers to particles in order to classify them as representations of the symmetries describing the dynamics of the underlying model. This classification allows us to extract a lot of information just from first principles. In nature there are several copies of the same fermionic gauge representation, i.e. several fields that are assigned the same quantum numbers. We then say that different copies belong to different flavours (or families). Flavour physics describes the interactions that distinguish between flavours, i.e. between the different copies. 

The fermions can interact through pure gauge interactions. These interaction are related to the unbroken symmetries and mediated therefore by massless gauge bosons. They do not distinguish among the flavours and do not constitute part of flavour physics. Fermions can also have Yukawa interactions, i.e. interactions where two fermions couple to a scalar. These interactions are source of flavour and $CP$ violation. Within the Standard Model (SM), flavour physics refers to the weak and Yukawa interactions.

Flavour physics can predict new physics (NP) before it's directly observed. Some examples are:
\begin{itemize}
\item[$\bullet$] The smallness of $\Gamma(K_L\rightarrow \mu^+\mu^-)/\Gamma(K^+\rightarrow \mu^+\nu)$ allowed for the prediction of the charm quark
\item[$\bullet$] The size of $\Delta m_K$ allowed for the charm mass prediction
\item[$\bullet$] The measurement of $\epsilon_K$ allowed for the prediction of the third generation
\item[$\bullet$] The size of $\Delta m_B$ allowed for a quite accurate top mass prediction $(\sim 150\,\text{GeV})$
\item[$\bullet$] The measurement of neutrino flavour transitions led to the discovery of neutrino masses
\end{itemize}

\subsection{Discrete symmetries in particle physics}

In this section we present the discrete symmetries $C$, $P$ and $T$, which play a leading role in the construction of the present model of particle physics. These three symmetries do not leave, separately, the SM Lagrangian invariant but their product $CPT$ does (at least everything points on that direction). These discrete symmetries give rise to multiplicative conservation laws. They have three levels of action: on the particle states, on the creation and annihilation operators, and on the fields. The action on one level determines the action on the other two. The main properties of these symmetries are: 
\begin{itemize}

\item[$\bullet$] \textbf{Charge Conjugation}

Charge conjugation on the states reverses the quantum numbers of particles  that are associated with internal symmetries. The charge conjugate of a particle is another particle with the same energy and momentum but opposite charges (anti-particle).  Charge conjugation on the fields converts a field $\psi(x)$ into a field $\psi^c(x)$ with opposite internal quantum numbers. If charge conjugation is a symmetry of the quantum field theory, there must exist a unitary operator $\mathcal{C}$ which represents it. 

We can use charge conjugation in order to eliminate final states for scattering and decay processes and to provide a link between different processes involving charged particles.

\item[$\bullet$] \textbf{Parity}

Classical parity is any element in the component of the Lorentz group that contains the matrix $P =\text{diag}(1, -1,-1,-1)$. Parity, like charge conjugation, gives rise to a multiplicative conservation law.  For example, the $\eta$ meson and the pions are pseudoscalars (eigenstates with eigenvalue $-1$ as opposed "$+1$ for scars), and so the decay $\eta\rightarrow \pi^+\pi^-$ is forbidden by conservation of parity. However, since parity transforms space, the eigenvalues of parity depend on the orbital angular momentum of a state and the intrinsic parity of a state is not in general conserved. 

\item[$\bullet$] \textbf{Time Reversal}

The idea of time reversal is to take the time evolution of some system and reverse it. To separate the effects of charge conjugation from those of time reversal, it is customary to assume that time reversal preserves the internal quantum numbers of all particles. In classical mechanics, time reversal can be implemented by changing the sign of the Hamiltonian. If we suppose that this effect is achieved in quantum theory by a unitary transformation $U_T$, we get
\begin{equation}
U_{T}^\dagger e^{iH t}U_T=e^{iHt}\quad\Rightarrow\quad U_T^\dagger H U_T=-H\quad\Rightarrow HU_{T}|n\rangle=-E_nU_T|n\rangle\,,
\end{equation}
for any state $|n\rangle$, entering in conflict with the principle that energy should be bounded from below. The way to solve this is by dropping the unitary operator and represent time reversal by 
an anti-unitary operator operator $\mathcal{T}$. 

\end{itemize}
\Tables~\ref{tab:CPTfields}--\ref{tab:Discrete} summarize some of the most important transformations under these symmetries.

\begin{table}[h]
\begin{center}
\caption{Discrete symmetry transformations for photon, gluon, complex scalar and fermion fields. We have defined: $\psi^c=C\overline{\psi}^T$ and $s^a$ is $+1$ for $a=1,3,4,6,8$ while $-1$ for $a,2,5,7$.}
\label{tab:CPTfields}
\begin{tabular}{l}
\hline\hline
\textbf{Fields transformations}\\
\hline
$
\begin{array}{l}
\text{Photon:}\\
\mathcal{P}A_\mu(t,\vec{r})\mathcal{P}^\dagger=A^\mu(t,-\vec{r})\\
\mathcal{T}A_\mu(t,\vec{r})\mathcal{T}^{-1}=A^\mu(-t,\vec{r})\\
\mathcal{C}A_\mu(t,\vec{r})\mathcal{C}^\dagger=-A_\mu(t,\vec{r})\\
\mathcal{CP}A_\mu(t,\vec{r})\mathcal{CP}^\dagger=-A_\mu(t,-\vec{r})
\end{array}
$
$
\begin{array}{l}
\text{Gluon:}\\
\mathcal{P}G_\mu^a(t,\vec{r})\mathcal{P}^\dagger=G^{a\mu}(t,-\vec{r})\\
\mathcal{T}G^a_\mu(t,\vec{r})\mathcal{T}^{-1}=s^aG^{a\mu}(-t,\vec{r})\\
\mathcal{C}G^a_\mu(t,\vec{r})\mathcal{C}^\dagger=-sG^a_\mu(t,\vec{r})\\
\mathcal{CP}G^a_\mu(t,\vec{r})\mathcal{CP}^\dagger=-s^aG^a_\mu(t,-\vec{r})
\end{array}
$
$
\begin{array}{l}
\text{Complex scalar:}\\
\mathcal{P}\phi(t,\vec{r})\mathcal{P}^\dagger=e^{i\alpha_p}\phi(t,-\vec{r})\\
\mathcal{T}\phi(t,\vec{r})\mathcal{T}^{-1}=e^{i\alpha_t}\phi(-t,\vec{r})\\
\mathcal{C}\phi(t,\vec{r})\mathcal{C}^\dagger=e^{i\alpha_c}\phi^\dagger(t,\vec{r})\\
\mathcal{CP}\phi(t,\vec{r})\mathcal{CP}^\dagger=e^{i\alpha}\phi^\dagger(t,-\vec{r})
\end{array}
$\\
\\
$
\begin{array}{l}
\text{Fermion:}\\
\mathcal{P}\psi(t,\vec{r})\mathcal{P}^\dagger=e^{i\beta_p}\gamma^0\psi(t,-\vec{r})\\
\mathcal{T}\psi(t,\vec{r})\mathcal{T}^{-1}=e^{i\beta_t}\gamma_0^\ast\gamma_5^\ast C^\ast \overline{\psi}^\dagger(-t,\vec{r})\\
\mathcal{C}\psi(t,\vec{r})\mathcal{C}^\dagger=e^{i\beta_c}\phi^c(t,\vec{r})\\
\mathcal{CP}\psi(t,\vec{r})\mathcal{CP}^\dagger=e^{i\alpha}\gamma^0C\overline{\psi}^T(t,-\vec{r})
\end{array}
$
$
\begin{array}{l}
\text{  }\\
\mathcal{P}\overline{\psi}(t,\vec{r})\mathcal{P}^\dagger=e^{-i\beta_p}\overline{\psi}(t,-\vec{r})\gamma^0\\
\mathcal{T}\overline{\psi}(t,\vec{r})\mathcal{T}^{-1}=e^{-i\beta_t}\psi^\dagger(-t,\vec{r})(C^{-1})^\ast \gamma_5^\ast\gamma_0^\ast\\
\mathcal{C}\overline{\psi}(t,\vec{r})\mathcal{C}^\dagger=e^{i\beta_c}\overline{\psi^c}(t,\vec{r})\\
\mathcal{CP}\overline{\psi}(t,\vec{r})\mathcal{CP}^\dagger=e^{-i\alpha}\psi^T(t,-\vec{r})C^{-1}\gamma^0
\end{array}
$
\\
\hline
\end{tabular}
\end{center}
\end{table}

\begin{table}[h]
\begin{center}
\caption{Symmetry transformation properties of some fermionic bilinears under the action of discrete symmetries. Overall phases and the coordinates have been omitted.}
\label{tab:Discrete}
\begin{tabular}{llllll}
\hline\hline
\textbf{Bilinear}&$\mathcal{P}$&$\mathcal{T}$&$\mathcal{C}$&$\mathcal{CP}$&$\mathcal{CPT}$\\
\hline
$\overline{\psi}\chi$&$\overline{\psi}\chi$&$\overline{\psi}\chi$&$\overline{\chi}\psi$&$\overline{\chi}\psi$&$\overline{\chi}\psi$\\
$\overline{\psi}\gamma_5\chi$&$-\overline{\psi}\gamma_5\chi$&$\overline{\psi}\gamma_5\chi$&$\overline{\chi}\gamma_5\psi$
&$-\overline{\chi}\gamma_5\psi$&$-\overline{\chi}\gamma_5\psi$\\
$\overline{\psi}P_{L,R}\chi$&$\overline{\psi}P_{R,L}\chi$&$\overline{\psi}P_{L,R}\chi$&$\overline{\chi}P_{L,R}\psi$
&$\overline{\chi}P_{R,L}\psi$&$\overline{\chi}P_{R,L}\psi$\\
$\overline{\psi}\gamma^\mu\chi$&$\overline{\psi}\gamma_\mu\chi$&$\overline{\psi}\gamma_\mu\chi$&$-\overline{\chi}\gamma^\mu\psi$
&$-\overline{\chi}\gamma_\mu\psi$&$-\overline{\chi}\gamma^\mu\psi$\\
$\overline{\psi}\gamma^\mu\gamma_5\chi$&$-\overline{\psi}\gamma_\mu\gamma_5\chi$&$\overline{\psi}\gamma_\mu\gamma_5\chi$&$\overline{\chi}\gamma^\mu\gamma_5\psi$
&$-\overline{\chi}\gamma_\mu\gamma_5\psi$&$-\overline{\chi}\gamma^\mu\gamma_5\psi$\\
$\overline{\psi}\gamma^\mu P_{L,R}\chi$&$\overline{\psi}\gamma_\mu P_{R,L}\chi$&$\overline{\psi}\gamma_\mu P_{L,R}\chi$&$-\overline{\chi}\gamma^\mu P_{R,L}\psi$
&$-\overline{\chi}\gamma_\mu P_{L,R}\psi$&$-\overline{\chi}\gamma^\mu P_{L,R}\psi$\\
$\overline{\psi}\sigma^{\mu\nu}\chi$&$\overline{\psi}\sigma_{\mu\nu}\chi$&$-\overline{\psi}\sigma_{\mu\nu}\chi$&$-\overline{\chi}\sigma^{\mu\nu}\psi$
&$-\overline{\chi}\sigma_{\mu\nu}\psi$&$\overline{\chi}\sigma^{\mu\nu}\psi$\\
\hline
\end{tabular}
\end{center}
\end{table}

\subsection{Basic Building Blocks of the SM}\label{subsec:1.2}
In this section we shall briefly present the building blocks of the SM, taking special attention to the relevant sector for flavour physics. Modern Quantum Field Theories are based on the gauge principle: \textit{The Lagrangian is invariant under a continuous group of local transformations. For each group generator there necessarily arises a corresponding vector field called the gauge field, responsible for ensuring the Lagrangian invariance under the local group transformations.}

Following the above principle, modern theories are developed through three simple steps: 
\begin{itemize}
\item[\textbf{(1)}] Define the gauge symmetry
\item[\textbf{(2)}] Choose the representations of the matter content under the symmetry
\item[\textbf{(3)}] Choose the way your original symmetry is broken 
\end{itemize}
The first two steps define the model in the unbroken phase. We then need a way to break this symmetry since at low energies we know that only charge (and colour) is manifestly preserved.

The best example satisfying the above three conditions and having an enormous success when confronting with data is the SM. The model construct upon the gauge group (step \textbf{(1)}) 
\begin{equation}
\mathcal{G}_{\mbox{\scriptsize{SM}}}=SU(3)_C\times SU(2)_L\times U(1)_Y\,.
\end{equation}
From the gauge principle, each generator of $\mathcal{G}_{\mbox{\scriptsize{SM}}}$ has a associated gauge vector field (first four lines of the table on the right in \Tref{tab:SM}). The known matter fields are embedded in irreducible representations of $\mathcal{G}_{\mbox{\scriptsize{SM}}}$ (step \textbf{(2)}) and are presented on the left table in \Tref{tab:SM}. 
\begin{table}[h]
\begin{center}
\caption{Standard model particle content, symmetry representations and forces.}
\label{tab:SM}
\begin{tabular}{cc}
\begin{tabular}{p{2.5cm}cl}
\hline\hline
\textbf{Matter}             & \textbf{Flavour}
                                                & $\mathcal{G}_{\mbox{\scriptsize{SM}}}$\\
\hline
$q_{L\alpha}\equiv
\begin{pmatrix}
u_{L\alpha}\\
d_{L\alpha}
\end{pmatrix}$  & 
$\begin{pmatrix}
\text{u}_{L}\\
\text{d}_{L}
\end{pmatrix}\,,\,
\begin{pmatrix}
\text{c}_{L}\\
\text{s}_{L}
\end{pmatrix}\,,\,
\begin{pmatrix}
\text{t}_{L}\\
\text{b}_{L}
\end{pmatrix}$             & $(\mathbf{3},\mathbf{2},1/6)$ \\
$u_{R\alpha}$  & 
$\text{u}_R\,,\,\text{c}_R\,,\,\text{t}_R$             & $(\mathbf{3},\mathbf{1},2/3)$ \\
$d_{R\alpha}$  & 
$\text{d}_R\,,\,\text{s}_R\,,\,\text{b}_R$             & $(\mathbf{3},\mathbf{1},-1/3)$ \\
$\ell_{L\alpha}\equiv
\begin{pmatrix}
\nu_{L\alpha}\\
e_{L\alpha}
\end{pmatrix}$&$
\begin{pmatrix}
\nu_{L\text{e}}\\
\text{e}_{L}
\end{pmatrix}\,,\,
\begin{pmatrix}
\nu_{L\mu}\\
\mu_{L}
\end{pmatrix}\,,\,
\begin{pmatrix}
\nu_{L\tau}\\
\tau_{L}
\end{pmatrix}$&$(\mathbf{1},\mathbf{2},-1/2)$\\
$e_{R\alpha}$  & 
$\text{e}_R\,,\,\mu_R\,,\,\tau_R$             & $(\mathbf{1},\mathbf{1},-1)$ \\
\hline\hline
\end{tabular}&
\begin{tabular}{ll}
\hline\hline
\textbf{Bosons}&\textbf{Force}\\
\hline
$G_\mu^a$&Strong\\
$W^\pm_\mu\,,\,Z^0_\mu$&Weak \\
$A_\mu$&EM\\
$\phi=
\begin{pmatrix}
\phi^+\\\phi^0
\end{pmatrix}$&$\begin{array}{l}\text{Yukawa-type}\\(\mathbf{1},\mathbf{2},1/2)\end{array}$\\
\hline\hline
\end{tabular}
\end{tabular}
\end{center}
\end{table}
The gauge fields interact with matter through the covariant derivative, which can be expressed in terms of the physical gauge bosons as  
\begin{equation}\label{niceD}
D_\mu=\partial_\mu-ig_sG_\mu^a\frac{\lambda_a}{2}-ig\left(W^+_\mu T_++W_\mu^- T_-\right)-ieA_\mu Q-\frac{ig}{c_W}Z_\mu^0\left(T_3-s_W^2 Q\right)\,,
\end{equation}
with $(T_\pm)_{ij}=(|\epsilon_{ij}|\pm\epsilon_{ij})/(2\sqrt{2})$ and $(T_{3})_{ij}=\delta_{ij}(-1)^{ij}/2$ for the $SU(2)$ doublet representations. The electric charge $Q$ is a linear combination of the generator of $U(1)_Y$ and the diagonal generator of $SU(2)_L$, and reads $Q=Y+T_3$. The full SM Lagrangian is now a combination of several ``distinct'' parts which can, in many scenarios, be studied separately. We write it as
\begin{equation}
\mathcal{L}_{{\mbox{\scriptsize{SM}}}}=\mathcal{L}_{\mbox{\scriptsize{Kin}}}^{\mbox{\scriptsize{gauge}}}
+\mathcal{L}_{\mbox{\scriptsize{Kin}}}^{\mbox{\scriptsize{fermion}}}+\mathcal{L}_{\mbox{\scriptsize{Higgs}}}
+\mathcal{L}_{\mbox{\scriptsize{Yukawa}}}+\mathcal{L}_{\mbox{\scriptsize{gf}}}
+\mathcal{L}_{\mbox{\scriptsize{FP}}}\,.
\end{equation}
The terms $\mathcal{L}_{\mbox{\scriptsize{gf}}}$ and $\mathcal{L}_{\mbox{\scriptsize{FP}}}$ denote the gauge fixing and Faddeev-Popov Lagrangian, respectively. While these contributions are very important for the self-consistency of the model, for flavour physics they play no role and, therefore, shall be ignored in these notes. The other Lagrangian terms are presented in \Tref{tab:Lag}. A useful summary of Feynman rules for the SM can be found in~\cite{Romao:2012pq}.
\begin{table}[h]
\begin{center}
\caption{Standard model Lagrangian equations for the four relevant sectors. 
With the following definitions: $G^a_{\mu\nu}=\partial_\mu G_\nu^a-\partial_\nu G_\mu^a+g_sf^{abc}G^a_\mu G^b_\nu\,,(a,b,c=1,...,8)$, $W^a_{\mu\nu}=\partial_\mu W_\nu^a-\partial_\nu W_\mu^a+g\epsilon^{abc}W^a_\mu W^b_\nu\,(a,b,c=1,...,3)$,
$B_{\mu\nu}=\partial_\mu B_\nu-\partial_\nu B_\mu$, $Y^{u,d,\ell}$ the up, down and charged-lepton Yuwaka coupling matrices and $\tilde{\phi}=i\tau_2\phi^\ast$.}
\label{tab:Lag}
\begin{tabular}{p{3cm}l}
\hline\hline
\textbf{Sector}&\textbf{Lagrangian}\\
\hline
$\mathcal{L}_{\mbox{\scriptsize{kin}}}^{\mbox{\scriptsize{gauge}}}\begin{array}{c}\\\\\end{array}$&$-\frac{1}{4}G^{a\mu\nu}G_{\mu\nu}^a-\frac{1}{4}W^{a\mu\nu}W_{\mu\nu}^{a}-\frac{1}{4}B^{\mu\nu}B_{\mu\nu}$\\
$\mathcal{L}_{\mbox{\scriptsize{kin}}}^{\mbox{\scriptsize{fermion}}}\begin{array}{c}\\\\\end{array}$&$\overline{q^0_{L\alpha}}iD\!\!\!\!/\, q^0_{L\alpha}+\overline{u^0_{R\alpha}}iD\!\!\!\!/\, u^0_{R\alpha}+\overline{d^0_{R\alpha}}iD\!\!\!\!/\, d^0_{R\alpha}+\overline{\ell^0_{L\alpha}}iD\!\!\!\!/\, \ell^0_{L\alpha}+\overline{e^0_{R\alpha}}iD\!\!\!\!/\, e^0_{R\alpha}$\\
$\mathcal{L}_{\mbox{\scriptsize{Higgs}}}\begin{array}{c}\\\\\end{array}$&$\left(D_\mu\phi\right)^\dagger\left(D^\mu\phi\right)
-V\left(\phi\right)$\\
$\mathcal{L}_{\mbox{\scriptsize{Yukawa}}}\begin{array}{c}\\\\\end{array}$&$-Y^d_{\alpha\beta}\,\overline{q^0_{L\alpha}}\phi d^0_{R\beta}-Y^u_{\alpha\beta}\,\overline{q^0_{L\alpha}}\tilde{\phi} u^0_{R\beta}-Y^\ell_{\alpha\beta}\,\overline{\ell^0_{L\alpha}}\phi e^0_{R\beta}+\text{h.c.}$\\
\hline\hline
\end{tabular}
\end{center}
\end{table}

In the SM, step \textbf{(3)} is is achieved through the scalar doublet field $\phi$, or Higgs field. In the Higgs sector, the Lagrangian $\mathcal{L}_{\mbox{\scriptsize{Higgs}}}$ contains the scalar potential $V(\phi)$ which has the general form
\begin{equation}\label{eq:sec1phipot}
V\left(\phi\right)=\mu^2_\phi\,\phi^\dagger\phi+\frac{\lambda_\phi}{2}(\phi^\dagger\phi)^2=\frac{\lambda_\phi}{2}\left(\phi^\dagger\phi+\frac{\mu_\phi^2}{\lambda_\phi}\right)^2+\text{const.}\,.
\end{equation}
The Higgs potential is responsible for the electroweak symmetry breaking $SU(2)_L\otimes U(1)_Y\rightarrow U(1)_Q$. This can be achieved spontaneously when the mass parameter $\mu_\phi^2$, in \Eref{eq:sec1phipot}, becomes negative. In this scenario $\langle\phi^\dagger\phi\rangle=0$ becomes a local maximum and the absolute minimum is shifted to the non-zero vacuum expectation value $\langle\phi^\dagger\phi\rangle\equiv v^2=-2\mu_\phi^2/\lambda_\phi$. The Higgs field can be rewritten in a more convenient basis, making use of the gauge freedom, in which only the physical components (the ones associated with physical particles) are present. This is known as the unitary gauge and the scalar doublet takes the form
\begin{equation}\label{eq:higgsunitary}
\phi=
\begin{pmatrix}
0\\
\dfrac{v+h}{\sqrt{2}}
\end{pmatrix}\,,\quad
\begin{array}{c}
\textbf{degrees}\\
\textbf{of}\\
\textbf{freedom}
\end{array}\textbf{:}
\left\{
\begin{array}{l}
\phi^+\text{ and }\text{Im}\{\phi^0\}\text{ are the Goldstone bosons. ``Rotated away'';}\\\\
\text{Re}\{\phi^0\}\text{ was shifted, such that }h\text{ represents the true}\\
\text{oscillations around the absolute minimum.}
\end{array}\right.
\end{equation}
In this basis it becomes clear that the gauge part of the kinetic term in the Higgs Lagrangian induces masses to some of the gauge bosons, i.e. to the ones associated with the broken generators,
\begin{equation}
\left(D_\mu\phi\right)^\dagger\left(D^\mu\phi\right)\sim m_W^2W^+_\mu W^{\mu-}+\frac{1}{2}m_Z^2Z_\mu^0Z^{\mu0}+\cdots\,,\quad\text{with:}\,
\left\{
\begin{array}{l}
m_W^2=\dfrac{g^2v^2}{4}\,,\quad m_Z^2=\dfrac{g^2v^2}{4c_W^2}\,,\\\\
m_A=0\quad\text{and}\quad m_G=0\,.
\end{array}\right.
\end{equation}
Before closing this short overview on the SM building blocks, it is useful to do a simple consistency check and look at the degrees of freedom in the process of spontaneous symmetry breaking (SSB). We can restrict ourself to the $SU(2)_L\otimes U(1)_Y\rightarrow U(1)_Q$ sector. Before SSB, the theory consists of one complex scalar doublet field (four degrees of freedom) and four gauge bosons (two degrees of freedom each); there are $4+2\times4=12$ degrees of freedom. After the SSB,  only $U(1)_{Q}$ remains as an explicit symmetry, i.e. only one generator leaves the vacuum invariant, so one would expect three Nambu-Goldstone bosons associated to the broken generators. Since we are working with a local gauge group, the Higgs mechanism allows these bosons to be absorbed as the longitudinal polarization of gauge bosons, $W^\pm$ and $Z^0$. So, in the end, we will have one real scalar field (one degree of freedom), three massive gauge bosons (three degrees of freedom each), and one massless gauge boson (the photon with two degrees of freedom). Summing up, after SSB there are $1+3\times3+2=12$ degrees of freedom, the same as in the unbroken phase.  

Note that no field except for the Higgs has a mass term in the unbroken phase. The Higgs mechanism is responsible for the mass generation of fermions and gauge bosons, but not of its own mass!

\subsection{The flavour structure of the SM}\label{subsec:1.3}
The origin of a non-trivial flavour structure in the SM is directly related with the presence of Yukawa interactions and gauge currents. The fermionic kinetic term is responsible for the weak charged currents (CC), weak neutral currents (NC) and for the electromagnetic neutral currents. They are given by 
\begin{subequations}
\begin{align}\label{LagraCC}
\text{\textbf{Charged Current:}}\quad&
\mathcal{L}_{\mbox{\scriptsize{CC}}}=\frac{g}{\sqrt{2}}\left(\overline{u^0_{L\alpha}}\gamma^\mu d^0_{L\alpha}W_\mu^++\overline{e^0_{L\alpha}}\gamma^\mu \nu^0_{L\alpha}W_\mu^-\right)+\text{h.c.}\,,\\
\nonumber\\
\label{LagraNC}
\text{\textbf{Neutral Current:}}\quad&\mathcal{L}_{\mbox{\scriptsize{NC}}}=eQ_f\overline{f^0}\gamma^\mu f^0A_\mu+\frac{g}{c_W}\overline{f^0}\gamma^\mu\left(g^f_V-g^f_A\gamma_5\right)f^0Z_\mu\,,
\end{align}
\end{subequations}
where
\begin{equation}
g^f_V=\frac{1}{2}T^f_3-s_W^2 Q_f\,,\quad g^f_A=\frac{1}{2}T^f_3\,,
\end{equation}
are the vector (V) and axial (A) couplings of the the gauge boson $Z^0$ to the fermions, respectively. The letter $f$ denotes any of the fermion fields. The charge of a fermion is denoted by $Q_f$, while $T_3^f$ denotes the weak isospin associated with the left-handed fermion.

When a theory has several fields with the same quantum numbers (flavours) one is free to rewrite the Lagrangian in terms of new fields, obtained from the original ones by means of a unitary transformation which mixes them. Why only unitary transformations? In principle, one can mix particles with the same quantum numbers in `any way' we want. However, by keeping it unitary we guarantee that the kinetic terms remain unaltered. This is important since having the kinetic Lagrangian with no cross terms, known as the canonical basis, allow us to easily identify our field content. We can define a set of transformations called weak basis transformations (WBTs) which are defined as transformations of the fermion fields which leave invariant the kinetic terms as well as the gauge interactions, i.e. they respect the gauge symmetry in the unbroken phase. The WBTs depend on the gauge theory that one is considering because, if there are more gauge interactions, then, in principle there will be less freedom to make WBTs. In the SM we define the WBTs as
\begin{equation}\label{eq:WBT}
\textbf{WBTs:}\quad
\left\{
\begin{array}{l}
q_L^0=W^q_L q_L^\prime\, ,\,u_R^0=W_R^u u_R^\prime\, ,\,d_R^0=W_R^d d_R^\prime\, ,\\\\
\ell_L^0=W^\ell_L \ell_L^\prime\, ,\,e_R^0=W^e_R e_R^\prime\, ,\\
\end{array}\right.\quad\longrightarrow\quad
\left\{
\begin{array}{l}
Y_u^{\prime}=W_L^\dagger Y_u W_R^u\, ,\\\\
Y_d^{\prime}=W_L^\dagger Y_d W_R^d\, ,\\\\
Y_e^{\prime}=W_L^{\ell\dagger} Y_e W_R^e\, .
\end{array}\right.
\end{equation}  
where $W_L^{q,\ell}$ and $W_R^{u,d,e}$ are $3\times 3$ unitary matrices acting in the flavour space. The transformed Yukawa matrices $Y_{u,d,e}^{\prime}$ have the same physical content as the original ones. To see the usefulness of WBTs let us start from a general basis where the mass matrix $Y_{u,d,e}$ have 18 free parameters each (9 modulus and 9 phases).  An arbitrary $n\times n$ complex matrix $A$ can be diagonalized by a bi-unitary transformation as $U_L^\dagger AV_R = \text{diag}$. This is known as single value decomposition. Using this information we can pass from a general basis to the new basis
\begin{equation}\label{eq:WBTIII}
\begin{array}{l}
\textbf{flavour basis I:}\\
\left\{
\begin{array}{l}
Y_u=U_L^u \lambda_u V_R^{u\dagger}\\\\
Y_d=U_L^d \lambda_d V_R^{d\dagger}\\\\
Y_e=U_L^e \lambda_e V_R^{e\dagger}
\end{array}\right.
\end{array}\quad
\begin{array}{c}
\textbf{WBTs}\\\\
W_L^q=U_L^d\,,\, W_R^u=V_R^u\,,\, W_R^d=V_R^d\\\\
W_L^\ell=U_L^e\,,\, W_R^e=V_R^e
\end{array}
\quad
\begin{array}{l}
\textbf{flavour basis II:}\\
\left\{
\begin{array}{l}
Y^\prime_u=V_{CKM}^\dagger \lambda_u \\\\
Y_d^\prime=\lambda_d \\\\
Y^\prime_e=\lambda_e \\
\end{array}\right.\,,
\end{array}
\end{equation}
with $\lambda_u=\text{diag}(y_u, y_c, y_t)$, $\lambda_d=\text{diag}(y_d, y_s, y_b)$ and $\lambda_e=\text{diag}(y_e, y_\mu, y_\tau)$ the real and positive fermion Yukawas (defined from the fermion masses, i.e. $y_f=\sqrt{2}m_f/v$), and $V_{\mbox{\scriptsize{CKM}}}=U_L^{u\dagger}U_L^d $.  This unitary matrix is the well known Cabbibo-Kobayashi-Maskawa (CKM) quark mixing matrix~\cite{Cabibbo:1963yz,Kobayashi:1973fv}. As we shall see in a while, this matrix only has four degrees of freedom. Therefore, in the flavour basis II we only have $6 \text{(masses)}+4\text{(mixing)}=10$ free parameters in the quark sector, mush less than in the general flavour basis I. Note that this is actually the minimal number of free parameters that one can have, since it is equal to the physical ones. Basis with less free parameters cannot be obtained by WBTs and they would have physical implications (correlations between physical observables). 

The WBTs become much a more fundamental aspect of the model when $Y^{u,d,\ell}\rightarrow 0$. In this limit the WBTs given in \Eref{eq:WBT} leave the whole Lagrangian invariant and therefore are promoted to symmetry generators of a global $U(3)^5$ symmetry 
\begin{equation}\label{eq:globalU3}
\mathcal{G}_{\mbox{\scriptsize{global}}}\equiv U(3)^5=SU(3)_q^{3}\times SU(3)^2_{l}\times U(1)^5\,,
\end{equation}
where 
\begin{equation}
SU(3)_q^3=SU(3)_{q_L}\times SU(3)_{u_R}\times SU(3)_{d_R}\quad\text{and}\quad
SU(3)_l^2=SU(3)_{\ell_L}\times SU(3)_{e_R}\,.
\end{equation}
In the presence of Yukawa terms only a reminiscent of the original global symmetry $\mathcal{G}_{\mbox{\scriptsize{global}}}$ remains unbroken. The easiest way to see which symmetry is left invariant is to look at the flavour basis II, introduced in \Eref{eq:WBTIII}, in which the number of parameters is reduced to the physical ones. In this basis the only field transformations that leave the Lagrangian invariant are rephasing rotations, and the presence of the $V_{\mbox{\scriptsize{CKM}}}$ matrix only allows one rotation in the quark sector. From this simple inspection we see that after the introduction of the Yukawa terms we are left with the residual symmetry
\begin{equation}\label{eq:residual}
\mathcal{G}_{\mbox{\scriptsize{global}}}\longrightarrow \mathcal{G}^{\mbox{\scriptsize{accidental}}}_{\mbox{\scriptsize{global}}}\equiv U(1)_B\times U(1)_{e}\times U(1)_{\mu}\times U(1)_{\tau}\,,
\end{equation}
with, of course, the gauge $U(1)_Y$ symmetry unbroken. These are called accidental symmetries, they were not imposed in the SM construction but end up appearing as a consequence of renormalizability and perturbativity. 

Looking at the WBTs as symmetry generators is actually very convenient in order to count the number of physical parameters present in the model. No mater which parameterization we choose for the SM flavour couplings $Y_{u,d,e}$, the number of physical parameters always remains unaltered. To learn how to count these parameters, let us first look at the charged lepton relevant flavour couplings $Y_{\alpha\beta}^e \overline{\ell^0_{L}}_\alpha \phi e^0_{R\beta}$. Our goal is to find out how many of the 18 real parameters are actually physical.  Now, if we look at the limit $Y^e\rightarrow 0$, we know that the Lagrangian will enjoy of a larger global symmetry, i.e. a $U(3)_{\ell_L} \times U(3)_{e_R}$ global symmetry. Another piece of information that is crucial is the residual symmetry of our model. Concerning the leptonic sector, as was seen above, we have the accidental $U(1)_e\times U(1)_\mu\times U(1)_\tau$. In other words, the presence of $Y_e$ induces the breaking
{\large\begin{equation}
\underbrace{\underbrace{\overbrace{U(1)_{\phi}}^{\text{Higgs}}\times\overbrace{U(3)_{\ell_L} \times U(3)_{e_R}}^{\text{Leptons}}}_{\text{1+9+9 generators}}\underset{Y_e}{\longrightarrow}\underbrace{U(1)_e\times U(1)_\mu\times U(1)_\tau\times U(1)_Y}_{\text{1+1+1+1 generators}}}_{\text{15 broken generators}},
\end{equation}}
leading to the existence of $15$ broken generators. We have included the Higgs and the hypercharge symmetries for completeness\footnote{Note that while in the SM these symmetries can be ignored in the process of counting broken generators, they play a crucial role in several extension of the SM.}. We can now use the broken generators to rotate $Y^e$ into a ``convenient'' symmetry-breaking direction. These rotations are nothing more that the WBTs described in \Eref{eq:WBTIII}, resulting in three physical parameters, i.e. the charged lepton masses. The result found in this simple exercise is actually more general and can be stated as follows
\begin{align}
\#\textbf{ Physical parameters }=\,\#\textbf{ Total parameters}-\#\textbf{ Broken generators}
\end{align}
Let us apply this result to the quark sector, we have
\begin{equation}
\begin{array}{l}
\#\textbf{ Total parameters: } \overbrace{(9+9)}^{Y_d} + \overbrace{(9+9)}^{ Y_u}=36\\\\
\#\textbf{ Broken generators: } \underbrace{3\times 9}_{U(3)^3}-\underbrace{1}_{U(1)_B}=26\\
\end{array}\quad\Longrightarrow\quad \#\textbf{ Physical parameters: } 10\,.
\end{equation}
Note that \Eref{eq:residual} is only true at the classical level since non-perturbative quantum effects break this down to just one abelian group $U(1)_{3B-L}$. However, this does not affect the parameter counting.

The Yukawa sector of the SM is responsible for the mass generation of the fermion species, after SSB. The fermion mass assignment in the SM is given by a Dirac mass term, $-m_f \bar{f}f=-m_f(\bar{f}_Lf_R+\bar{f}_Rf_L)$. Although it is invariant under $U(1)_{Q}$, the fermion mass term is not invariant under $SU(2)_L\otimes U(1)_Y$. Indeed, a fermion mass term is not a singlet under $SU(2)_L$, and, besides, the right- and left-handed components of $f$ have different weak hypercharges. As a result, no pure fermionic mass terms can be constructed consistently with gauge invariant principles, as it was mention in the prevous section. In the SM fermion masses can arise from Yukawa interactions with the scalar Higgs doublet, i.e the Lagrangian part $\mathcal{L}_{\mbox{\scriptsize{Yukawa}}}$. Using the Higgs filed given in \Eref{eq:higgsunitary}, one can see that the Yukawa Lagrangian splits into two parts, one relative to the fermion masses, $\mathcal{L}_{\mbox{\scriptsize{mass}}}$, and another corresponding to the interaction of the Higgs field with the fermions, $\mathcal{L}_{\mbox{\scriptsize{hff}}}$, 
\begin{subequations}
\begin{align}\label{eq:Lmass}
\textbf{Mass:}\quad&-\mathcal{L}_{\mbox{\scriptsize{mass}}}=M^e_{\alpha\beta}\,\overline{e^0_{L\alpha }} e^0_{R\beta}+ M^u_{\alpha\beta}\,\overline{u^0_{L\alpha }} u^0_{R\beta}+ M^d_{\alpha\beta}\,\overline{d^0_{L\alpha}}d^0_{R\beta }+\text{h.c.}\, ,\\
\nonumber\\\label{eq:Lhff}
\textbf{hff:}\quad&-\mathcal{L}_{\mbox{\scriptsize{hff}}}=\frac{1}{\sqrt{2}}Y^\ell_{\alpha\beta}\,\overline{e^0_{L\alpha }} e^0_{R\beta}\,h+ \frac{1}{\sqrt{2}}Y^u_{\alpha\beta}\,\overline{u^0_{L\alpha }} u^0_{R\beta}\,h+ \frac{1}{\sqrt{2}}Y^d_{\alpha\beta}\,\overline{d^0_{L\alpha}}d^0_{R\beta }\,h+\text{h.c.}\,,
\end{align}   
\end{subequations}
with the fermion mass matrices given by
\begin{equation}
M^{f}=\frac{v }{\sqrt{2}}Y^f\,,\quad \text{with}\quad f=\{u,d,e\}\,.
\end{equation}
At this stage it is worth pointing out that, in the SM, no renormalizable mass term for neutrinos can be constructed due to the absence of the right-handed fields $\nu_R$. Also, a particular feature of the SM is to have the mass terms proportional to the Yukawa couplings, leading to the absence of flavour changing neutral currents (FCNC) in the scalar sector. Extensions beyond SM in general ``struggle'', i.e. need additional assumptions beyond new particles, in order to reproduce this alignment~\cite{Glashow:1976nt}.
 
The Higgs mechanism breaks the $SU(2)_L$ group, which means that in the broken phase we are able to rotate the fields in the same $SU(2)_L$ multiplet through different unitary transformations. Therefore, we see from the new weak basis defined in \Eref{eq:WBTIII} that we can redefine the field $d_L$ as $d_L^\prime=V_{\mbox{\scriptsize{CKM}}}d_L$ such that the mass matrices are both diagonal and charged current sector becomes
\begin{equation}\label{eq:LCC}
\mathcal{L}_{\mbox{\scriptsize{CC}}}=\frac{g}{\sqrt{2}}\left(\overline{u_{L\alpha}}\left(V_{\mbox{\scriptsize{CKM}}}\right)_{
\alpha\beta}\gamma^\mu d_{L\beta}W_\mu^++\overline{e_{L\alpha}}\gamma^\mu \nu_{L\alpha}W_\mu^-\right)+\text{h.c.}\,,
\end{equation}
with
\begin{equation}\label{eq:CKMdef}
V_{\mbox{\scriptsize{CKM}}}\equiv U_L^{u\dagger}U_L^d=
\begin{pmatrix}
V_{ud}&V_{us}&V_{ub}\\
V_{cd}&V_{cs}&V_{cb}\\
V_{td}&V_{ts}&V_{tb}
\end{pmatrix}\,.
\end{equation}
The unitary matrix present in the leptonic sector is the identity matrix since $\nu_L^0$ can be rotated freely through a unitary transformation, due to the absence of a mass term. Therefore, in the SM the only tree-level flavour-changing interactions are present in the charged currents. Since the matrix $V_{\mbox{\scriptsize{CKM}}}$ is a $3\times 3$ unitary matrix, it has $9$ free parameters. However, the additional freedom
\begin{equation}
V_{\mbox{\scriptsize{CKM}}}\longrightarrow K_u^\dagger V_{\mbox{\scriptsize{CKM}}}K_d\,,
\end{equation} 
with $K_{u,d}$ phase diagonal matrices, reflecting the freedom in redefining the phases of the quarks in the mass basis, leads to $4$ mixing parameters. Therefore, as stated before the weak basis in \Eref{eq:WBTIII} has $4\,\text{mixing}+6\,\text{masses}=10$ parameters. This is known as the quark physical basis, since the number of free parameters coincides with the number of physical ones. Working in the mass eigenbasis, i.e. in the basis where the mass matrix of the fermions are real and positive, one can shift all the non-trivial flavour structure into the charged current sector. This is a very convenient basis to work in, since the fermion propagation gets quite simple. Still, we could opt to work in another basis at the cost of introducing extra complexity in the model. 

In the SM $CP$ violation shows up in the complex Yukawa couplings. If we $CP$ conjugate a typical Yukawa term we get, see \Tref{tab:Discrete},
\begin{equation}
\mathcal{CP}\,\left( \overline{\psi_{L\alpha}}\phi\psi_{R\beta}\right)\mathcal{CP}^\dagger=\overline{\psi_{R\beta}}\phi^\dagger \psi_{L\alpha}\,.
\end{equation} 
We then see that by requesting $CP$ invariance in the Yukawa sector we get
\begin{equation}
\mathcal{CP}\mathcal{L}_{\mbox{\scriptsize{Yuk}}}\mathcal{CP}^\dagger=\mathcal{L}_{\mbox{\scriptsize{Yuk}}}\quad
\Rightarrow\quad Y_{\alpha\beta}=Y_{\alpha\beta}^\ast\,,
\end{equation}
i.e. real Yukawa couplings are the necessary condition for $CP$-invariance. We can do the same exercise but now for the charged current Lagrangian, in the mass eigenbasis,
\begin{equation}
\mathcal{CP}\mathcal{L}_{\mbox{\scriptsize{CC}}}\mathcal{CP}^\dagger=\mathcal{L}_{\mbox{\scriptsize{Yuk}}}\quad
\Rightarrow\quad V_{\alpha\beta}=V_{\alpha\beta}^\ast\,,
\end{equation}
i.e. real CKM mixing matrix as the necessary condition. Therefore, the complex nature of the Yukawa couplings (or CKM mixing matrix) is the origin of CP violation in the SM. The above results are basis dependent. We know, that there are always phases that can be rotated away. So the question is whether we have a basis independent way of checking for $CP$ violation. The answer is yes, the above conclusions can be formulated in a basis invariant way through the quantity~\cite{Bernabeu:1986fc}
\begin{equation}
\text{Tr}[H_u,H_d]^3=6i\sum_{\alpha,\beta=u,c,t,...}\sum_{\alpha^\prime,\beta^\prime=d,s,b,...}=m_\alpha^4 m_\beta^2 m_{\alpha^\prime}^4 m_{\beta^\prime}^2\, \text{Im}Q_{\alpha \alpha^\prime \beta \beta^\prime}
\end{equation}
where
\begin{equation}
Q_{\alpha\alpha^\prime \beta\beta^\prime}\equiv V_{\alpha\alpha^\prime}V_{\beta\beta^{\prime}}
V_{\alpha\beta^\prime}^\ast V_{\beta\alpha^{\prime}}^\ast
\end{equation}
is the rephasing-invariant quartet. For three generations, the above invariant reads
\begin{equation}
\text{Tr}[H_u,H_d]^3=6i(m_t^2-m_c^2)(m_t^2-m_u^2)(m_c^2-m_u^2)(m_b^2-m_s^2)(m_b^2-m_d^2)(m_s^2-m_d^2)\, J\,,
\end{equation}
with $J\equiv \text{Im} Q_{uscb}=\text{Im}[ V_{us} V_{cb}V_{ub}^\ast V_{cs}^\ast]$ known as the Jarlskog invariant~\cite{Jarlskog:1985ht}. The CKM-mechanism is the origin of $CP$ violation in the SM and lead to the nobel prize attribution in 2008 to Kobayashi and Maskawa who were the first to propose three flavours of quarks as the origin of CP violation~\cite{Kobayashi:1973fv}.

Different parametrizations for the CKM mixing matrix can be used. We shall follow the standard procedure and use the Particle Data Group (PDG) parametrization~\cite{Agashe:2014kda} 
\begin{align}\label{eq:CKMPDG}
\begin{split}
V_{\mbox{\scriptsize{CKM}}}&=R_1(\theta_{23})\Gamma(\delta)R_2(\theta_{13})
\Gamma(-\delta)R_3(\theta_{12})\\
&=
\begin{pmatrix}
c_{12}c_{13} & s_{12}c_{13} &s_{13} e^{-i\delta} \\
-s_{12}c_{23}-c_{12}s_{23}s_{13}e^{i\delta}&c_{12}c_{23}-s_{12}s_{23}s_{13}e^{i\delta}&s_{23}c_{13}\\
s_{12}s_{23}-c_{12}c_{23}s_{13}e^{i\delta} & -c_{12}s_{23}-s_{12}c_{23}s_{13}e^{i\delta}& c_{23}c_{13}
\end{pmatrix}
\end{split}
\end{align}
where $c_{ij}\equiv \cos{\theta_{ij}}$, $s_{ij}\equiv \sin{\theta_{ij}}$, $R(\theta_{ij})$ is the rotation in the plane $i-j$ and $\Gamma(\delta)=\text{diag}(1,1,e^{i\delta})$. The three $s_{ij}$ are the real mixing parameters and $\delta$ is the Kobayashi-Maskawa phase. While the range of this phase is $0\leq \delta <2\pi$, the measurements of $\mathcal{CP}$ violation in $K$
decays force it to be in the range $0<\delta<\pi$. From experiments we know that there exists a strong hierarchy on the mixing angles, i.e. $s_{13}\ll s_{23}\ll s_{12}\ll 1$. We can write the mixing angles as
\begin{align}
\begin{split}
s_{12}=&\lambda=\frac{|V_{us}|}{\sqrt{|V_{ud}|^2+|V_{us}|^2}}\,,\quad s_{23}=A\lambda^2=\lambda\left|\frac{V_{cb}}{V_{us}}\right|\,,\\
s_{13}e^{i\delta}=&V_{ub}^\ast=A\lambda^3(\rho+i\eta)=\frac{A\lambda^3(\bar{\rho}+i\bar{\eta})\sqrt{1-A^2\lambda^4}}{\sqrt{1-\lambda^2}[1-A^2\lambda^4(\bar{\rho}+i\bar{\eta})]}\,.
\end{split}
\end{align}
With these relations we ensure that
\begin{equation}
\bar{\rho}+i\bar{\eta}=-\frac{V_{ud}V_{ub}^\ast}{V_{cd}V_{cb}^\ast}
\end{equation}
is independent of any phase convention. The above expression allows us to express the CKM matrix in terms of: $\lambda$, $A$, $\bar{\rho}$ and $\bar{\eta}$. While the parametrization in term of these parameter is exact, it is common to approximate this result for small $\lambda$. Up to fourth power corrections, we can expand the bar parameters as $\bar{\rho}=\rho(1-\lambda^2/2)$ and $\bar{\eta}=\eta(1-\lambda^2/2)$ known as Wolfenstein parametrization~\cite{Wolfenstein:1983yz}
\begin{equation}
V_{\mbox{\scriptsize{CKM}}}=
\begin{pmatrix}
1-\lambda^2/2&\lambda&A\lambda^3(\rho-i\eta)\\
-\lambda&1-\lambda^2/2&A\lambda^2\\
A\lambda^3(1-\rho-i\eta)&-A\lambda^2&1
\end{pmatrix}+\mathcal{O}(\lambda^4)\,.
\end{equation}
The unitarity on the CKM matrix implies relations between its entries:
\begin{align}
\begin{split}
\textbf{Columns Orthogonality:}&\quad \sum_i V_{ij}V_{ik}^\ast=\delta_{jk}\,,\\
\textbf{Rows Orthogonality:}&\quad \sum_i V_{ij}V_{kj}^\ast=\delta_{ik}\,.
\end{split}
\end{align}
The six vanishing combinations are sums of complex number, so that they can be represented as triangles in the complex plane. The most used triangle is given by
\begin{figure}[h]
\centering\includegraphics[width=.7\linewidth]{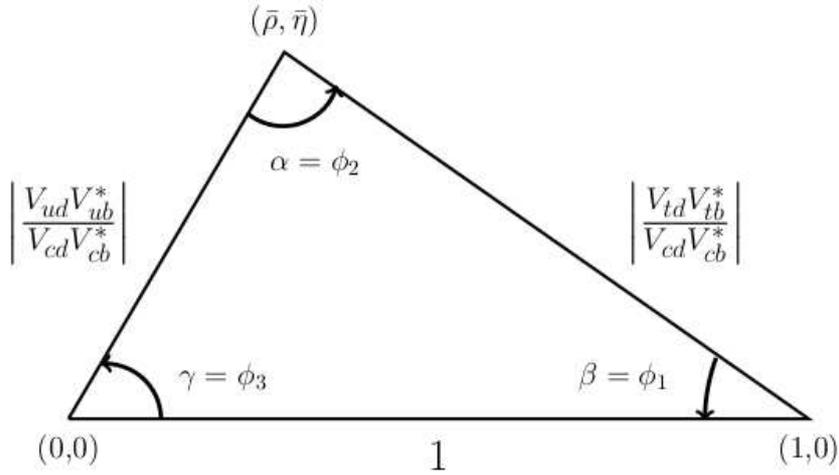}
\caption{Unitary triangle representation in the complex plane $\bar{\rho},\bar{\eta}$}
\label{fig:1}
\end{figure}
\begin{equation}
V_{ud}V_{ub}^\ast+V_{cd}V_{cb}^\ast+V_{td}V_{tb}^\ast=0\,,.
\end{equation}
In \Fref{fig:1} we have divided each side by the best-known value, i.e. $V_{cd}V_{cb}^\ast$. The angles of the unitary triangle are also represented in \Fref{fig:1} and are given by
\begin{equation}
\beta=\phi_1=\text{arg}\left(-\frac{V_{cd}V_{cb}^\ast}{V_{td}V_{tb}^\ast}\right)\,,\quad
\alpha=\phi_2=\text{arg}\left(-\frac{V_{td}V_{tb}^\ast}{V_{ud}V_{ub}^\ast}\right)\,,\quad
\gamma=\phi_3=\text{arg}\left(-\frac{V_{ud}V_{ub}^\ast}{V_{cd}V_{cb}^\ast}\right)\,.
\end{equation} 
Measurements of $CP$-violating observables can constraint these angles and also the parameters $\bar{\eta}$, $\bar{\rho}$. Using the Wolfenstein parametrization as a give line, we can get simpler expressions for the unitary triangle angles
\begin{align}
\begin{split}
\beta=&\pi+\text{arg}(V_{cd}V_{cb}^\ast)-\text{arg}(V_{td}V_{tb}^\ast)\simeq -\text{arg}(V_{td})\,,\\\\
\gamma=&\pi+\text{arg}(V_{ud}V_{ub}^\ast)-\text{arg}(V_{cd}V_{cb}^\ast)\simeq -\text{arg}(V_{ub})\,.
\end{split}
\end{align} 
With the help of the unitary triangle where the $d$-quark is replaced by the $s$-quark, i.e.
\begin{equation}
V_{us}V_{ub}^\ast+V_{cs}V_{cb}^\ast+V_{ts}V_{tb}^\ast=0\,,
\end{equation}
we can define another angle
\begin{equation}
\beta_s=\text{arg}\left(-\frac{V_{ts}V_{tb}^\ast}{V_{cs}V_{cb}^\ast}\right)=\pi+\text{arg}(V_{ts}V_{tb}^\ast)-\text{arg}(V_{cs}V_{cb}^\ast)\simeq \pi+\text{arg}(V_{ts})\,.
\end{equation}
This allow us to write the CKM mixing matrix up to $\mathcal{O}(\lambda^5)$ as 
\begin{equation}
V_{\mbox{\scriptsize{CKM}}}\simeq 
\begin{pmatrix}
|V_{ud}|&|V_{us}|&|V_{ub}|e^{i\gamma}\\
-|V_{cd}|&|V_{cs}|&|V_{cb}|\\
|V_{td}|e^{i\beta}&-|V_{ts}|e^{i\beta}&|V_{tb}|
\end{pmatrix}\,.
\end{equation}
The area of all triangles is the same and is given by half of the absolute value of the Jarlskog invariant, i..e $\text{Area}_\Delta=|J|/2$. The Jarlskog invariant in the parametrizations presented above take the form
\begin{equation}
J=\text{Im}\left[V_{ud}V_{cs}V_{us}^\ast
V_{cd}^\ast\right]=\frac{1}{8}\sin(2\theta_{12})\sin(2\theta_{13})\sin(2\theta_{23})\sin\delta\simeq A^2\lambda^6\eta\,.
\end{equation}
The absolute values of the CKM matrix can be found in the following processes:
\begin{itemize}
\item[$\bullet$] $|V_{ud}|$: $\beta$-decay $(A,Z)\rightarrow (A,Z+1)+e^-+\bar{\nu}_e$;

\item[$\bullet$] $|V_{us}|$: $K$-decay $K^+\rightarrow \pi^0+\ell^++\nu_\ell$;

\item[$\bullet$] $|V_{cd}|$: $\nu$-production of $c$'s $\nu_\ell+d\rightarrow \ell^-+c$;

\item[$\bullet$] $|V_{cs}|$: charm decay $D^\pm\rightarrow K^0+\ell^\pm+\nu_\ell$;

\item[$\bullet$] $|V_{ub}|$: $B$-decay $b\rightarrow u+\ell^-+\bar{\nu}_\ell$;

\item[$\bullet$] $|V_{cb}|$: $B$-decay $b\rightarrow c+\ell^-+\bar{\nu}_\ell$;

\item[$\bullet$] $|V_{td}|$ and $|V_{ts}|$ : $\Delta m$ in $B^0-\overline{B^0}$;

\item[$\bullet$] $|V_{tb}|$: top decays.
\end{itemize}
The result of a global fit gives~\cite{Agashe:2014kda}
\begin{equation}
|V_{CKM}|=
\begin{pmatrix}
0.97427\pm 0.00014&0.22536\pm0.00061&0.00355\pm0.00015\\
0.22522\pm0.00061&0.97343\pm0.00015&0.0414\pm 0.0012\\
0.00886^{+0.00033}_{-0.00032}&0.0405^{0.0011}_{-0.0012}&0.99914\pm0.00005
\end{pmatrix}
\end{equation}
or in terms of the Wofenstein parameters
\begin{equation}
\lambda=0.22537\pm0.00061\,,\quad A=0.814^{+0.023}_{-0.024}\,,\quad \bar{\rho}=0.117\pm 0.021\quad\text{and}\quad \bar{\eta}=0.353\pm 0.013\,.
\end{equation}
The Jarlskog invariant is $J=(3.06^{+0.21}_{-0.20})\times 10^{-5}$. The angles of the unitary triangle can be tested in $B$-decays:
\begin{itemize}
\item[$\bullet$] $\sin 2\beta$: $B_d^0\rightarrow J/\Psi K_S$

\item[$\bullet$] $\sin 2\alpha$: $B_d^0\rightarrow \pi^+\pi^-$

\item[$\bullet$] $\sin 2\gamma$: $B_s^0\rightarrow D^{\pm}_S K^{\mp}$
\end{itemize}

\subsection{GIM mechanism}\label{subsec:1.4}

We have learned that the structure of the SM is such that it ensures the absence of the tree level flavour changing neutral currents. Both neutral gauge boson and Higgs boson couplings are diagonal in the flavour mass eigenstate basis. Thus, the flavour changing neutral-current processes involving quarks are generated in higher orders in the electroweak interactions. Since they are strongly suppressed in Nature, it is interesting to discuss the predictions for them in the electroweak theory. For the quark sector, the generic examples of flavour changing neutral-current transitions are the reactions:
\begin{itemize}
\item[$\bullet$] $d \bar{s}\rightarrow \bar{d}s$ $(\Delta S = 2)$,\quad $b\bar{d}\rightarrow \bar{b}d$ $(\Delta B= 2 )$;
\item[$\bullet$] $s \rightarrow d \gamma$ $(\Delta S = 1)$,\quad $b \rightarrow s \gamma$ $(\Delta B = 1)$. 
\end{itemize}
Such transitions are responsible for physical processes like $K^0-\overline{K^0}$ and $B^0-\overline{B^0}$ mixing, for radiative flavour changing decays of strange and bottom mesons and for decays like $K\rightarrow \pi e^+ e^-$ or $B \rightarrow K^\ast e^+ e^-$.
\begin{figure}[h]
\centering\includegraphics[width=.9\linewidth]{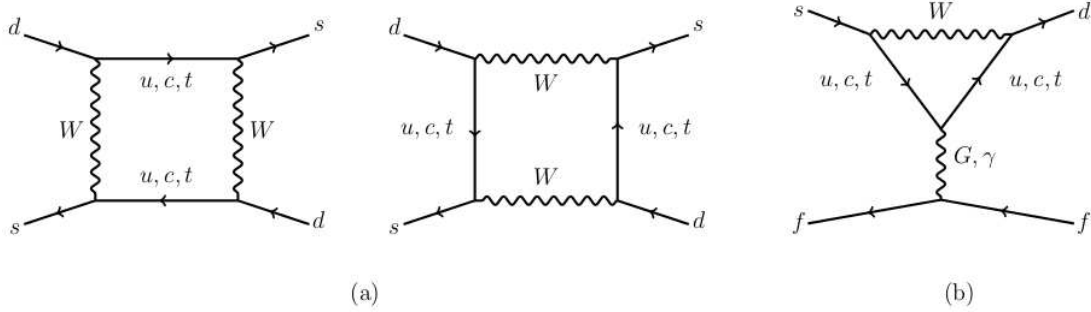}
\caption{In (a) $\Delta S=2$ box diagrams. In (b) $\Delta S=1$ penguin contribution.}
\label{fig:2}
\end{figure}
On dimensional grounds, we then get the following estimate for the $\bar{s}d\rightarrow \bar{s}d$
transition amplitude, depicted in \Fref{fig:2}a , with double $W$-boson, $u$- and/or $c$-quark exchange (the contribution from the top quark exchange is strongly suppressed by its very small mixing with the first two generations of quarks):
\begin{align}
\begin{split}
A\sim&\left(\frac{e}{\sqrt{2}s_W}\right)^4\frac{1}{M_W^2}\sum_{i,j=u,c}V_{is}^\ast V_{id}V_{js}^\ast V_{jd}\left[1+\mathcal{O}\left(\frac{m_{q_i}^2}{M_W^2},\frac{m_{q_j}^2}{M_W^2}\right)\right]\\
\sim&\alpha G_F\left[(V_{ts}^\ast V_{td})^2+\mathcal{O}\left(\sum_{i,j=u,c}V_{is}^\ast V_{id}V_{js}^\ast V_{jd}\frac{m_q^2}{M_W^2}\right)\right]
\end{split}
\end{align}
In the last step we have used the CKM unitarity condition: $\sum_{i,j=u,c}V_{is}^\ast V_{id}=-V_{ts}^\ast V_{td}$. We then see that the leading term is suppressed by very small CKM angles as the double top quark exchange contribution. The remaining terms, which are proportional to larger CKM angles, are in turn suppressed by light quark masses.

Such a mechanism of suppression of the flavour changing neutral-current amplitudes is known as the Glashow-Iliopoulos-Maiani (GIM) mechanism~\cite{Glashow:1970gm}. The strong suppression of the flavour changing neutral-current transitions is indeed a SM prediction. However, this follows not only from the structure of the theory but also depends on the empirical pattern of the quark masses and mixing angles. Therefore, from the SM point of view, the successful predictions for the flavour changing neutral-current processes are rather accidental. 
%and probably is the new physics the one responsible for it. 

Let us now look at the $\Delta F= 1$ transitions at the qualitative level. At one-loop, they receive contributions from box diagrams and also from the so-called penguin diagrams like in \Fref{fig:2}b. The corresponding amplitude goes as
\begin{equation}
A\sim\alpha G_F\sum_{i,j=u,c}V_{id}^\ast V_{is}\ln\frac{m_{q_i}^2}{M_W^2}+\mathcal{O}(V_{td}^\ast V_{ts})=\alpha G_F V_{ud}^\ast V_{us} \ln\frac{m_u^2}{m_c^2}+\mathcal{O}(V_{td}^\ast V_{ts})\,.
\end{equation}
Note that the dimensionless coefficient of the first term contains logarithms of light quark masses. Since the masses of the up and charm quarks are quite different, there is no additional suppression except for the usual one in this case (unlike the previously considered box diagrams). We can then say that the GIM mechanism is power-like in the case of box diagrams, but only logarithmic in the case of certain penguin diagrams.

\section{Effective theories and their use in flavour physics}\label{sec:2}

\begin{figure}[h]
\centering\includegraphics[width=.9\linewidth]{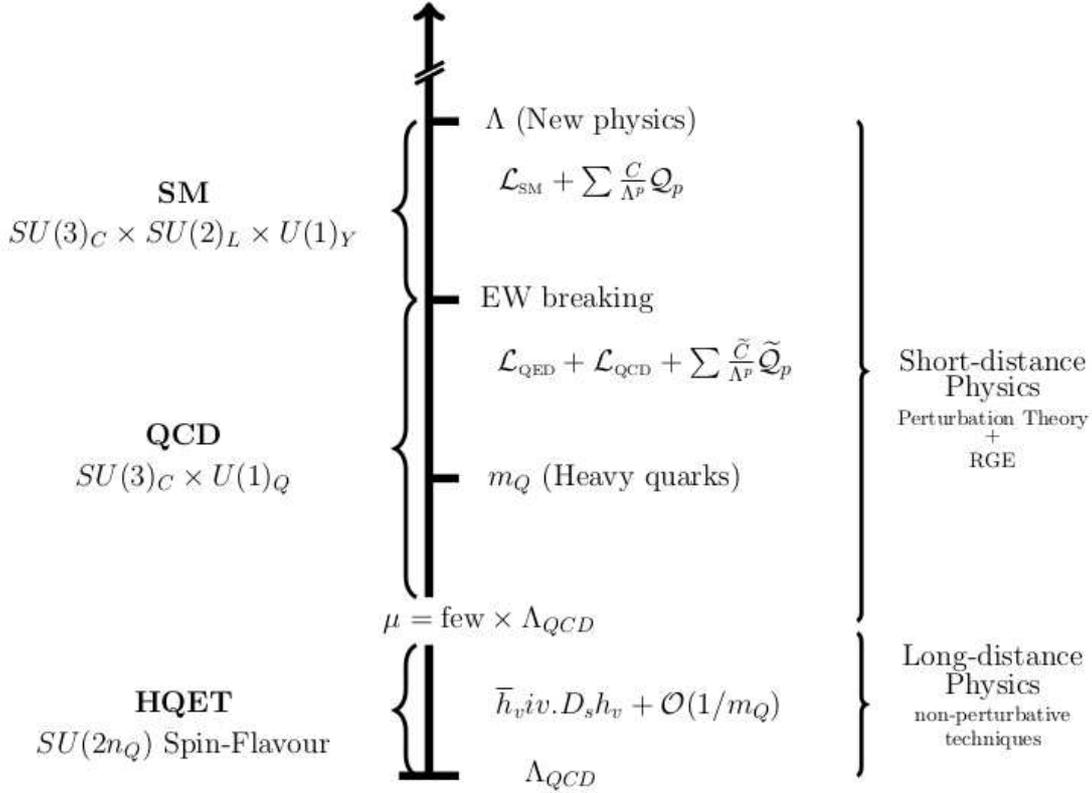}
\caption{General schematic idea behind effective field theories}
\label{fig:4}
\end{figure}
Effective field theory formalism is a very powerful tool when several scales are present in a quantum field theory. The principle in effective field theories is to just include the appropriate degrees of freedom to describe physical phenomena occurring at a given scale. By
%ignoring substructure and
integrating out
degrees of freedom at shorter distances we
% are hopefully simplifying
try to simplify 
the model at longer distances. This approach works best when there is a large separation between length scale of interest and the length scale of the underlying dynamics. Figure~\ref{fig:4} summarizes the general philosophy behind this approach. We summarize the effective field theory formalism in three simple steps~\cite{Neubert:2005mu}:
\begin{itemize}
\item[$\bullet$] \textbf{Step 1:} Choose a cutoff scale $\Lambda \lesssim M$ (with $M$ some fundamental scale) and divide the field into high- and low-frequency modes, i.e.
\begin{equation}
\phi=\underbrace{\phi_H}_{
\begin{array}{c}
\text{Fourier modes}\\
\omega>\Lambda
\end{array}}+\underbrace{\phi_L}_{
\begin{array}{c}
\text{Fourier modes}\\
\omega<\Lambda
\end{array}}\,.
\end{equation}
The component $\phi_L$ describes the low-energy physics through the correlation functions
\begin{equation}
\langle 0|T\{\phi_L(x_1)\cdots \phi_L(x_n)\}|0\rangle=\frac{1}{Z[0]}\left.\left(-i\frac{\delta}{\delta J_L(x_1)}\right)\cdots \left(-i\frac{\delta}{\delta J_L(x_n)}\right)Z[j_L]\right|_{J_L=0}\,,
\end{equation} 
where the generating functional is
\begin{equation}
Z[J_L]=\int \mathcal{D}\phi_L\mathcal{D}\phi_H\, e^{i\mathcal{S}(\phi_L,\phi_H)+i\int d^Dx J_L(x)\phi_L(x)}\quad\text{and}\quad \mathcal{S}(\phi_L,\phi_R)=\int d^Dx\mathcal{L}(x)\,.
\end{equation}
We have used $D$ for the space-time dimension and only the external source of the low-frequency modes is relevant for the correlation functions computed at low energy.

\item[$\bullet$] \textbf{Step 2:} Integrate out the high-frequency modes below the scale $\Lambda$, i.e.
\begin{equation}
Z[J_L]\equiv \underbrace{\int\mathcal{D}\phi_L\, e^{i\mathcal{S}_\Lambda (\phi_L)+i\int d^Dx J_L(x)\phi_L(x)}}_{\text{No $\phi_H$ dependence}}\quad\text{and}\quad e^{iS_\Lambda (\phi_L)}=\int\mathcal{D}\phi_H e^{i\mathcal{S}(\phi_L,\phi_H)}\,.
\end{equation}
The action $S_\Lambda(\phi_L)$ is known as ``Wilsonian effective action'', which is non-local on scale $\Delta x^\mu\sim 1/\Lambda$ and depends on the choice made for the cutoff scale $\Lambda$.

\item[$\bullet$] \textbf{Step 3:} Expand the non-local action in terms of local operators composed of light fields, which is known as operator-product expansion (OPE). This expansion is possible in the low-energy regime, i.e. $E\ll \Lambda$, and leads to
\begin{equation}
S_{\mbox{\scriptsize{$\Lambda$}}}(\phi_L)=\int d^Dx \mathcal{L}_{\mbox{\scriptsize{$\Lambda$}}}^{\mbox{\scriptsize{eff}}}(x)\,,\quad \text{with}\quad \underbrace{\mathcal{L}_\Lambda^{\mbox{\scriptsize{eff}}}(x)=\sum_i \overbrace{C_i}^{
\begin{array}{c}
\text{Wilson}\\
\text{coeff.}
\end{array}} \overbrace{\mathcal{Q}_i(\phi_L(x))}^{
\begin{array}{c}
\text{local}\\
\text{operator}
\end{array}}}_{\text{Effective Lagrangian}}
\end{equation}

\end{itemize} 
The procedure described above is quite general and powerful, allowing us to obtain the Lagrangian relevant for a given scale. However, the effective Lagrangian is a sum of infinite operators which would naively destroy the predictability of the effective theory. In order to understand why this is not the case one can use the remarkably simple and powerful ``naive dimensional analysis'' (NDA) approach:
\begin{equation}
\begin{array}{c}
\textbf{NDA:}\\
(c=\hbar=1)
\end{array}
\quad \left\{
\begin{array}{l}
[m]=[E]=[p]=[x^{-1}]=[t^{-1}]=1\\\\
\text{Assuming }[C_i]=-\gamma_i
\end{array}\right.\quad\text{then}\quad C_i=g_iM^{-\gamma_i}\,.
\end{equation}
The coupling $g_i$ is dimensionless and form ``naturalness'' $\mathcal{O}(1)$, while $M$ is the fundamental energy scale of the theory. Taken for simplicity the effective Lagrangian dimensionless, the effective operator $\mathcal{Q}_i$ scales for $E\ll \Lambda <M$ as
\begin{equation}
g_i\left(\frac{E}{M}\right)^{\gamma_i}=
\left\{
\begin{array}{ll}
\mathcal{O}(1)&\quad \text{if }\gamma_i=0\\
\ll1&\quad \text{if }\gamma_i>0\\
\gg1&\quad \text{if }\gamma_i<0
\end{array}
\right.
\end{equation}
This tell us that only the couplings that have $\gamma_i<0$ are relevant. Therefore, given a precision goal we can truncate the series in $\mathcal{L}_\Lambda$ in  a given order in $E/M$. This implies a finite number of operators, which brings back the predictability of the effective theory. The dimension $\gamma_i$ can change due to interactions, this is known as anomalous dimension. We can be more formal and require the action to be dimensionless. In this case if $\delta_i=[\mathcal{O}_i]$ the coefficient dimension is $\gamma_i=\delta_i-D$. We summarize the operator relevance classification in \Tref{tab:dimension}.
\begin{table}[h]
\begin{center}
\caption{Classification of operators based on their dimension.}
\label{tab:dimension}
\begin{tabular}{lcl}
\hline\hline
\textbf{Dimension}&\textbf{Importance for $E\rightarrow 0$}&\textbf{Terminology}\\
\hline
\multirow{3}{*}{$\delta_i<D,\, \gamma_i<0$}&\multirow{3}{*}{grows}&Relevant operators (super-renormalizable)\\
&&$\bullet$ usually unimportant;\\
&&$\bullet$ protected by symmetries\\
\hline
\multirow{2}{*}{$\delta_i=D,\, \gamma_i=0$}&\multirow{2}{*}{constant}&Marginal operators (renormalizable)\\
&&$\bullet$ renormalizable QFT\\
\hline
\multirow{3}{*}{$\delta_i>D,\, \gamma_i>0$}&\multirow{3}{*}{falls}&Irrelevant operators (non-renormalizable)\\
&&$\bullet$ the most important (relevant)\\
&&$\bullet$ sensitive to fundamental scale \\
\hline\hline
\end{tabular}
\end{center}
\end{table}

As a final comment note that while most of the time $\phi_H$ is identified with a heavy particle, the method presented above is much more general. As opposed to integrate out some heavy particle, we can work on a scenario where only light particles are present. In this case we can lower the cutoff scale $\Lambda$ by a small amount $\Lambda-\delta\Lambda$ and integrate out high frequencies of the light particle. This implies that the operators $\mathcal{O}_i(\phi_L)$ will remain the same, as no contribution from extra particles are present. And the effects of lowering the cutoff scale must enter into the effective couplings $C_i(\Lambda)$. This approach gives an intuitive understanding of the running of the coupling constants.

\subsection{Weak currents and OPE}\label{subsec:2.1}
Hadrons can decay through weak interaction mediation, between their quark constituents. The typical binding energy of quarks in hadrons is $\mathcal{O}(1\, \text{GeV})$, much below the weak scale $\mathcal{O}(M_{W,Z})$. The idea behind the OPE treatment is to start from short-distance dynamics and refine it step-by-step with non-perturbative corrections. Let us look at the part of generating functional containing the $W$ boson~\cite{Buras:1998raa}, i.e.
\begin{equation}
Z_{\mbox{\scriptsize{W}}}\sim \int [dW^+][dW^-]\text{Exp}\left(i\int d^4x \mathcal{L}_W\right)\,,
\end{equation} 
with
\begin{align}
\begin{split}
\mathcal{L}_{\mbox{\scriptsize{W}}}=&-\frac{1}{2}\left(\partial_\mu W_\nu^+-\partial_\nu W_\mu^+\right)\left(\partial^\mu W^{-\nu}-\partial^\nu W^{-\mu}\right)+M_W^2W^+_\mu W^{-\mu}\\
&+\frac{g_2}{2\sqrt{2}}(J_\mu^+W^{+\mu}+J_\mu^-W^{-\mu})
\end{split}
\end{align}
the Lagrangian density containing the kinetic terms of the $W$ boson and its interactions with charged currents. These interactions can be extracted from \Eref{eq:LCC}. Since we are not interested in $W$ as external sources, we have omitted gauge self-interactions. Following the usual procedure in QFT, we can perform a Gaussian functional integration which leads us to a non-local action for quarks
\begin{equation}
\mathcal{S}_{\mbox{\scriptsize{nl}}}=\int d^4x \mathcal{L}_{kin}-\frac{g_2^2}{8}\int d^4xd^4y\,J_\mu^-(x)\Delta^{\mu\nu}(x,y)J_\nu^+(y)\,,
\end{equation}
where $\Delta^{\mu\nu}(x,y)$ is the $W$ boson propagator. In the unitary gauge it reads
\begin{equation}
\Delta^{\mu\nu}(x,y)=\int \frac{d^4k}{(2\pi)^4}\Delta_{\mu\nu}(k)e^{-ik(x-y)}\,,\quad 
\Delta^{\mu\nu}(k)=\frac{-1}{k^2-M_W^2}\left(g_{\mu\nu}-\frac{k_\mu k_\nu}{M_W^2}\right)\,.
\end{equation}
The idea now is to formally expand in $1/M_W^2$ powers the propagator, which allows us to get a local action. To lowest order the propagator becomes
\begin{equation}
\Delta^{\mu\nu}(x,y)\simeq \frac{g^{\mu\nu}}{M_W^2}\delta^{(4)}(x-y)\,,
\end{equation} 
which in turns lead to the effective Hamiltonian
\begin{equation}
\mathcal{H}_{\mbox{\scriptsize{eff}}}=-\frac{G_F}{\sqrt{2}}J_\mu^-J^{+\mu} (x)=-\frac{G_F}{\sqrt{2}}V^\ast_{\alpha\beta}V_{\alpha^\prime\beta^\prime}(\overline{d_\alpha}u_{\beta})_{V-A}(\overline{d_{\alpha^\prime}}u_{\beta^\prime})_{V-A}\,.
\end{equation} 
\begin{figure}[h]
\centering\includegraphics[width=.7\linewidth]{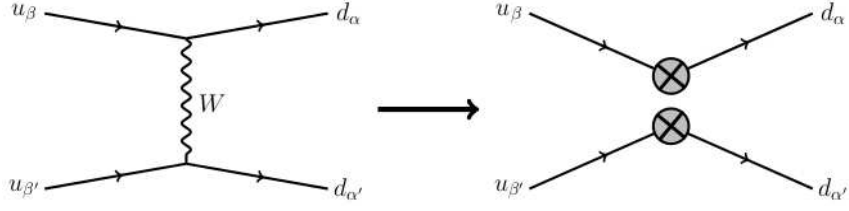}
\caption{Diagrammatic representation of the new local operators obtain from OPE formalism}
\label{fig:5}
\end{figure}
We have adopt the notation $(\bar{\psi}\chi)_{V\mp A}\equiv\bar{\psi}\gamma^\mu (1\mp\gamma_5)\chi$. This simple example introduces the main idea behind OPE, as already mentioned in the previous section. The above computation is nothing more than the usual `integrating out' in effective theories. While we have used a path integral approach, the computation done is equivalent to the expansion of the $W$ boson propagator in the amplitude matrix element, obtained from the usual Feynman rules approach (\Fref{fig:5}).    

Therefore, in general the OPE allows us to write an effective Hamiltonian of the form
\begin{equation}
\mathcal{H}_{\mbox{\scriptsize{eff}}}=\frac{G_F}{\sqrt{2}}\sum_i \lambda_{\mbox{\scriptsize{CKM}}}^i C_i(\mu)\mathcal{Q}_i\,,
\end{equation} 
where $\lambda_{\mbox{\scriptsize{CKM}}}^i$ contains CKM factors (1 for semi-leptonic operators, 2 for quark operators), $C_i(\mu)$ are the Wilson coefficients and $\mathcal{Q}_i$ is a local operator governing the process in question. The coefficients $C_i(\mu)$ are weights of the operators $\mathcal{Q}_i$ on the effective Hamiltonian, i.e. they describe the strength with which a given operator contributes to the Hamiltonian. These are scale dependent couplings and can be calculated using perturbative methods (as long the scale $\mu$ is not too small). The operators $\mathcal{O}_i$ are the leading terms in the short-distance expansion described above; in the cases we are interested in, these will correspond to four-fermion operators. Therefore, at short distances we see processes mediated by heavy particles as point-like interactions.  

We are interested in evaluating decay amplitude for a given type of meson $P$. With the help of the effective Hamiltonian this can be done quite `easily' using
\begin{equation}\label{secOPE: APF}
A(P\rightarrow F)=\langle F|\mathcal{H}_{eff}|P\rangle=\frac{G_F}{\sqrt{2}}\sum_i\lambda_{\mbox{\scriptsize{CKM}}}^i C_i(\mu)\langle F|\mathcal{Q}_i(\mu)|P\rangle\,,
\end{equation}
where $F$ denotes the final state, i.e we are looking at $P\rightarrow F$. The matrix element $\langle F|\mathcal{Q}_i(\mu)|P\rangle$ is evaluated at the renormalization scale $\mu$ and is the step that in general requires non-perturbative methods. 

\Eq[b]~(\ref{secOPE: APF}) and \Fref{fig:6} compiles the essence of the OPE method which allow the calculation of an amplitude $A(P\rightarrow F)$ to be factorize into two contributions:

\begin{figure}[h]
\centering\includegraphics[width=.9\linewidth]{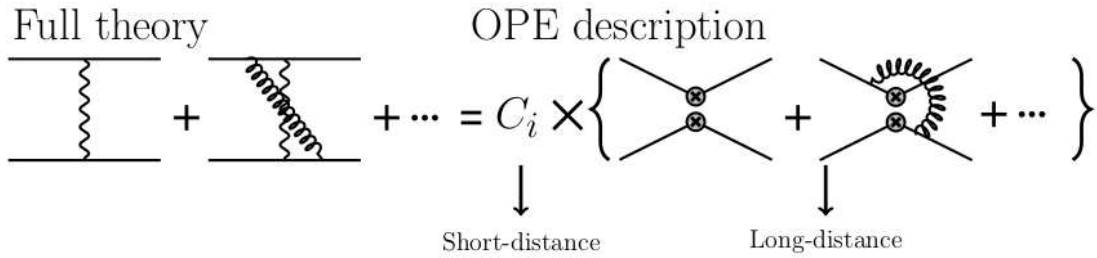}
\caption{Typical full theory description vs. OPE description}
\label{fig:6}
\end{figure}

\begin{itemize}
\item[]\textbf{Short-distance effects}

 The computation of short-distance effects, or perturbative calculation, are all contained in the Wilson coefficients $C_i(\mu)$. These coefficients will include the contributions from integrating out the heavy particles such as top quarks, gauge bosons $W$ and $Z$, and any new heavy field present in SM extensions. All effects of QCD interactions above the factorization scale $\mu$ are contained in these coefficients. $C_i(\mu)$ are independent of external states. This means that they are always the same
no matter we consider the physical amplitudes where quarks are bound inside mesons, or any other unphysical amplitude with on-shell or off-shell quarks in the external lines. 
 
 \item[]\textbf{Long-distance effects}
 
The computation of long-distance effects is present in the calculation of the matrix element $\langle\mathcal{Q}_i(\mu)\rangle$. This means that all low-energy contributions below the factorization scale $\mu$ are encoded in the matrix element. The task is then to evaluate local operators between hadron states. This is the hardest task to do in the OPE treatment, since
it requires in general a non-perturbative analysis.  
\end{itemize}
As we saw, the most difficult aspect of OPE is the non-perturbative computation of $\langle\mathcal{Q}_i(\mu)\rangle$. Still the method offers a considerably simplified approach to the full amplitude computation. Next we shall illustrate the OPE in the context of $K^0\rightarrow \pi^+\pi^-$ decay.
\begin{figure}[h]
\centering\includegraphics[width=.9\linewidth]{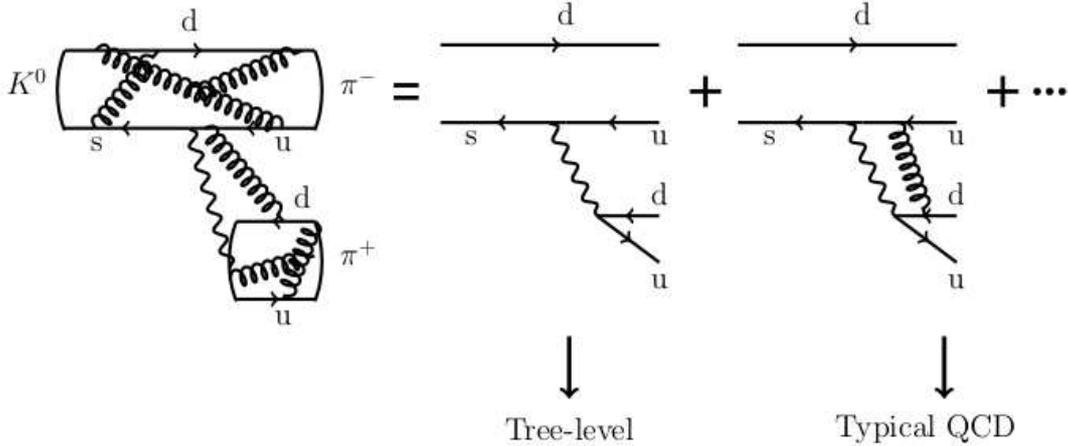}
\caption{General representation of $K^0\rightarrow \pi^+\pi^-$ decay. The two diagrams on the right are the typical leading contributions.}
\label{fig:7}
\end{figure}
We are, therefore, interested in the transition $s\rightarrow uud$ as shown in \Fref{fig:7}. A convenient choice is to take all the light quarks to be massless and with the same off-shell momentum $p$. The Wilson coefficients $C_i(\mu)$ can then be found in perturbation theory from the 3 simple steps:
\begin{itemize}
\item[(1)] Compute the amplitude ($A_{full}$) of the process in the full theory, i.e. in the presence of the $W$ propagator, for arbitrary external states

\item[(2)] Compute the matrix element $\langle Q_i(\mu)\rangle$ with the same treatment for external states

\item[(3)] Compute $C_i(\mu)$ from the relation $A_{full}=A_{eff}=\frac{G_F}{\sqrt{2}}\sum_i \lambda_{CKM}^iC_{i}(\mu)\langle Q_i(\mu)\rangle$; this is known as matching of the full theory onto the effective one
\end{itemize} 
Note that the choice of momenta leads to a gauge dependent amplitude. However, this cancels out with the gauge dependence from $\langle Q_i(\mu)\rangle$ such that $C_i(\mu)$ is physical. To order $\mathcal{O}(\alpha_S)$ we have four diagrams contributing: 1 with just $W$ propagator; 1 ($\times$ 3 combinations) with $W$ and gluon. Without QCD corrections we get the effective dimension 6 operator
\begin{equation}
\mathcal{Q}_2=(\overline{s}_iu_i)_{V-A}(\overline{u}_jd_j)_{V-A}\,,
\end{equation}
with $i,j$ color indices (the notation $\mathcal{Q}_2$ is for historical reasons.). 
\begin{figure}[h]
\centering\includegraphics[width=.25\linewidth]{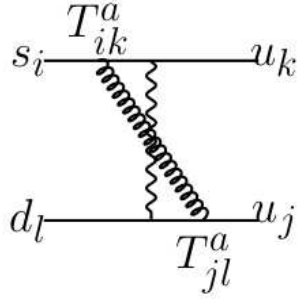}
\caption{Colour structure of typical QCD correction}
\label{fig:8}
\end{figure}
When QCD corrections are taken into account we at at order $\mathcal{O}(\alpha_S)$ the effective operator
\begin{equation}
\mathcal{Q}_1=(\overline{s}_iu_j)_{V-A}(\overline{u}_jd_i)_{V-A}\,,
\end{equation}
which resembles $\mathcal{Q}_2$ apart from the different color structure (see \Fref{fig:8}). This structure is obtained with the help of the $SU(N)$ Gell-Mann matrices identity
\begin{equation}
(\overline{s}_i T_{ik}^au_k)(\overline{u}_jT^a_{jl}d_l)=-\frac{1}{2N}(\overline{s}_iu_i)(\overline{u}_jd_j)+\frac{1}{2}(\overline{s}_iu_j)(\overline{u}_jd_i)\,.
\end{equation}
Gluonic corrections to the matrix element of the original operator $\mathcal{Q}_2$ involve not just contributions from itself but additional structure from $\mathcal{Q}_1$. We say that
the operators $\mathcal{Q}_1$ and $\mathcal{Q}_2$ mix under renormalization. Therefore, a convenient basis for the above operators is
\begin{equation}
\mathcal{\mathcal{Q}}_{\pm}=\frac{\mathcal{Q}_2\pm \mathcal{Q}_1}{2}\,,\quad C_{\pm}=C_2\pm C_1\,,
\end{equation}
where the renormalization of $+$ and $-$ are independent. We can then evaluate the full amplitude, which gives
\begin{equation}\label{eq:Afull}
-iA_{full}=-i\frac{G_F}{\sqrt{2}}V^\ast_{us}V_{ud}\left[\left(1+\gamma_+\alpha_s\ln\frac{M_W^2}{-p^2}\right)S_++\left(1+\gamma_-\alpha_s\ln\frac{M_W^2}{-p^2}\right)S_-\right]\,,
\end{equation}
where $S_{\pm}$ is the tree-level matrix elements of $\mathcal{Q}_\pm$ and $\gamma_\pm$ some numbers to be specify. This ends our first step. Next, we compute the matrix elements in the effective theory, which is given by
\begin{equation}\label{secQPE: Qpm}
-i\langle\mathcal{Q}_{\pm} \rangle=-i\frac{G_F}{\sqrt{2}}V^\ast_{us}V_{ud}\left[1+\gamma_\pm\alpha_s\left(\frac{1}{\epsilon}+\ln\frac{\mu^2}{-p^2}\right)\right]S_\pm\,.
\end{equation}
The last step is matching. From \Eref{eq:Afull} and \Eref{secQPE: Qpm} one easily reads the Wilson coefficient to be
\begin{equation}\label{secOPE: Cpm}
C_{\pm}=1+\gamma_\pm\alpha_s\ln\frac{M_W^2}{\mu^2}\,.
\end{equation}
A note of caution is in order. In the computation of the amplitude we did not perform any quark field renormalization. However, the renormalization in the effective theory can be explicitly seen in Eq.~\eqref{secQPE: Qpm}. Having divergent Wilson coefficients would be a clearly signal of inconsistency. Therefore, the above result was obtained after a renormalization on $\langle Q_{\pm}\rangle$ and using the MS scheme~\cite{Buras:1998raa}. The presence of this divergence in Eq.~\eqref{secQPE: Qpm} is directly linked to the $\ln M_W$ dependence of the decay amplitude in the full theory, which diverges in the limit $M_W\rightarrow \infty$. 

Summing up, the effective Hamiltonian describing $K^0\rightarrow \pi^+\pi^-$ decay is given by
\begin{equation}\label{secOPE: Heff}
\mathcal{H}_{\mbox{\scriptsize{eff}}}=\frac{G_F}{\sqrt{2}}V_{us}^\ast V_{ud}\left(C_+(\mu)\mathcal{Q}_++C_-(\mu)\mathcal{Q}_-\right)
\end{equation}
up to $\mathcal{O}(\alpha_s\text{log})$ and with $C_{\pm}$ given by Eq.~\eqref{secOPE: Cpm}. In obtaining the decay amplitude from Eq.~\eqref{secOPE: Heff}, the  matrix elements $\langle 2\pi| \mathcal{Q}_\pm|K\rangle$ have to be taken, normalized at an appropriated scale $\mu$. A typical scale for $K$ decays is $\mu\simeq 1\,\text{GeV}\ll M_W$. Going beyond leading logarithmic approximation $\mathcal{O}(\alpha_s\text{log})$ makes the Wilson coefficients and matrix elements scheme dependent. This scheme dependence is unphysical and cancels out in the product of Wilson coefficient and matrix elements, as long as both quantities are evaluated with the same scheme. 

 In the example above we have whitenessed in first hand the OPE factorization. Schematically, its has the following structure
 \begin{equation}
 \left(1+\alpha_S\gamma_{\pm}\ln \frac{M_W^2}{-p^2}\right)\rightarrow \left(1+\alpha_S\gamma_{\pm}\ln \frac{M_W^2}{\mu^2}\right) \left(1+\alpha_S\gamma_{\pm}\ln \frac{\mu^2}{-p^2}\right)\,,
 \end{equation}
which is achieved from the splitting of the logarithm into the sum of two terms. From the integration over virtual moment point of view this splitting reads
\begin{equation}
\int_{-p^2}^{M_W^2}\frac{dk^2}{k^2}=\underbrace{\int_{\mu^2}^{M_W^2} \frac{dk^2}{k^2}}_{
\begin{array}{c}
\text{Short-distance effects}\\
\text{or}\\
\text{large virtual momenta}
\end{array}
}+\underbrace{\int_{-p^2}^{\mu^2} \frac{dk^2}{k^2}}_{
\begin{array}{c}
\text{Long-distance effects}\\
\text{or}\\
\text{low virtual momenta}
\end{array}
}\,.
\end{equation}

At this stage it is important to have a closer look to the Wilson coefficients found above. We can rewrite them, for convenience, as
\begin{equation}
C_{\pm}=1+\frac{\gamma_{\pm}(\alpha_S)}{2}\ln\frac{\mu^2}{M_W^2}\,,\quad \text{with}\quad \gamma_{\pm}(\alpha_S)=\frac{\alpha_S(\mu)}{4\pi}\gamma_\pm^{(0)}\,,\quad \text{and}\quad\gamma_\pm^{(0)}=\left\{
\begin{array}{l}
4\\
-8
\end{array}
\right.\,.
\end{equation}
The factor multiplying the logarithm is $\mathcal{O}(1/10)$ for $\mu=1\, \text{GeV}$ and therefore sizeable for perturbation theory; the logarithm itself is large $\mathcal{O}(10)$ making perturbation theory to fail. We then have the scenario where the coupling constant is small, but we have large logarithms. This is actually a common situation in QFTs. The naive perturbation done in terms of the coupling constant is no longer enough, and we must resum the terms $(\alpha_S \ln\mu/M_W)^n$ to all orders $n$. This procedure reorganizes the pertubation series by solving the renormalization group equation (RGE) for the Wilson coefficients. The RGE for the Wilson coefficients follows from the fact that the unrenormalized coefficients  $C^{(0)}_{\pm}=Z_cC_{\pm}$ are $\mu$ independent. This then leads us to 
\begin{equation}\label{secOPE: RGECpm}
\frac{d}{d\ln\mu}C_{\pm}(\mu)=\gamma_\pm(\alpha_S) C_\pm(\mu)\quad\text{with}\quad
\gamma_{\pm}=-Z_c^{-1}\frac{d}{d\ln\mu}Z_c\,.
\end{equation} 
The parameters $\gamma_\pm(\alpha_S)$ are also known as anomalous dimension of $C_\pm$. The Wilson coefficients are dimensionless numbers in the usual sense. However, because of the presence of the scale $M_W$ in the logarithm, these coefficients will depend on the energy scale $\mu$. Therefore,  $\gamma_\pm(\alpha_S)$ are scaling dimensions, measuring the rate of change of these coefficients with a changing scale $\mu$. In general, when not working in the diagonal basis, these scaling dimensions are matrices mixing all Wilson coefficients. Using the RGE for the coupling constant
\begin{equation}
\frac{d\alpha_S}{d\ln\mu}=-2\beta_0\frac{\alpha_S^2}{4\pi}\,,
\end{equation}
we can solve Eq.~\eqref{secOPE: RGECpm}
\begin{equation}
C_\pm(\mu)=\left[\frac{\alpha_S(M_W)}{\alpha_S(\mu)}\right]^{\gamma_\pm^{(0)}/2\beta_0}C_\pm(M_W)=\left[\frac{1}{1+\beta_0(\alpha_S(\mu)/4\pi)\ln(M_W^2/\mu^2)}\right]^{\gamma_\pm^{(0)}/2\beta_0}\,,
\end{equation}
where we have used the condition $C_\pm(M_W)=1$, since no large logarithms should be present at $\mu=M_W$. The expression above contains the logarithmic corrections $\alpha_S\ln M_W/\mu$ to all orders in $\alpha_S$. This shows the general result that renormalization group method allows us to go beyond the naive perturbation theory. 

Two final remarks are in order. This approach can be generalized to go from $M_W$ down to $m_c$, for example. Then we can do this by steps, first evolving down to the scale $m_b$ and then see the theory below this scale as an effective theory where the $b$ quark has been integrated out. One should satisfy the continuity of the running coupling at the threshold, also known as threshold effects. These effects should be, in general, taken in consideration in the running. The second important effect is the generation of QCD penguin operators.

\subsection{Effective Hamiltonians: Some examples}\label{subsec:2.2}
In this section we summarize the Standard Model operator basis for FCNC processes, which is useful when computing quantities based on the OPE formalism. We use the notation $q=u,d,s,c,b$. The loop functions appearing in the Wilson coefficients are given by
\begin{align}
\begin{split}
 \tilde{E}_0(x)=&-\frac{7}{12}+\mathcal{O}(1/x)\\
 f(x)=&\frac{x}{2}+\frac{4}{3}\ln x-\frac{125}{36}+\mathcal{O}(1/x)\\
 g(x)=&-\frac{x}{2}-\frac{3}{2}\ln x +\mathcal{O}(1/x) .
\end{split}
\end{align}

\begin{itemize}
\item[$\bullet$] \textbf{Current-current operators}
\begin{figure}[h]
\centering\includegraphics[width=.5\linewidth]{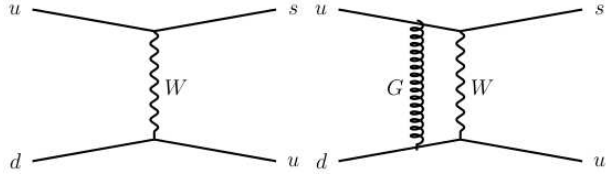}
\caption{Tree-level contribution and typical QCD correction topologies}
\label{fig:9}
\end{figure}
\begin{align}
\begin{split}
\mathcal{Q}^p_1=&(\overline{s}_i p^i)_{V-A}(\overline{p}_j  b^j)_{V-A}\,,\quad C_1(M_W)=1-\frac{11}{6}\frac{\alpha_s(m_W)}{4\pi}\\
\mathcal{Q}^p_2=&(\overline{s}_i  p^j)_{V-A}(\overline{p}_j  b^i)_{V-A}\,,\quad C_2(M_W)=\frac{11}{2}\frac{\alpha_s(m_W)}{4\pi}
\end{split}
\end{align}

\item[$\bullet$] \textbf{QCD Penguin operators:}

\begin{figure}[h]
\centering\includegraphics[width=.25\linewidth]{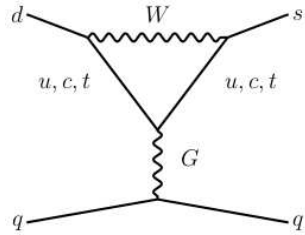}
\caption{QCD penguin topology}
\label{fig:10}
\end{figure}

\begin{align}
\begin{split}
\mathcal{Q}_{3(5)}=&(\overline{s}_i b^i)_{V- A}\sum_{q}(\overline{q}_j q^j)_{V\mp A}\,,\quad C_{3(5)}=-\frac{1}{6}\tilde{E}_0\left(\frac{m_t^2}{m_W^2}\right)\frac{\alpha_s(m_W)}{4\pi}\\
\mathcal{Q}_{4(6)}=&(\overline{s}_i b^j)_{V-A}\sum_{q}(\overline{q}_j q^i)_{V\mp A}\,,\quad C_{4(6)}=\frac{1}{2}\tilde{E}_0\left(\frac{m_t^2}{m_W^2}\right)\frac{\alpha_s(m_W)}{4\pi}
\end{split}
\end{align}

\item[$\bullet$] \textbf{Electroweak Penguin operators:}

\begin{figure}[h]
\centering\includegraphics[width=.5\linewidth]{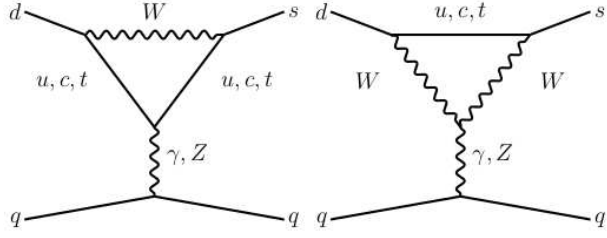}
\caption{Electroweak penguin topologies}
\label{fig:11}
\end{figure}

\begin{align}
\begin{split}
\mathcal{Q}_{7(9)}=&(\overline{s}_ib^i)_{V-A}\sum_{q}\frac{3}{2}Q_q(\overline{q}_j q^j)_{V\pm A}\,,\\
&C_7=f\left(\frac{m_t^2}{m_W^2}\right)\frac{\alpha(m_W)}{6\pi}\,,\quad C_9=\left[f\left(\frac{m_t^2}{m_W^2}\right)+\frac{1}{s_W^2}g\left(\frac{m_t^2}{m_W^2}\right)\right]\frac{\alpha(m_W)}{4\pi}\\
\mathcal{Q}_{8(10)}=&(\overline{s}_i b^j)_{V-A}\sum_{q}\frac{3}{2}Q_q(\overline{q}_j q^i)_{V\pm A}\,,\quad C_{8(10)}=0
\end{split}
\end{align}

\item[$\bullet$] \textbf{Electromagnetic and chromo-magnetic dipole operators:}

\begin{figure}[h]
\centering\includegraphics[width=.25\linewidth]{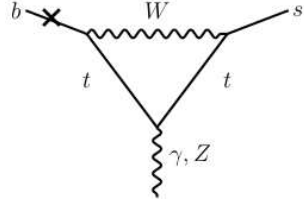}
\caption{Topology for electro- and chromo-magnetic dipoles. The cross means mass insertion.}
\label{fig:12}
\end{figure}

\begin{align}
\begin{split}
\mathcal{Q}_{7\gamma}=&-\frac{e}{8\pi^2}m_b\,\overline{s_L}_i\sigma^{\mu\nu}b_{R}^i F_{\mu\nu}\,,\quad C_{7\gamma}=-\frac{1}{3}+\mathcal{O}\left(\frac{m_W^2}{m_t^2}\right)\\
\mathcal{Q}_{8g}=&-\frac{g}{8\pi^2}m_b\,\overline{s_L}_i\sigma^{\mu\nu}(T^a)^{i}_{\,j}b_{R}^j G^a_{\mu\nu}\,,\quad C_{8g}=-\frac{1}{8}+\mathcal{O}\left(\frac{m_W^2}{m_t^2}\right)
\end{split}
\end{align}

\item[$\bullet$] \textbf{$\mathbf{\Delta S=2}$ and $\mathbf{\Delta B=2}$ operators}

\begin{figure}[h]
\centering\includegraphics[width=.35\linewidth]{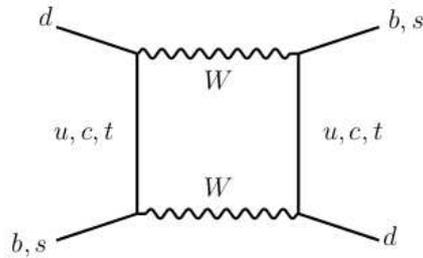}
\caption{Box topology}
\label{fig:13}
\end{figure}

\begin{align}
\begin{split}
\mathcal{Q}(\Delta S=2)=&(\overline{s}_id^i)_{V_A}(\overline{s}_jd^j)_{V-A}\,,\quad \mathcal{Q}(\Delta B=2)=(\overline{b}_id^i)_{V_A}(\overline{b}_jd^j)_{V-A}
\end{split}
\end{align}

\item[$\bullet$] \textbf{Semileptonic operators:}

\begin{figure}[h]
\centering\includegraphics[width=.3\linewidth]{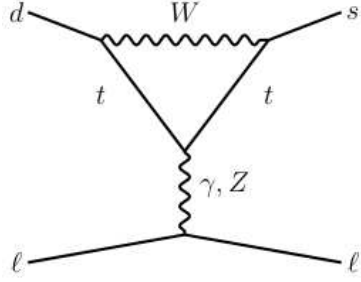}
\caption{Semileptonic penguin topology}
\label{fig:14}
\end{figure}

\begin{align}
\begin{split}
\mathcal{Q}_{7V,A}=&(\overline{s}_id^i)_{V-A}(\overline{e} e)_{V,A}\,,\quad \mathcal{Q}_{9V,10A}=(\overline{s}_ib^i)_{V-A}(\overline{\mu} \mu)_{V,A}\,,\\
\mathcal{Q}_{\bar{\nu}\nu}=&(\overline{s}_i d^i)_{V-A}(\overline{\nu}\nu)_{V-A}\,,\quad\mathcal{Q}_{\bar{\mu}\mu}=(\overline{s}_i d^i)_{V-A}(\overline{\mu} \mu)_{V-A}
\end{split}
\end{align}
\end{itemize}

With the list of $d=6$ operators we are able to describe several SM flavour changing processes. For example, the relevant interactions to the parton process $b\rightarrow s+\bar{q}q$ can be parametrized though the Hamiltonian
\begin{equation}
\mathcal{H}_{\mbox{\scriptsize{SM}}}^{b\rightarrow s+q\bar{q}}=-\frac{G_F}{\sqrt{2}}\left[\sum_{p=u,c}V_{pb}^\ast V_{ps}\sum_{i=1,2}C_i(\mu)\mathcal{Q}_i^{p}+
V_{tb}^\ast V_{ts}\sum_{i=3,\cdots,10}C_i(\mu)Q_i\right]\,.
\end{equation}
If we are also interested in $b\rightarrow s$ transitions with a photon or a lepton pair in the final state, additional dimension-six operators must be included. We then get,
\begin{equation}
\mathcal{H}_{\mbox{\scriptsize{SM}}}^{b\rightarrow s+\gamma(l\bar{l})}=\mathcal{H}_{SM}^{b\rightarrow s+q\bar{q}}-\frac{G_F}{\sqrt{2}}V_{tb}^\ast V_{ts}\left[C_{7\gamma}(\mu)Q_{7\gamma}+C_{8g}(\mu)Q_{8g}+C_{9V}(\mu)Q_{9V}+C_{10A}(\mu)Q_{10A}\right]\,.
\end{equation}

\subsection{Effective theories for heavy flavours: a brief introduction}\label{subsec:2.3}

What is there to integrate out, when there are no heavy particles? The answer to this question is in looking for different scales, e.g. in $B-$physics $m_b\gg \Lambda_{\mbox{\scriptsize{QCD}}}$. Then we can use the effective theory approach and integrate out all short-distance fluctuations associated with scales $\gg \Lambda_{\mbox{\scriptsize{QCD}}}$. In this scenario physics at the $m_b$ scale are short-distance effects, while heavy quark related hadronic physics governed at confinement scale $\Lambda_{\mbox{\scriptsize{QCD}}}$ reflect long-distance effects. The separation of the short-distance and long-distance effects associated with these two scales is vital for any quantitative description in heavy-quark physics. 

The prime example of this separation is on heavy quark effective field theory (HQET)~\cite{Georgi:1990um}. What is the physical picture behind HQET?
\begin{itemize}
\item[$\bullet$] Scale hierarchy $m_b\gg \Lambda_{\mbox{\scriptsize{QCD}}}$, $\alpha_2(m_B)$ is perturbative (asymptotic freedom)
\item[$\bullet$] Heavy quark - heavy quark system is perturbative
\item[$\bullet$] Heavy-light bound states are not perturbative
\item[$\bullet$] Characterized by a small Compton wavelength;  $\lambda_Q\sim1/m_Q\ll 1/\Lambda_{QCD}\sim R_{had}$(typical hadronic size)
\end{itemize}
These requirements simplify the physics of hadrons made up of a heavy quark. In mesons composed of a heavy quark, $Q$, and a light antiquark, $\bar{q}$ (and gluons and $q\bar{q}$ pairs), the heavy quark acts as a static color source with fixed four-velocity, $v_\mu$ , and the wave function of the light degrees of freedom becomes insensitive the mass (flavour) of the heavy quark. Since the magnetic moment of a heavy quark scales like $\mu_Q\sim 1/m_Q$, its spin also decouples. This results in
\begin{itemize}
\item[]\textbf{$\mathbf{SU(2n_Q)}$ spin-flavour symmetry: }\textit{In heavy-quark limit $(m_Q\rightarrow \infty)$, configuration  of light degrees of freedom is independent of the spin and flavour of the heavy quark.}
\end{itemize}

\begin{figure}[h]
\centering\includegraphics[width=.3\linewidth]{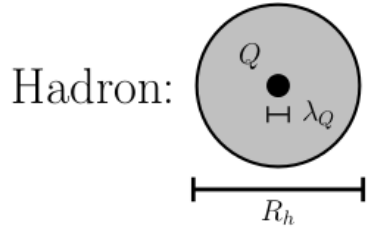}
\caption{Pictorial representation of the Hadron. The black central dot represents the heavy quark and the gay are the light degrees of freedom. $R_h$ is the size of the hadron while $\lambda_Q$ the Compton wave length of the heavy quark.}
\label{fig:15}
\end{figure}

In the effective description that we are looking there are some other important aspects:
\begin{itemize}
\item[$\bullet$] Heavy quarks carries almost all momentum;
\item[$\bullet$] The momentum exchange between heavy quark and light degrees of freedom is predominantly soft (soft gluon exchange):
\begin{equation}
\Delta P_Q=-\Delta P_{light}=\mathcal{O}(\Lambda_{\mbox{\scriptsize{QCD}}})\quad\Rightarrow\quad \Delta v_Q=\mathcal{O}(\Lambda_{\mbox{\scriptsize{QCD}}}/m_Q)\,;
\end{equation}
\item[$\bullet$] Heavy-quark velocity becomes a conserved quantum number in $m_Q\rightarrow \infty$ limit. This is known as the Georgi ``velocity superselection rule''; 

\item[$\bullet$] Spin doublets such as $(B,B^\ast)$ should be degenerate in the heavy quark limit: $m_{B^\ast}-m_B=46\,\text{MeV}\ll\Lambda_{\mbox{\scriptsize{QCD}}}$;

\item[$\bullet$] Away from the heavy-quark limit, $1/m_Q$ corrections are expected: $m_{B^\ast}-m_{B}=(c_1-c_0)\lambda_2/m_b+\mathcal{O}(1/m_b^2)$;

 \item[$\bullet$] The approach gives a prediction for $(m_{B^\ast}-m_B)/(m_{D^\ast}-m_D)\simeq m_c/m_b\simeq 1/3$; Not far from the experimental value of 0.32.
\end{itemize}

\begin{figure}[h]
\centering\includegraphics[width=.6\linewidth]{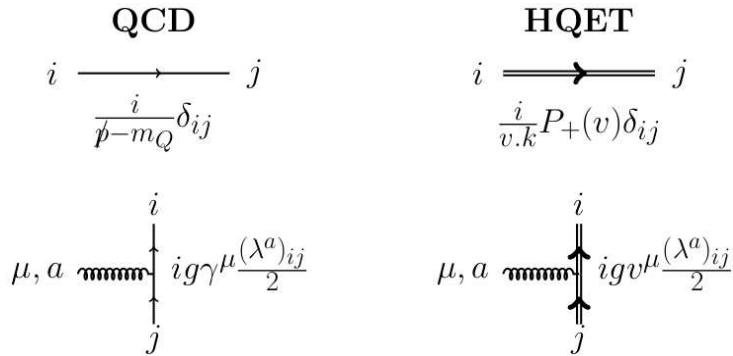}
\caption{Feynman rules QCD vs. HQET}
\label{fig:16}
\end{figure}

We can now construct an effective theory that makes the effects of the heavy-quark symmetry explicit, i.e. the HQET. The heavy quark $Q$ in the interactions with soft partons (ligh quark $q$ and gluon $g$) is almost on-shell, such that we can expand the momentum as
\begin{equation}
p_Q^\mu=\underbrace{m_Qv^\mu}_{
\begin{array}{c}
\text{hadron}\\
\text{rest frame}\\
v^\mu=(1,0,0,0)
\end{array}
}+\underbrace{k^\mu}_{
\begin{array}{c}
\text{residual off-shell}\\
\text{momentum}\\
|k|=\mathcal{O}(\Lambda_{QCD})
\end{array}
}
\end{equation}
Expanding the heavy quark propagator we get
\begin{equation}
\frac{i}{p\!\!\!/-m_Q}=\frac{i(p\!\!\!/+m_Q)}{p^2-m_Q^2}=\frac{i(m_Qv\!\!\!/+k\!\!\!/+m_Q)}{2m_Qv.k+k^2}=\frac{i}{v.k}\frac{1+v\!\!\!/}{2}+\cdots\,.
\end{equation}
We can see that in this expansion the propagator is no longer dependent on the mass of the heavy quark, a clear manifestation of the heavy quark flavour symmetry. To derive the effective Lagrangian is convenient to decompose the Dirac spinor components into `upper' (large) and `lower' (small) pieces
\begin{equation}
Q(x)=e^{-im_Q c.x}[\underbrace{h_v(x)+H_v(x)}_{
\begin{array}{c}
\text{carry the}\\
\text{residual }k
\end{array}
}]\,,\quad \text{with}\quad\left[
\quad
\begin{array}{l}
h_v(x)=e^{im_Q v.x}P_+Q(x)\\\\
H_v(x)=e^{im_Q v.x}P_-Q(x)
\end{array}\right.
\end{equation}
and $P_{\pm}=(1\pm v\!\!\!/)/2$ are projector operators. In the rest frame of the heavy quark $P_+=(1+\gamma^0)/2$ project onto the heavy quark components. An useful identity of these projectors is
\begin{equation}
P_+\gamma^\mu P_+=P_+v^\mu P_+=v^\mu P_+\,.
\end{equation}
Note that $h_v(x)$ and $H_v(x)$ are eigenstates of the velocity operator, i.e. $v\!\!\!/h_v(x)=h_v(x)$ and $v\!\!\!/H_v(x)=-H_v(x)$. In terms of these fields the QCD Lagrangian can now be written as
\begin{align}
\begin{split}
\mathcal{L}_{\mbox{\scriptsize{Q}}}=&\overline{Q}(iD\!\!\!\!/-m_Q)Q\\
=&\overline{h}_{v} iD\!\!\!\!/ h_v+\overline{H}_v(iD\!\!\!\!/-2m_Q)H_v+\overline{h}_viD\!\!\!\!/ H_v+\overline{H}_viD\!\!\!\!/ h_v\\
=&\overline{h}_{v} iv.D h_v+\overline{H}_v(-iv.D-2m_Q)H_v+\overline{h}_viD\!\!\!\!/_\bot H_v+\overline{H}_viD\!\!\!\!/_\bot h_v
\end{split}
\end{align}
where we defined $i\vec{D}^\mu_\bot=iD^\mu-v^\mu iv.D$, orthogonal to the heavy-quark velocity $v.D_\bot=0$. In the rest frame, $D^\mu_\bot=(0,\vec{D})$ contains the spatial components of the covariant derivative. We see from the Lagrangian above that the component $h_v(x)$ is a massless mode describing a quantum fluctuation around mass-shell, while $H_v(x)$ is a massive mode with mass $2m_Q$ describing a hard quantum fluctuation. This heavy component can be integrated out by using the classical equation of motion
\begin{equation}
H_v=\frac{1}{2m_Q+iv.D}iD\!\!\!\!/_\bot h_v=\frac{1}{2m_Q}\sum_{n=0}^{\infty}\underbrace{\left(-\frac{iv.D}{2m_Q}\right)^n}_{
\begin{array}{c}
\text{small}\\
k\ll m_Q
\end{array}}iD\!\!\!\!/_\bot h_v\longrightarrow \, H_v\simeq \left(\frac{D}{m_Q}\right)h_v\sim \left(\frac{\Lambda_{QCD}}{m_Q}\right)h_v\,.
\end{equation}
The  effective Lagrangian can then be written as
\begin{equation}
\begin{array}{rll}
\mathcal{L}_{\mbox{\scriptsize{HQET}}}=&\overline{h}_v iv.D_sh_v+\overline{h}_v iD\!\!\!\!/_\bot\dfrac{1}{2m_Q+iv.D}iD\!\!\!\! /_\bot h_v&\longrightarrow\quad\textbf{non-local}\\\\
=&\overline{h}_v iv.D_sh_v+\dfrac{1}{2m_Q}\sum_{n=0}^\infty\overline{h}_v iD\!\!\!\!/_\bot\left(-\dfrac{iv.D}{2m_Q}\right)^niD\!\!\!\! /_\bot h_v&\longrightarrow\quad\textbf{local}
\end{array}
\end{equation}
Therefore at leading only $h_v(x)$ contributes, and the effects of $H_v(x)$ are suppressed by powers of $\Lambda_{\mbox{\scriptsize{QCD}}}/m_Q$, i.e.
\begin{equation}
\mathcal{L}_{\mbox{\scriptsize{HQET}}}=\overline{h}_v iv.D_sh_v+\mathcal{O}(1/m_Q)\,,\quad\text{with}\quad iD^\mu_s=i\partial^\mu+\underbrace{g_sG_s^\mu}_{
\text{soft gluons}
}\,.
\end{equation}
It is straightforward to extend the above result for higher order of power corrections. At the next to leading order we get
\begin{equation}
\mathcal{L}_{\mbox{\scriptsize{HQET}}}=\underbrace{\overline{h}_viv.D_sh_s}_{
\begin{array}{c}
SU(2n_Q)\\
\text{spin-flavour}\\
\text{symmetry}
\end{array}
}+\frac{1}{2m_Q}[\underbrace{\overline{h}_v(iD_{s\bot})^2h_v}^{
\begin{array}{c}
-\overline{h}_v(i\vec{D}_{s})^2h_v\\
\uparrow
\end{array}
}_{
\begin{array}{c}
\text{kinetic-energy}\\
\text{operator}
\end{array}
}+\underbrace{C_{\mbox{\scriptsize{mag}}}(\mu)\frac{g_s}{2}\overline{h}_v\sigma_{\mu\nu}G_s^{\mu\nu}h_v}^{
\begin{array}{c}
\hspace{0.8cm}-4\overline{h}_v\vec{S}.\vec{B}_ch_v\\
\hspace{0.8cm}\uparrow
\end{array}
}_{
\begin{array}{c}
\text{chromo-magnetic}\\
\text{from pert. theo.}
\end{array}
}]+\cdots\,,
\end{equation}
where we have make use of the identity
\begin{equation}
P_+iD\!\!\!\!/_\bot iD\!\!\!\!/_\bot P_+=P_+\left[(iD_\bot)^2+\frac{g_s}{2}\sigma_{\mu\nu}G^{\mu\nu}\right]P_+
\end{equation}
and $i[D^\mu,D^\nu]=g_sG^{\mu\nu}$ is the gluon fields-strength tensor. Here $\vec{S}$ is the spin operator and $B_c^i=-1/2\epsilon^{ijk}G^{jk}$ are the components of the colour-magnetic field. The Wilson coefficient is computed through RGE-improved perturbation theory~\cite{Amoros:1997rx}. The leading term is $SU(2n_Q)$ spin-flavour invariant, i.e. no reference to the heavy-quark mass (flavour symmetry) and invariant under the spin rotations $h_v\rightarrow (1+i/2\vec{\epsilon}.\vec{\sigma})h_v$. The flavour symmetry is broken by the operators arising at order $1/m_Q$ and higher. Note, however, that at this order the kinetic term conserves the spin symmetry, while the chromo-magnetic operator breaks the both flavour and spin symmetry. Figure~\ref{fig:16} shows the changes in the Feynman rules in the new formalism.

Up to now we have integrated out small components in the heavy-quark fields and obtained an effective local Lagrangian that describes the long-distance physics in the full theory. The way heavy-quarks participate in the strong interaction is through their couplings to gluons. These can be soft (virtual momentum small, of the order of the confining scale) or hard (virtual momentum large, of the order of the heavy quark mass). In the approach used above we have integrated out the hard gluons as they, contrarily to the soft ones, break the heavy-quark symmetries. However, hard gluons are important once we decide to add short-distance effects. Their effects lead to a renormalization of the coefficients of the operators in the HQET Lagrangian, which are calculable in perturbation theory. There is no renormalization at leading order. Nor renormalization of the kinetic operator due to Lorentz invariance (``reparametrization invariance''). However, the chromo-magnetic interaction will be affected.

\begin{figure}[h]
\centering\includegraphics[width=.5\linewidth]{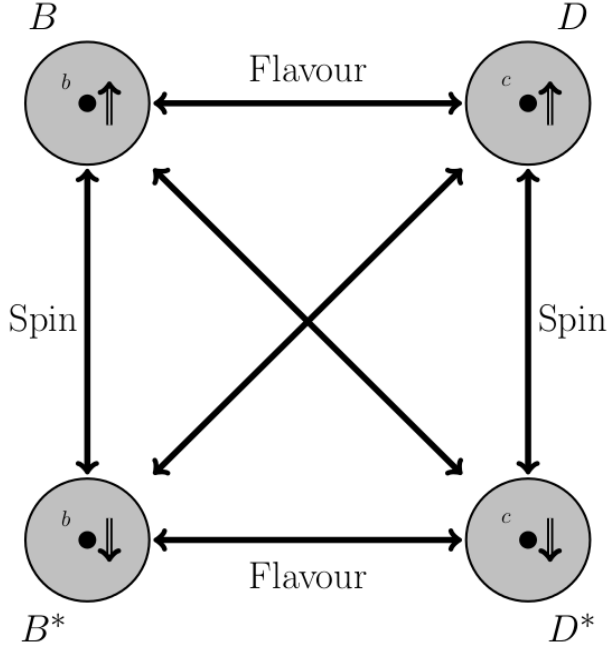}
\caption{Spin-flavour symmetry between  $B$- and $D$-system}
\label{fig:17}
\end{figure}

Heavy-quark symmetry is particularly predictive for exclusive semi-leptonic $B$ decays such as $B\rightarrow D^{(\ast)}\ell\bar{\nu}$. It allow us to extract the CKM matrix elements $|V_{cb}|$ and $|V_{ub}|$ with controlled theoretical uncertainties, through the correlations shown in \Fref{fig:17} .

\begin{figure}[h]
\centering\includegraphics[width=.8\linewidth]{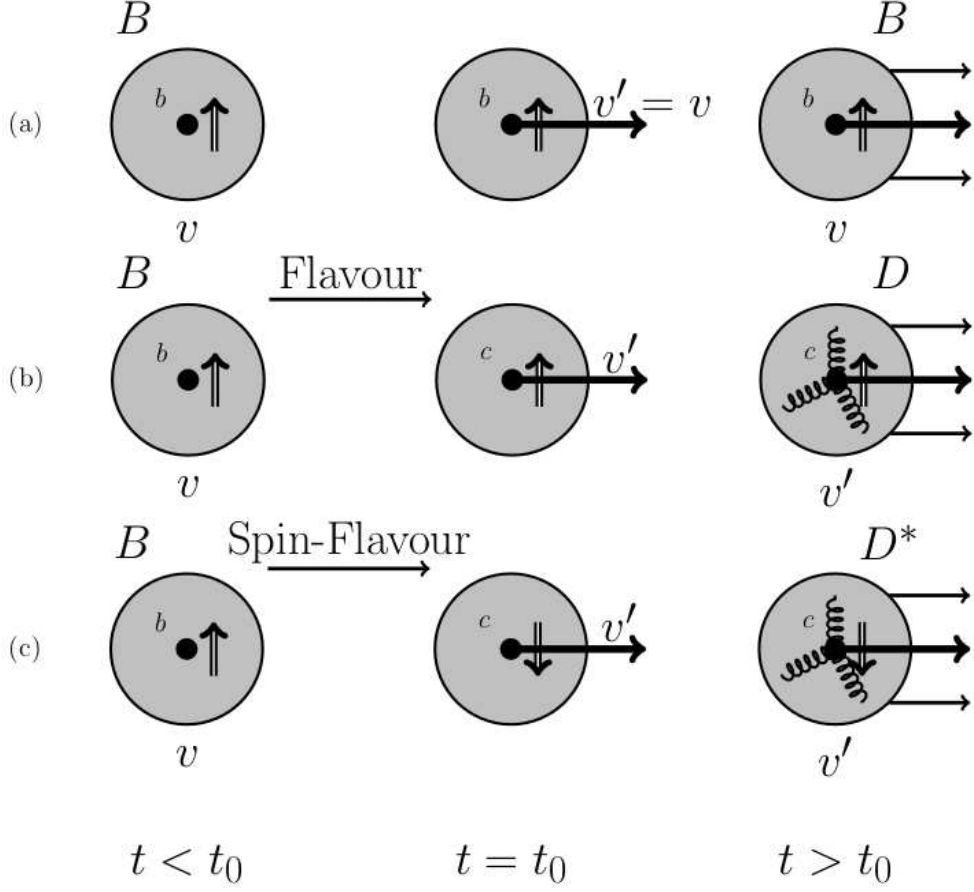}
\caption{Evolution with time of the hadron for the different scenarios where spin-flavour symmetry is applied.}
\label{fig:19}
\end{figure}

A clever use of heavy-quark symmetries allows us to calculate the decay rate at the special kinematic point of maximum momentum transfer to the leptons $(v=v^\prime)$, i.e. ``zero recoil'' point. How can we deal with confinement effects in this hadronic process? We can consider elastic scattering of a $B$ meson, $\bar{B}(v)\rightarrow \bar{B}(v^\prime)$, induced by the vector current $J^\mu=\bar{b}\gamma^\mu b$.
%\begin{figure}[h]
%\centering\includegraphics[width=.8\linewidth]{fig/Plot18.pdf}
%\caption{Description of my figure}
%\label{fig:18}
%\end{figure}
The heavy quark acts as a static source of color, and the light quarks orbit around it before the action of the vector current. On average, the $b$ quark and the $B$ meson have the same velocity. The action of the current is to replace instantaneously (at $t=t_0$) the color source by one moving at speed $v^\prime$. Nothing happens if $v=v^\prime$, i.e. the final state remains a $B$ meson with probability 1 (case (a) in \Fref{fig:19}). However, for $v\neq v^\prime$, the probability for an elastic transition is less than 1. The light constituents find them selfs interacting with moving source. Soft gluons will have to be exchanged in order to rearrange them and form a $B$ meson moving at a different speed, leading to a form factor suppression. In the Heavy-quark mass limit, i.e. $m_b\rightarrow \infty$, the process is described by a dimensionless probability function $\xi(v.v^\prime)$ called the Isgur-Wise function. The hadronic matrix elements describing the scattering process is then
\begin{equation}
\frac{1}{m_B}\langle  \bar{B}(v^\prime)|\overline{b}_{v^\prime}\gamma^\mu b_v|\bar{B}(v)\rangle=\xi(v.v^\prime)(v+v^\prime)^\mu\,,\quad\text{with}\quad \xi(v.v^\prime)\leq 1\,,\, \xi(1)=1\,.
\end{equation}
The $1/m_B$ factor on the left-hand side of the equation compensates the normalization of the meson state, i.e. $\langle \bar{B}(p^\prime)|\bar{B}(p) \rangle=2m_Bv^0(2\pi)^3\delta(\vec{p}-\vec{p}^\prime)$. We can then use the flavour symmetry to replace $b-$ by $c-$quark in the final state, thereby obtaining a $B\rightarrow D$ transition. This transforms the scattering process into a weak decay process.

Nothing will happen to the matrix element since in the heavy-quark limit the Lagrangian is invariant under the $b_{v^\prime}\rightarrow c_{v^\prime}$ replacement (case (b) in \Fref{fig:19}), i.e.
\begin{equation}
\frac{1}{\sqrt{m_Bm_D}}\langle  \bar{D}(v^\prime)|\overline{c}_{v^\prime}\gamma^\mu b_v|\bar{B}(v)\rangle=\xi(v.v^\prime)(v+v^\prime)^\mu\,.
\end{equation}
This is a very interesting prediction of the heavy-quark symmetry. Since in general the matrix element of a flavour-changing current between two pseudo-scalar mesons is given by
\begin{equation}
 \langle\bar{D}(v^\prime)|\overline{c}_{v^\prime}\gamma^\mu b_v|\bar{B}(v)\rangle=f_+(q^2)(p+p^\prime)^\mu-f_-(q^2)(p-p^\prime)^\mu\,,
\end{equation}
with $f_{\pm}(q^2)$ the form factors and $q=p-p^\prime$. The heavy-quark symmetry relates the two a priori independent form factors to one and the same function, i.e. the Isgur-Wise function $(f_{\pm}(q^2)\propto \xi(v.v^\prime))$.

Next, we can use the spin symmetry to flip the spin of $c-$quark in final state, thereby obtaining a $B\rightarrow D^\ast$ transition (case (c) in \Fref{fig:19}). The current gets transformed to
\begin{equation}
\langle D^\ast (v^\prime,\epsilon)|\overline{c}_{v^\prime}\gamma^\mu(1-\gamma_5)b_v|B(v)\rangle=
\langle D^\ast(v^\prime,\epsilon)|\overline{c}_{v^\prime}\gamma^\mu b_v|B(v)\rangle-
\langle D^\ast(v^\prime,\epsilon)|\overline{c}_{v^\prime}\gamma^\mu\gamma_5 b_v|B(v)\rangle
\end{equation}
with
\begin{align}
\begin{split}
\frac{1}{\sqrt{m_Bm_{D^\ast}}}\langle D^\ast(v^\prime,\epsilon)|\overline{c}_{v^\prime}\gamma^\mu b_v|B(v)\rangle=&i\epsilon^{\mu\nu\alpha\beta}\epsilon_\nu^\ast v^\prime_\alpha v_\beta \xi(v.v^\prime)\\
\frac{1}{\sqrt{m_Bm_{D^\ast}}}\langle D^\ast(v^\prime,\epsilon)|\overline{c}_{v^\prime}\gamma^\mu \gamma_5b_v|B(v)\rangle=&[\epsilon^{\ast \mu}(v.v^\prime+1)-v^{\prime \mu}\epsilon^\ast.v]\xi(v.v^\prime)
\end{split}
\end{align} 
where $\epsilon$ denotes the polarization of the $D^\ast$ meson. The general Lorentz-invariant matrix elements of these hadron currents are given by
\begin{align}
\begin{split}
\langle D^\ast(v^\prime,\epsilon)|\overline{c}_{v^\prime}\gamma^\mu b_v|B(v)\rangle=&\frac{2i}{m_B+m_{D^\ast}}\epsilon_{\mu\nu\alpha\beta}\epsilon^{\ast \nu}p^{\prime\alpha}p^\beta V(q^2)\\
\langle D^\ast(v^\prime,\epsilon)|\overline{c}_{v^\prime}\gamma^\mu \gamma_5b_v|B(v)\rangle=&(m_B+m_{D^\ast})\epsilon_\mu^\ast A_1(q^2)-\frac{\epsilon^\ast.p}{m_B+m_{D^\ast}}(p+p^\prime)_\mu A_2(q^2)\\
&-2m_{D^\ast}\frac{\epsilon^\ast.q}{q^2}q_\mu A_{3}(q^2)+2m_{D^\ast}\frac{\epsilon^\ast.q}{q^2}q_\mu A_0(q^2)
\end{split}
\end{align}
with
\begin{equation}
A_3(q^2)=\frac{m_B+m_D}{2m_{D^\ast}}A_1(q^2)-\frac{m_B-m_{D^\ast}}{2m_{D^\ast}}A_2(q^2)\,.
\end{equation}
In general, these exclusive semileptonic decays processes can be described by six a priori independent hadronic form factors
\begin{equation}
\begin{array}{l}
\left[\begin{array}{l}
\textbf{For }\mathbf{0^-\rightarrow 0^-}\textbf{ transition: }I\rightarrow F\ell\nu_\ell\\\\
\langle F(v^\prime)|V_\mu^{cb}|I(v)\rangle=\sqrt{m_Im_F}\left[\xi_+(v.v^\prime)(v+v^\prime)_\mu+\xi_-(v.v^\prime)(v-v^\prime)_\mu \right]
\end{array}\right.\\\\
\left[
\begin{array}{l}
\textbf{For }\mathbf{0^-\rightarrow 1^-}\textbf{ transition: }I\rightarrow F^\ast \ell\nu_\ell\\\\
\langle F^\ast(v^\prime)|V_\mu^{cb}|I(v)\rangle=i\sqrt{m_Im_{F^\ast}}\xi_V(v.v^\prime)\epsilon_{\mu\nu\alpha\beta}\epsilon^{\mu\nu}
v^{\prime\alpha}v^\beta\\
\langle F^\ast(v^\prime)|A_\mu^{cb}|I(v)\rangle=\sqrt{m_Im_{F^\ast}}[\xi_{A_1}(v.v^\prime)(v.v^\prime+1)\epsilon^\ast_\mu-\xi_{A_2}(v.v^\prime)\epsilon^\ast.v v_\mu\\
 \hspace{3.5cm}-\xi_{A_3}(v.v^\prime)\epsilon^\ast.v v^\prime_\mu]\,.
\end{array}\right.
\end{array}
\end{equation}
with $V_\mu$ and $A_\mu$ the vector- and axial-currents, respectively. The heavy-quark limit imposes the relations:
\begin{equation}
\xi_+(v.v^\prime)=\xi_V(v.v^\prime)=\xi_{A_1}(v.v^\prime)=\xi_{A_3}(v.v^\prime)=\xi(v.v^\prime)\quad
\text{and}\quad \xi_-(v.v^\prime)=\xi_{A_2}(v.v^\prime)=0\,.
\end{equation}
These relations are model independent and are a consequence of QCD in the limit $m_b,m_c\gg \Lambda_{\mbox{\scriptsize{QCD}}}$. For the processes described below the form factor correlations read
\begin{align}
\begin{split}
\xi(v.v^\prime)&=\frac{2\sqrt{m_Bm_D}}{m_B\pm m_D}f_\pm(q^2)=\frac{2\sqrt{m_Bm_{D^\ast}}}{m_B+ m_{D^\ast}}V(q^2)=\frac{2\sqrt{m_Bm_{D^\ast}}}{m_B+ m_{D^\ast}}A_0(q^2)\\
&=\frac{2\sqrt{m_Bm_{D^\ast}}}{m_B+ m_{D^\ast}}A_2(q^2)=\frac{2\sqrt{m_Bm_{D^\ast}}}{m_B+ m_{D^\ast}}\left[1-\frac{q^2}{(m_B+m_{D^\ast})^2}\right]^{-1}A_1(q^2)\,,
\end{split}
\end{align}
with $ q^2=m_B^2+m_{D^\ast}^2-2m_Bm_{D^\ast}v.v^\prime$. These from factors play an important role in describing semileptonic decays as $\bar{B}\rightarrow D^{(\ast)}\ell \nu$. In terms of the recoil variable $\omega=v.v^\prime$, the differential decay rate in the heavy quark limit for these processes is given by
\begin{equation}
\frac{d\Gamma(B\rightarrow D^{(\ast)}\ell\bar{\nu})}{d\omega}=\frac{G_F^2\eta_{ew}^2}{48\pi^3}|V_{cb}|^2\times F\times
\left\{
\begin{array}{ll}
(\omega^2-1)^{1/2}\mathcal{F}_\ast^2(\omega)\,,&\text{for }B\rightarrow D^\ast\\\\
(\omega^2-1)^{3/2}\mathcal{F}^2(\omega)\,,&\text{for }B\rightarrow D
\end{array}
\right.\,,
\end{equation} 
with $\eta_{ew}\simeq 1$ a parameter accounting for the electroweak corrections to the four-fermion operator mediating the decay and
\begin{equation}
F=\left\{
\begin{array}{ll}
m_B^5r^3(1-r)^2(\omega+1)^2\left(1+\dfrac{4\omega}{\omega+1}\dfrac{1-2r\omega+r^2}{(1-r)^2}\right)\,,\,r=\dfrac{m_{D^\ast}}{m_B}&\text{for }D^\ast\\\\
(m_B+m_D)^2m_D^3&\text{for }D
\end{array}
\right.
\end{equation}
Both $\mathcal{F}(\omega)$ and $\mathcal{F}_\ast(\omega)$ are equal in the heavy-quark mass limit and are normalized such that $\mathcal{F}_{(\ast)}(1)=1$, allowing a model independent extraction of $|V_{cb}|$. The above differential decay rate expressions receive symmetry-breaking corrections, since the mass of the heavy quark is not infinitely large:
\begin{itemize}
\item[$\bullet$] Corrections of order $\mathcal{O}(\alpha_s^n(m_Q))$ (hard gluons) can be calculated perturbatively;
\item[$\bullet$] Power corrections of order $\mathcal{O}((\Lambda_{\mbox{\scriptsize{QCD}}}/m_Q)^n)$ are non-perturbative and more difficult to control.
\end{itemize}
These corrections have been estimated and schematically give
\begin{align}
\begin{split}
\mathcal{F}_\ast(1)\simeq& 1+\underbrace{c_A(\alpha_s)}+\overbrace{0\times \frac{\Lambda_{\mbox{\scriptsize{QCD}}}}{m_Q}}^{
\begin{array}{c}
\text{Luke}\\
\text{Theorem}
\end{array}
}+\overbrace{\text{cons}\times\frac{\Lambda^2_{\mbox{\scriptsize{QCD}}}}{m^2_Q}}^{
\begin{array}{c}
\text{lattice/}\\
\text{models}
\end{array}
}+\cdots\\
&\quad\text{Perturbative}\\
\mathcal{F}(1)\simeq& 1+\overbrace{c_V(\alpha_s)}+\underbrace{\text{const}\times \frac{\Lambda_{\mbox{\scriptsize{QCD}}}}{m_Q}}_{
\begin{array}{c}
\text{lattice/}\\
\text{models}
\end{array}
}+\cdots
\end{split}
\end{align}
The absence of the $\mathcal{O}(\Lambda_{\mbox{\scriptsize{QCD}}}/m_Q)$ term for $B\rightarrow D^\ast \ell\bar{\nu}_\ell$ at the zero-recoil limit, i.e. $\omega=1$, is a consequence of the Luke theorem:
\begin{itemize}
\item[]\textit{The matrix elements describing the leading $1/m_Q$ corrections to weak decay amplitudes vanish at zero recoil, to all order in perturbation theory.}
\end{itemize}
The reason why in the semi-leptonic decay $B\rightarrow D\ell \bar{\nu}_\ell$ this is no longer true is more subtle and can be found in~\cite{Neubert:1991xw}. Therefore, from the value of $\mathcal{F}_\ast(1)$ the value of $|V_{cb}|$ is estimated to be
\begin{equation}
\begin{array}{ll}
|V_{cb}|=(39.48\pm0.5_{exp}\pm0.74_{theo})\times 10^{-3}&\text{from lattice QCD}\,,\\
|V_{cb}|=(41.4\pm0.5_{exp}\pm1.0_{theo})\times 10^{-3}&\text{from QCD sum rules}\,,
\end{array}
\end{equation}
showing the power of HQET in describing non-pertubative systems.

\section{Some aspect of $CP$ violation}\label{sec:3}

\subsection{$CP$ violation in the Universe}\label{sec:3.1}

One of the currents issues related with flavour physics and $CP$ violation is the Baryon asymmetry of the Universe. Our understanding of the Universe is based on the Standard Cosmological Model, where the Universe expanded from a primordial hot and dense initial state at some finite time in the past (the so-called Big Bang) and is then followed by a period of inflationary expansion that ensured the curvature to become approximately zero~\cite{Linde:2007fr}. After this inflationary epoch, the Universe continued to expand but at a low rate. 
The rate of expansion is determined by the component of energy density that dominates the total energy density; at the present time this is the so-called dark energy component, which causes the expansion to accelerate due to its negative pressure. 

In our surroundings the objects are mostly made of matter, e.g. planets, stars, etc.. The present value of the baryon asymmetry of the Universe inferred from WMAP seven-year data combined with baryon acoustic oscillations is~\cite{Bennett:2012zja}
\begin{equation}
\eta_B\equiv \frac{n_B-n_{\bar{B}}}{n_\gamma}=(6.19\pm 0.14)\times 10^{-10}\,,
\end{equation} 
where $n_B$, $n_{\bar{B}}$ and $n_\gamma$ are the number density of baryons, antibaryons and photons at present time, respectively. The smallness of this quantity poses a challenge to both particle physics and cosmology. If we take inflation for granted, then in the early Universe any primordial cosmological asymmetry would be erased during the inflationary period. This is one argument that strongly suggests this asymmetry to be dynamically generated, instead of being an initial accidental state. Sakharov realized the need of three ingredients in order to create a baryon asymmetry from an initial state with baryon number equal to zero~\cite{Sakharov:1967dj}. The three conditions can be stated as follows:
\begin{itemize}
\item[i)] Baryon number violation;
\item[ii)] $C$ and $CP$ violation;
\item[iii)] Departure from thermal equilibrium.
\end{itemize} 

The first condition is rather obvious. If there is no $B$ violation, the baryon number is conserved in all interactions and, therefore, commutes with the Hamiltonian at any time, i.e.
\begin{equation}
[B,\mathcal{H}]=0\quad\Rightarrow\quad B(t)=\int_0^t[B,\mathcal{H}] dt^\prime=0\,.
\end{equation}

The second condition is a little more delicate. Let us start by writing the baryon number operator
\begin{equation}
\hat{B}=\frac{1}{3}\sum_i\int d^3x\,:\psi_i^\dagger (\vec{x},t)\psi_i (\vec{x},t):\,,
\end{equation}
where $\psi_{i}(\vec{x},t)$ denotes the quark field of flavour $i$ and $::$ denote the normal ordering. The $C$, $P$ and $T$ transformations of these fields are given in \Tables~\ref{tab:CPTfields}--\ref{tab:Discrete}. Thus, the fermionic number satisfies the following transformations
\begin{align}
\begin{split}
\mathcal{P}:\psi_i^\dagger(\vec{x},t)\psi_i(\vec{x},t):\mathcal{P}^{-1}=&:\psi_i^\dagger(-\vec{x},t)\psi_i(-\vec{x},t):\,,\\
\mathcal{C}:\psi_i^\dagger(\vec{x},t)\psi_i(\vec{x},t):\mathcal{C}^{-1}=&-:\psi_i^\dagger(\vec{x},t)\psi_i(-\vec{x},t):\,,\\
\mathcal{T}:\psi_i^\dagger(\vec{x},t)\psi_i(\vec{x},t):\mathcal{T}^{-1}=&:\psi_i^\dagger(\vec{x},-t)\psi_i(\vec{x},-t):\,.\\
\end{split}
\end{align}
We can, therefore, find how the baryon number operator transforms under these operators. One gets
\begin{equation}
\mathcal{C}\hat{B}\mathcal{C}^{-1}=-\hat{B}\,,\quad (\mathcal{CP})\hat{B}(\mathcal{CP})^{-1}=-\hat{B}\,,\quad (\mathcal{CPT})\hat{B}(\mathcal{CPT})^{-1}=-\hat{B}\,.
\end{equation}
Now, if $C$ is conserved, then $[\mathcal{C},\mathcal{H}]=0$ and the expectation value of the baryon number is given by
\begin{align}
\begin{split}
\left<\hat{B}(t)\right>=&\left<e^{i\mathcal{H}t}\hat{B}(0)e^{-i\mathcal{H}t}\right>=\left<\mathcal{C}^{-1}\mathcal{C}e^{i\mathcal{H}t}\hat{B}(0)e^{-i\mathcal{H}t}\right>=\left<e^{i\mathcal{H}t}\mathcal{C}\hat{B}(0)\mathcal{C}^{-1}e^{-i\mathcal{H}t}\right>\\
=&-\left<e^{i\mathcal{H}t}\hat{B}(0)e^{-i\mathcal{H}t}\right>=-\left<\hat{B}(t)\right>\,.
\end{split}
\end{align}
We see that the expectation value $\left<\hat{B}(t)\right>$ is only different from zero if $C$ is not a symmetry of the Hamiltonian. The same is true for $CP$.

The last condition can be understood as follows. In thermal equilibrium, the thermal average are weighted by the density operator $\rho=e^{-\beta\mathcal{H}}$, with $\beta=1/T$. Assuming $CPT$ invariance of the Hamiltonian we get
\begin{align}
\begin{split}
\left<\hat{B}(t)\right>_T=&\text{Tr}\left[e^{\beta\mathcal{H}}\hat{B}\right]=\text{Tr}\left[(\mathcal{CPT})^{-1}(\mathcal{CPT})e^{\beta\mathcal{H}}\hat{B}\right]=\text{Tr}\left[e^{\beta\mathcal{H}}(\mathcal{CPT})\hat{B}(\mathcal{CPT})^{-1}\right]\\
=&-\left<\hat{B}(t)\right>_T\,.
\end{split}
\end{align}
This means that, within a $CPT$ invariant Hamiltonian, the thermal average is zero and no net baryon asymmetry is generated since the inverse processes will destroy the asymmetry generated in the direct decays. Departure from thermal equilibrium is very common in the early Universe when interaction rates cannot keep up with the expansion rate of the Universe.

All three of these condition can be found in the SM, however the amount of $CP$ violation from the CKM mechanisms is to small in order to generate such an asymmetry.

\subsection{Weak and strong phases}\label{subsec:3.2}
$CP$ is violated in nature by the weak interactions. The imposition of $CP$ invariance in a transition amplitude is expressed as
\begin{equation}
\left(\mathcal{CP}\right)\widehat{T}\left(\mathcal{CP}\right)^\dagger=\widehat{T}\,.
\end{equation} 
In classical physics, the square of the \textit{CP} transformation is identical to the identity transformation, and therefore $\left(\mathcal{CP}\right)^2$ corresponds to a conserved quantum number. The value of $\left(\mathcal{CP}\right)^2$ for initial and final states must be identical, and is a purely arbitrary phase. Without loss of generality one can choose $\left(\mathcal{CP}\right)^2=1$. The \textit{CP} transformations read
\begin{equation}
\mathcal{CP}\left|i\right>=e^{i\xi_i}\left|\overline{i}\right>\,,\quad \mathcal{CP}\left|\overline{i}\right>=e^{-i\xi_i}\left|i\right>\,,
\end{equation}
with $\xi_i$ an arbitrary phase. The $CP$ constraints on the transition amplitudes from an initial state $i$ to the final states $f$ and $g$ are
\begin{equation}\label{eq:TCP}
\begin{array}{c}
\textbf{Final state}\\
f/\bar{f}
\end{array}
\left[
\begin{array}{l}
\left<f\right|\widehat{T}\left|i\right>=e^{i(\xi_i-\xi_f)}\left<\overline{f}\right|\widehat{T}\left|\overline{i}\right>\\\\
\left<\overline{f}\right|\widehat{T}\left|i\right>=e^{i(\xi_i+\xi_f)}\left<f\right|\widehat{T}\left|\overline{i}\right>
\end{array}\right.\,,\,
\begin{array}{c}
\textbf{Final state}\\
g/\bar{g}
\end{array}
\left[
\begin{array}{l}
\left<g\right|\widehat{T}\left|i\right>=e^{i(\xi_i-\xi_g)}\left<\overline{g}\right|\widehat{T}\left|\overline{i}\right>\\\\
\left<\overline{g}\right|\widehat{T}\left|i\right>=e^{i(\xi_i+\xi_g)}\left<g\right|\widehat{T}\left|\overline{i}\right>
\end{array}\right.\,.
\end{equation}
From these transition amplitudes one sees that the modulus of each process is equal to the modulus of the $CP$ conjugated one. Therefore, the $CP$-violating quantities are
\begin{equation}\label{eq:modulus}
\begin{array}{c}
\begin{array}{c}
\textbf{Final state}\\
f/\bar{f}
\end{array}
\left[
\begin{array}{l}
\left|\left<f\right|\widehat{T}\left|i\right>\right|-\left|\left<\overline{f}\right|\widehat{T}\left|\overline{i}\right>\right|\\\\
\left|\left<\overline{f}\right|\widehat{T}\left|i\right>\right|-\left|\left<f\right|\widehat{T}\left|\overline{i}\right>\right|
\end{array}\right.\,,
\begin{array}{c}
\textbf{Final state}\\
b/\bar{b}
\end{array}
\left[
\begin{array}{l}
\left|\left<g\right|\widehat{T}\left|i\right>\right|-\left|\left<\overline{g}\right|\widehat{T}\left|\overline{i}\right>\right|\\\\
\left|\left<\overline{g}\right|\widehat{T}\left|i\right>\right|-\left|\left<g\right|\widehat{T}\left|\overline{i}\right>\right|
\end{array}\right.\\
\hspace{2.5cm}\downarrow \hspace{6.4cm}\downarrow
\\
\hspace{2.5cm}\neq 0\hspace{0.5cm}\Longrightarrow \hspace{0.5cm}\textbf{CP violation}\hspace{1cm}\Longleftarrow\hspace{0.5cm}\neq 0
\end{array}
\end{equation}
If we only had one final state, say $f$, the relevant expressions would be the ones presented in the first line of \Eref{eq:TCP} and \eqref{eq:modulus}. In Eq.~\eqref{eq:TCP}, we only have two phases for two complex equations and therefore no other quantity beyond the one presented in \Eref{eq:modulus} would violate $CP$. The fact that we have two final states, $f$ and $g$, leads to three arbitrary phases but four complex equations. Since we only have four real $CP$-violating quantities in \Eref{eq:modulus}, a physical $CP$ condition on the phases of the decay amplitudes must remain. One can find that the quantity
\begin{equation}
\left<f\right|\widehat{T}\left|i\right>\left<\overline{f}\right|\widehat{T}\left|i\right>\left<g\right|\widehat{T}\left|\overline{i}\right>\left<g\right|\widehat{T}\left|\overline{i}\right>-\left<g\right|\widehat{T}\left|i\right>\left<\overline{g}\right|\widehat{T}\left|i\right>\left<f\right|\widehat{T}\left|\overline{i}\right>\left<f\right|\widehat{T}\left|\overline{i}\right>
\end{equation}
must vanish if $CP$ invariance holds. 

The presence of complex phases is closely related with $CP$ violation. One simple argument to support this statement is due to $CPT$ invariance. If $CPT$ is conserved then $CP$ violation is the same as $T$ violation. Since $T$ transforms a number into its complex conjugate, the $CP$ violation must be related to the presence of complex numbers. One should stress, however, that the phase of a transition amplitude is arbitrary and non-physical, due to the freedom of phase redefinition of the kets and bras. Only phases which are rephasing invariant can lead to $CP$ violation. These are in general relative phases of transition amplitudes. There are three types of phases that can arise in transitions amplitudes:
\begin{itemize}
\item[$\bullet$] \textbf{`weak' or $CP$-odd phases.}

The weak phases are defined as the phases that change sign under $CP$ conjugation, and usually originate from complex couplings in the Lagrangian. 

\item[$\bullet$] \textbf{`strong' or $CP$-even phases.}

The strong phases are the ones that remain unchanged under $CP$ conjugation. They may arise from the trace of products of an even number of $\gamma$ matrices together with $\gamma_5$, or final-state-interaction scatterings from on-shell states. The last one appears when the total amplitude for the decay $i\rightarrow f$ includes contributions from $i\rightarrow f^\prime \rightarrow f$, where the decay $i\rightarrow f^\prime$ is through weak interactions and $f^\prime\rightarrow f$ through strong or electromagnetic ones. If the intermediate states are on mass shell this creates an absorptive part. These are also typical phases appearing on absorptive parts of loops diagrams in perturbation theory.

\item[$\bullet$] \textbf{`spurious' $CP$-transformation phases.}

The spurious phases are global, purely conventional relative phases between the amplitude of a process and the amplitude for the $CP$-conjugate process. These phases do not originate in any dynamics, they just come from the assumed $CP$ transformation of the field operators and on the kets and bras they act upon~\cite{Branco:1999fs}.

\end{itemize}

\subsection{Types of $CP$ Violation}\label{subsec:3.3}

\begin{itemize}
\item[$\bullet$] \textbf{$CP$-violation in Decays (direct $CP$ violation)}

This type of $CP$-violation occurs when a meson $P$ and its $CP$-conjugate decay at different rates to the same final state (up to $CP$ conjugacy). This can be characterized by the relation
\begin{equation}
\left|\frac{\overline{A}_{\bar{f}}}{A_f}\right|\neq 1\,.
\end{equation}
In charged meson decays, where mixing is not present, this is the only source of $CP$ violation:
\begin{equation}
a_{f^{\pm}}=\frac{\Gamma(P^-\rightarrow f^-)-\Gamma(P^+\rightarrow f^+)}{\Gamma(P^-\rightarrow f^-)+\Gamma(P^+\rightarrow f^+)}=\frac{|\overline{A}_{\bar{f}}/A_f|^2-1}{|\overline{A}_{\bar{f}}/A_f|^2+1}\,.
\end{equation}
In order to have $CP$ violation in transition amplitudes from $i\, (\bar{i})$ to $f\,(\bar{f})$, the transition amplitudes need to be a sum of two or more interfering amplitudes. The way we can see this is through an explicit example. Consider for instance
\begin{equation}
\left<f\right|T\left|i\right>=Ae^{i(\delta+\phi)}\,,\quad \left<\bar{f}\right|T\left|\bar{i}\right>=Ae^{i(\delta-\phi+\theta)}\,,
\end{equation}
with $A$ a real positive number, $\delta$ a strong phase, $\phi$ a weak phase and $\theta$ a spurious one. It is easy to see that these transition amplitudes satisfy the first equation of \Eref{eq:TCP} with
\begin{equation}
\xi_i-\xi_f=2\phi-\theta\,,
\end{equation}
leading to 
\begin{equation}
\left|\left<f\right|T\left|i\right>\right|-\left|\left<\bar{f}\right|T\left|\bar{i}\right>\right|=A-A=0\,.
\end{equation}
Therefore, no $CP$ violation is generated in such a transition. This is no longer true when there is interference. For that, we consider
\begin{align}
\begin{split}
\left<f\right|T\left|i\right>=&A_1e^{i(\delta_1+\phi_1)}+A_2e^{i(\delta_2+\phi_2)}\,,\\
\left<\bar{f}\right|T\left|\bar{i}\right>=&A_1e^{i(\delta_1-\phi_1+\theta_1)}+A_2e^{i(\delta_2-\phi_2+\theta_2)}\,,
\end{split}
\end{align}
where $\delta_i$, $\phi_i$ and $\theta_i$ are the strong, weak and spurious phases, respectively. Now, it is no longer possible to satisfy \Eref{eq:TCP}. We can evaluate the $CP$-violating quantity
\begin{equation}\label{CPA}
\frac{\left|\left<f\right|T\left|i\right>\right|^2-\left|\left<\bar{f}\right|T\left|\bar{i}\right>\right|^2}{\left|\left<f\right|T\left|i\right>\right|^2+\left|\left<\bar{f}\right|T\left|\bar{i}\right>\right|^2}=\frac{-4A_1A_2\sin(\delta_1-\delta_2)\sin(\phi_1-\phi_2)}{2A_1^2+2A_2^2+4A_1A_2\cos(\delta_1-\delta_2)\cos(\phi_1-\phi_2)}\,.
\end{equation}
This expression will be used later on (in a different form) and, therefore, it is useful to make a few remarks:
\begin{itemize}
\item The existence of both weak and strong phases is crucial for $CP$ violation;
\item Only relative phases (weak and strong) are relevant in physical processes;
\item The limiting case $|\phi_1-\phi_2|=|\delta_1-\delta_2|=\pi/2$ and $A_1=A_2$ gives the maximum value of the $CP$ asymmetry;
\end{itemize}
It is possible to have $CP$ violation without strong phases, if we have more than one final state and its $CP$ conjugate. For example, having the transition amplitudes
\begin{align}
\begin{split}
\left<f\right|T\left|i\right>=A_1e^{i(\delta_1+\phi_1)}\,,\quad \left<f\right|T\left|\bar{i}\right>=A_1e^{i(\delta_1-\phi_1+\theta)}\,,\\
\left<g\right|T\left|i\right>=A_2e^{i(\delta_2+\phi_2)}\,,\quad \left<g\right|T\left|\bar{i}\right>=A_2e^{i(\delta_2-\phi_2+\theta)}\,,
\end{split}
\end{align}
with $f=\bar{f}$ and $g=\bar{g}$, we can build the quantity
\begin{equation}
\left<f\right|T\left|\bar{i}\right>\left<g\right|T\left|\bar{i}\right>-\left<g\right|T\left|i\right>\left<f\right|T\left|i\right>=
2iA_1A_2e^{i(\delta_1+\delta_2+\theta)}\sin(\phi_1-\phi_2)\,.
\end{equation}
In this quantity the strong phases are basically irrelevant and $CP$ violation is dictated by the weak phases. However, these two distinct final states must be correlated such that the decay involve both simultaneously, otherwise this can not be an observable. This is actually the case in kaon de decays to $\pi^+\pi^-$ and $\pi^0\pi^0$ (see Sec.~\ref{subsec:3.4}).

\item[$\bullet$] \textbf{$CP$-violation in mixing (indirect $CP$ violation)}

This type of $CP$ violation occurs when degenerated neutral mesons are not the $CP$ eigenstates. This can be characterized by the relation
\begin{equation}
\left|\frac{q}{p}\right|\neq 1\,.
\end{equation}
This is the only source of $CP$ violation in semileptonic final states such as $P^0\rightarrow l^+ X$. In such a scenario the asymmetry can be observed in
\begin{equation}
a_{SL}=\frac{\Gamma (\overline{P^0}_{phys}(t)\rightarrow l^+ X)-\Gamma (P^0_{phys}(t)\rightarrow l^- \bar{X})}{\Gamma (\overline{P^0}_{phys}(t)\rightarrow l^+ X)+\Gamma (P^0_{phys}(t)\rightarrow l^- \bar{X})}=\frac{1-|q/p|^2}{1+|q/p|^2}\,.
\end{equation}
The meson $P_{phys}^0(t)$ represents the time evolved state. As we shall see in Sec.~\ref{subsec:3.4}, 
\begin{equation}
a_{SL}=\text{Im}\left(\frac{\Gamma_{12}}{M_{12}}\right).
\end{equation}
This means that in our model we just need to know $M_{12}$ and $\Gamma_{12}$, in order to compute the $CP$ violating observable. However, in general $\Gamma_{12}$ is plagued with large hadronic uncertainties, making this computation more cumbersome. 

\item[$\bullet$] \textbf{$CP$-violation in interference decays}

This type of $CP$ violation only occurs in decays where the final state $f$ is common for both $P^0$ and $\overline{P^0}$. This can be characterized by the relation 
\begin{equation}
\text{Im}\,\lambda_f\neq 0\,,
\end{equation}
where $\lambda_f=(q/p)(A(\overline{P^0}\rightarrow f_{CP})/A(P^0\rightarrow f_{CP}))$. One example is where this asymmetry can be observed is in decays involving $\mathcal{CP}$ eigenstates with $\pm1$ eigenvalues. Then we have the $CP$ violating observable
\begin{equation}
a_{f_{CP}}(t)=\frac{\Gamma(\overline{P^0}\rightarrow f_{CP})-\Gamma(P^0\rightarrow f_{CP})}{\Gamma(\overline{P^0}\rightarrow f_{CP})+\Gamma(P^0\rightarrow f_{CP})}\,.
\end{equation}
In the $B$-system this leads to
\begin{equation}
a_{f_{CP}}(t)=-\frac{1-|\lambda_{f_{CP}}|^2}{1+|\lambda_{f_{CP}}|^2}\cos(\Delta m_B t)+\frac{2\text{Im}\,\lambda_{f_{CP}}}{1+|\lambda_{f_{CP}}|^2}\sin(\Delta m_B t)
\end{equation}
The fist term on th l.h.s. corresponds to $\mathcal{CP}$ violation through mixing, while the last term is due to interference. In decays with $|\lambda_{CP}|=1$ only the interference effect survives
\begin{equation}
a_{f_{CP}}(t)=\text{Im}\,\lambda_{f_{CP}}\sin(\Delta m_B t)\,.
\end{equation}
We know $\Delta m_B$ so we can measure $\text{Im}\,\lambda_{f_{CP}}$. This quantity is the phase between mixing and decay amplitudes. To a good approximation $|A(\overline{P^0}\rightarrow f_{CP})|=|A(P^0\rightarrow f_{CP})|$ and since in the standard parametrization $q/p=e^{i2\beta}$, we have to a good approximation
\begin{equation}
\text{Im}\,\lambda_{f_{CP}}=\text{Im}\left[\frac{q}{p}\frac{A(\overline{P^0}\rightarrow f_{CP})}{A(P^0\rightarrow f_{CP})}\right]\simeq \sin 2\beta\,.
\end{equation}

\end{itemize}

\subsection{Neutral Meson Mixing: General description}\label{subsec:3.4}

In this section we shall follow closely the the discussions in~\cite{Silva:2004gz}. We are interested in describing how $CP$ violation arises from the mixing of a neutral meson $P_0$ with its antiparticle $\overline{P^0}$. Consider the simplest scenario where the two states $|P^0\rangle$ and $|\overline{P^0}\rangle$ that are degenerated can neither decay or transform into each other. In such a system an arbitrary state can then be represented as
\begin{equation}
|\psi(t)
\rangle=a(t)|P^0\rangle+b(t)|\overline{P^0}\rangle
\end{equation} 
and evolve through the Schrodinger equation with diagonal Hamiltonian. This scenario is exactly what happens in the neutral meson system when only QCD interactions are active. Turning on the electroweak interactions will induce, even if small, off-diagonal Hamiltonian entries mixing both states leading to the breaking of the
%system 
degeneracy. In general, to describe the time evolution of this new state we would require the state
\begin{equation}
|\psi(t)\rangle=a(t)|P^0\rangle+b(t)|\overline{P^0}\rangle+\sum_ic_i(t)|n_i\rangle\,,
\end{equation}
where $n_i$ are final states of the $P^0$ and $\overline{P^0}$ decays. However, we may study the mixing in this particle-antiparticle system separately from its subsequent decay if the following conditions are satisfied: $a(0),\,b(0)\neq0$ and $c_i(0)=0$; time scale larger than the typical strong-interaction scale; no interactions between final states (Weisskopf-Wigner approximation). In this way the neutral meson mixing is described by two-component wave function
\begin{equation}
\psi(t)=
\begin{pmatrix}
a(t)\\
b(t)
\end{pmatrix}
\end{equation}
evolving according to a Schrodinger equation
\begin{equation}\label{eq:NMMtimeevol}
i\frac{d}{dt}
\psi(t)\rangle
= \underbrace{\left(M-\dfrac{i}{2}\Gamma\right)}_{H}
\psi(t)
=
\begin{pmatrix}
M_{11}-\dfrac{i}{2}\Gamma_{11}&M_{12}-\dfrac{i}{2}\Gamma_{12}\\\\
M_{21}-\dfrac{i}{2}\Gamma_{21}&M_{22}-\dfrac{i}{2}\Gamma_{22}
\end{pmatrix}
\psi(t)\,,
\end{equation}
with $t$ the proper time, $H$ a $2\times2$ matrix written in the $P^0 - \overline{P^0}$ rest frame and $M, \,\Gamma$ its Hermitian parts. The meson flavour basis $\left\{|P^0\rangle,\, |\overline{P^0}\rangle\right\}$ satisfies the following relations:
\begin{itemize}
\item \textbf{Orthogonality:} $\langle P^0|\overline{P^0}\rangle=\langle\overline{P^0}|P^0\rangle=0$ and $\langle P^0|P^0\rangle=\langle\overline{P^0}|\overline{P^0}\rangle=1$.
\item \textbf{Completeness:} $|P^0\rangle\langle P^0|+|\overline{P^0}\rangle\langle \overline{P^0}|=1$.
\item \textbf{Effective Hamiltonian decomposition:} $\mathcal{H}=\begin{pmatrix}|P^0\rangle,&|\overline{P^0}\rangle\end{pmatrix}\mathbf{H}\begin{pmatrix}\langle P^0|\\
\langle\overline{P^0}|\end{pmatrix}$
\end{itemize}
In terms of the total Hamiltonian
\begin{equation} 
\mathcal{H}=\overbrace{\mathcal{H}_{\mbox{\scriptsize{QCD}}}+\mathcal{H}_{\mbox{\scriptsize{QED}}}}^{
\begin{array}{c}
\textbf{CP}
\end{array}
}+
\overbrace{\mathcal{H}_{\mbox{\scriptsize{EW}}}}^{
\begin{array}{c}
\textbf{CPV}
\end{array}
}
\end{equation} 
we can have the usual perturbation expansion, up to second order,
\begin{equation}
\left(M-\frac{i}{2}\Gamma\right)_{ij}=\langle i|\mathcal{H}|j\rangle+\sum_n\frac{\langle i|\mathcal{H}|n\rangle\langle n|\mathcal{H}|j\rangle}{m_{0}-(E_n-i\epsilon )}
\end{equation} 
where $i$ and $j$ can be $K^0$ or $\overline{K^0}$ and $|n\rangle$ any eigenstate of $\mathcal{H}_{\mbox{\scriptsize{QCD}}}+\mathcal{H}_{\mbox{\scriptsize{QED}}}$ with eigenvalue $E_n$, but with $n\neq K^0,\,\overline{K^0}$. Using the identity
\begin{equation}
\frac{1}{m_0-(E_n-i\epsilon)}=P\frac{1}{m_0-E_n}-i\pi \delta(m_0-E_n)
\end{equation}
we can find the Hermitian matrices $M$ and $\Gamma$ up to second order in perturbation theory. They are given by
\begin{align}
\begin{split}
M_{ij}=&\overbrace{\langle i|\mathcal{H}|j\rangle}^{m_0\delta_{ij}+\langle i|\mathcal{H}_{\mbox{\scriptsize{EW}}}|j\rangle}+\sum_n P\frac{\langle i|\mathcal{H}_{\mbox{\scriptsize{EW}}}|n\rangle\langle n|\mathcal{H}_{\mbox{\scriptsize{EW}}}|j\rangle}{m_0-E_n}\,,\\
\Gamma_{ij}=&2\pi \sum_n\delta(m_0-E_n)\langle i|\mathcal{H}_{\mbox{\scriptsize{EW}}}|n\rangle\langle n|\mathcal{H}_{\mbox{\scriptsize{EW}}}|j\rangle\,,
\end{split}
\end{align}
with $P$ projecting out the principal part. The general $CP$ transformation of the states is given by
\begin{equation}
\mathcal{CP}|P^0(\vec{p})\rangle= -e^{i\xi}|\overline{P^0}(-\vec{p})\rangle\quad \text{and}\quad \mathcal{CP}|\overline{P^0}(\vec{p})\rangle= -e^{-i\xi}|P^0(-\vec{p})\rangle\,.
\end{equation}
We then see that the $CP$-invariant combinations are given by
\begin{equation}\label{eq:CPeigen}
|P_1\rangle=\frac{1}{\sqrt{2}}\left(|P^0\rangle-e^{i\xi}|\overline{P^0}\rangle\right)\,,\quad |P_2\rangle=\frac{1}{\sqrt{2}}\left(|P^0\rangle+e^{i\xi}|\overline{P^0}\rangle\right)\,,
\end{equation}
in such a way that 
\begin{equation}
\mathcal{CP}|P_1\rangle=|P_1\rangle\quad \text{and}\quad\mathcal{CP}|P_2\rangle=-|P_2\rangle\,.
\end{equation} 
Requesting $CP$ invariance is equivalent to the Hamiltonian condition $\mathcal{H}=(\mathcal{CP})\mathcal{H}(\mathcal{CP})^\dagger$. This in turns imply $H_{12}=e^{-2i\xi}H_{21}$ and $H_{11}=H_{22}$. Note that $\xi$ is a spurious phase without any physical relevance, therefore, we conclude that the phases of $H_{12}$ and $H_{21}$ also lack meaning. We can then summarize, in \Tref{tab:CPT}, the physical conditions given the present discrete symmetries.
\begin{table}[h]
\begin{center}
\caption{Constrains on the mixing matrix when the system respect some or no discrete symmetries.}
\label{tab:CPT}
\begin{tabular}{cc}
\hline\hline
\textbf{Conservation}&\textbf{Constraints}\\
\hline
$\mathcal{CPT}$&$H_{11}=H_{22}\quad (M_{11}=M_{22}\text{ and }\Gamma_{11}=\Gamma_{22})$\\
$\mathcal{CP}$&$H_{11}=H_{22}$ and $|H_{12}|=|H_{21}|$\\
$\mathcal{T}$&$|H_{12}|=|H_{21}|$\\
None&$H$ is general\\
\hline\hline
\end{tabular}
\end{center}
\end{table}
In these notes we are interested in $\mathcal{CPT}$-invariant theories.\footnote{the general framework can be found in~\cite{Branco:1999fs}, for example.} As a result, the matrix responsible by the evolution of our system is given by
\begin{equation}
H=
\begin{pmatrix}
M_{11}-\frac{i}{2}\Gamma_{11}&M_{12}-\frac{i}{2}\Gamma_{12}\\
M_{12}^\ast-\frac{i}{2}\Gamma_{12}&M_{11}-\frac{i}{2}\Gamma_{11}
\end{pmatrix}\,.
\end{equation}
If $CP$ was a symmetry of the system, i.e. $[\mathcal{CP},\mathcal{H}]=0$, the states $|P_{1,2}\rangle$ would be the true eigenstates of \Eref{eq:NMMtimeevol}.  The presence of $CP$-violating terms will destroy this result, in order to see this we go to the mass basis. The time evolution in \Eref{eq:NMMtimeevol} becomes trivial in the mass basis where the Hamiltonian $H$ is diagonal. The complex eigenvalues ($\mu_{L,H}$) and corresponding eigenvectors ($|P_{L,H}\rangle$) of $H$ are given by (using the phase convention $\xi=0$)
\begin{align}\label{eq:masseigen}
\begin{split}
&\textbf{Eigenvalues:}\hspace{5.2cm} \textbf{Eigenvectors:}\\
&\left[
\begin{array}{l}
M_{11}-\frac{i}{2}\Gamma_{11}-pq=\mu_L=m_L-\dfrac{i}{2}\Gamma_L\,,\\\\
M_{11}-\frac{i}{2}\Gamma_{11}+pq=\mu_H=m_H-\dfrac{i}{2}\Gamma_H
\end{array}\right.\quad
\left[\begin{array}{l}
|P_{L}\rangle=\dfrac{1}{\sqrt{|p|^2+|q|^2}}\left(p|P^0\rangle-q|\overline{P^0}\rangle\right)\,,\\\\
|P_{H}\rangle=\dfrac{1}{\sqrt{|p|^2+|q|^2}}\left(p|P^0\rangle+q|\overline{P^0}\rangle\right)
\end{array}\right.
\end{split}
\end{align}
where
\begin{equation}
p^2=M_{12}-\frac{i}{2}\Gamma_{12}\,,\quad q^2=M_{12}^\ast-\frac{i}{2}\Gamma_{12}^\ast\,.
\end{equation}
Note that $m_{H,L}$ and $\Gamma_{H,L}$ are not eigenvalues of $M$ and $\Gamma$ but, nevertheless, satisfy the relations $\text{Tr}\,M=m_H+m_L=2M_{11}$ and $\text{Tr}\,\Gamma=\Gamma_H+\Gamma_L=2\Gamma_{11}$.  They can also be written as
\begin{align}
\begin{split}
m_L=&M_{11}-\text{Re}\,pq\,,\quad m_H=M_{11}+\text{Re}\, pq\,,\\
\Gamma_L=&\Gamma_{11}+2\text{Im}\, pq\,,\quad \Gamma_H=\Gamma_{11}-2\text{Im}\,pq\,.
\end{split}
\end{align}
We are using a convention in which $\Delta m = m_H - m_L >0$. It is also convenient to define
\begin{equation}
\mu\equiv \frac{\mu_H+\mu_L}{2}\equiv m-\frac{i}{2}\Gamma\,,\quad \Delta\mu\equiv\mu_H-\mu_L\equiv \Delta m-\frac{i}{2}\Delta\Gamma\,.
\end{equation}
with
\begin{align}
\begin{split}
\Delta m=&m_H-m_L=2\text{Re}\,pq\,,\quad m=\frac{m_H+m_L}{2}=M_{11}\,,\\
\Delta \Gamma=&\Gamma_H-\Gamma_L=-4\text{Im}\,pq\,,\quad \Gamma=\frac{\Gamma_H+\Gamma_L}{2}=\Gamma_{11}\,.
\end{split}
\end{align}
The relation between these parameters and the elements of $H$ in the flavour basis can be found through the diagonalization procedure, leading to
\begin{equation}
\mu=H_{11}=H_{22}\,,\quad \Delta\mu=2\sqrt{H_{12}H_{21}}\,,\quad \frac{q}{p}=\sqrt{\frac{H_{21}}{H_{12}}}=\frac{2H_{21}}{\Delta \mu}
\end{equation}
Which in a more familiar form can be written as
\begin{align}
\begin{split}
&\left(\Delta m\right)^2-\frac{1}{4}\left(\Delta \Gamma\right)^2=4|M_{12}|^2-|\Gamma_{12}|^2\,,\quad \left(\Delta m\right)\left(\Delta \Gamma\right)=4\text{Re}\left(M_{12}^\ast\Gamma_{12}\right)\,,\\
&\frac{1-\bar{\epsilon}}{1+\bar{\epsilon}}=\frac{q}{p}=\sqrt{\frac{M_{12}^\ast-\frac{i}{2}\Gamma_{12}^\ast}{M_{12}-\frac{i}{2}\Gamma_{12}}}=\frac{2M_{12}^\ast-i\Gamma_{12}^\ast}{\Delta m-\frac{i}{2}\Delta\Gamma}=\frac{\Delta m-\frac{i}{2}\Delta\Gamma}{2M_{12}-i\Gamma_{12}}\equiv r e^{i\kappa}\,.
\end{split}
\end{align}
The small complex parameter $\bar{\epsilon}$ depends on the phase convention chosen
for the $P^0-\overline{P^0}$ system. Therefore, as a spurious phase, it shall not be taken as a physical measure of $CP$ violation. Nevertheless, the quantities $\text{Re}\,\bar{\epsilon}$ and $r$ are independent of phase conventions. Therefore, departures of $r$ from 1 are a measure of $CP$ violation. If $r=1$ ($\bar{\epsilon}=0$) then $p=q$ and the mass eigenstates in \Eref{eq:masseigen} coincide with the $CP$ eigenstates in \Eref{eq:CPeigen}. When this parameter is not $1$ there is a small admixture of the $CP$ eigenstates in the final (mass) eigenstates, i.e.
\begin{equation}
|P_L\rangle=\frac{1}{\sqrt{1+|\bar{\epsilon}|^2}}\left(|P_1\rangle+\bar{\epsilon} |P_2\rangle\right)\,,\quad |P_H\rangle=\frac{1}{\sqrt{1+|\bar{\epsilon}|^2}}\left(|P_2\rangle+\bar{\epsilon} |P_1\rangle\right)
\end{equation}
The physical observables measured in neutral meson oscillations can be parametrized by the dimensionless parameters
\begin{equation}
x=\frac{\Delta m}{\Gamma}\,,\quad y=\frac{\Delta \Gamma}{2\Gamma}\,,\quad r-1=\left|\frac{q}{p}\right|-1\,.
\end{equation}  
On can check, after some algebra, that 
\begin{equation}\label{eq:ortho}
\frac{|p|^2-|q|^2}{|p|^2+|q|^2}=\frac{1-r^2}{1+r^2}=\frac{\text{Im}(M_{12}^\ast\Gamma_{12})}{|M_{12}|^2+|\Gamma_{12}/2|^2+\frac{1}{4}[(\Delta m)^2+(\Delta\Gamma/2)^2]}\,,
\end{equation}
which is actually the quantity which measures the non-orthogonality between $P_{L,H}$, i.e.
\begin{equation}
\langle P_H|P_L\rangle=\frac{1-r^2}{1+r^2}=\frac{2\text{Re}\,\bar{\epsilon}}{1+|\bar{\epsilon}|^2}\,.
\end{equation}

Concerning time evolution. For the $|P^0_{L,H}\rangle$ states the solutions is rather trivial
\begin{equation}\label{eq:tPHL}
|P_{L,H}(t)\rangle=T_{L,H}(t)|P_{L,H}\rangle\,,\quad \text{with}\quad T_{X}(t)=e^{-i\mu_X t}
=e^{-\Gamma_{X}t/2}e^{-im_X t}\,.
\end{equation}  
The states produced in strong interactions are the $|P^0\rangle$ and $|\overline{P^0}\rangle$. It turns then useful to look at the times evolutions for these states. Using \Eref{eq:tPHL} and \Eref{eq:masseigen}, we find
\begin{align}
\begin{split}
|P^0(t)\rangle=&\frac{\sqrt{|p|^2+|q|^2}}{2p}\left[T_H(t)|P_H\rangle+T_L(t)|P_L\rangle\right]\,,\\|\overline{P^0}(t)\rangle=&\frac{\sqrt{|p|^2+|q|^2}}{2p}\left[T_H(t)|P_H\rangle-T_L(t)|P_L\rangle\right]\,.
\end{split}
\end{align}
This form is useful for studies in the $K^0-\overline{K^0}$ system. An alternative expression, useful in the $B^0-\overline{B^0}$ system
\begin{equation}
|P^0(t)\rangle=f_+(t)|P^0\rangle+\frac{q}{p}f_-(t)|\overline{P^0}\rangle\,,\quad |\overline{P^0}(t)\rangle=f_+(t)|\overline{P^0}\rangle+\frac{p}{q}f_-(t)|P^0\rangle\,,
\end{equation}
where
\begin{equation}
f_{\pm}(t)=\frac{T_{H}(t)\pm T_L(t)}{2}=\frac{1}{2}\left[e^{-im_H t}e^{-\Gamma_H t/2}\pm e^{-im_L t}e^{-\Gamma_L t/2}\right]\,.
\end{equation}
One see right away that for $t=0$ one has, for example, a pure $|P^0\rangle$ state, which as time evolves mixes with $|\overline{P^0}\rangle$. The probabilities of finding these states at later time are then given by
\begin{align}
\begin{split}
\mathcal{P}(P^0\rightarrow P^0;t)=&\mathcal{P}(\overline{P^0}\rightarrow \overline{P^0};t)=|f_+(t)|^2=\frac{1}{2}\text{exp}\left[-\frac{\Gamma t}{2}\right]\left(\cos(\Delta m t)+\cosh (\Delta \Gamma/2)\right)\\
\mathcal{P}(P^0\rightarrow \overline{P^0};t)=&\left|\frac{q}{p}\right|^2|f_-(t)|^2=\frac{1}{2}\left|\frac{q}{p}\right|^2\text{exp}\left[-\frac{\Gamma t}{2}\right]\left(-\cos(\Delta m t)+\cosh (\Delta \Gamma/2)\right)\\
\mathcal{P}(\overline{P^0}\rightarrow P^0;t)=&\left|\frac{p}{q}\right|^2|f_-(t)|^2=\frac{1}{2}\left|\frac{p}{q}\right|^2\text{exp}\left[-\frac{\Gamma t}{2}\right]\left(-\cos(\Delta m t)+\cosh (\Delta \Gamma/2)\right)
\end{split}
\end{align}

Note that several important aspects in meson oscillations were not covered here. For example, the existence of a reciprocal basis and its importance, this topic and many others can be found in~\cite{Branco:1999fs,Silva:2004gz}.

\subsection{Neutral Meson Mixing: The $K^0-\overline{K^0}$ and $B^{0}_{d,s}-\overline{B^0}_{d,s}$ systems}\label{subsec:3.5}

The general formalism for meson oscillations, shortly described in the previous section, can now be applied to the particular systems which we are interested in. 

\subsubsection{The $K^0-\overline{K^0}$ system: $|K^0\rangle=|d\bar{s}\rangle\,,\,|\overline{K^0}\rangle=|\bar{d}s\rangle$}\label{subsubsec:3.4.1}
In this system instead of using the notation heavy (H) of light (L) for the mass eigenstates we change it to the standard notation of long (L) and short (S) life time particle. This means
\begin{equation}
|K_S\rangle\equiv|P_L\rangle\quad\text{and}\quad |K_L\rangle\equiv|P_H\rangle \,.
\end{equation}
From the calculation of the $K_L-K_S$ mass difference, Gaillard and Lee [133] were able to estimate the value of the charm quark mass
before its discovery. Also,  kaon oscillation offers, within the Standard Model, a viable description of $\mathcal{CP}$ violation in $K_L\rightarrow \pi\pi$ decay.

In the kaon system we have
\begin{equation}
\tau_L\equiv\frac{1}{\Gamma_L}=51.16\pm0.21\,\text{ps}\,,\quad 
\tau_S\equiv\frac{1}{\Gamma_S}=\left(0.8954\pm0.0004\right)\times 10^{-1}\text{ps}
\end{equation}
\begin{equation}
m_K=497.614\pm 0.024\,\text{MeV}\,,\quad
\Delta m_K=\left(3.484\pm0.006\right)\times 10^{-12}\,\text{MeV}
\end{equation}
end up having to a good approximation
\begin{equation}\label{eq:sec3MGrelations}
\Delta m_K\simeq 2|M_{12}|\simeq -\dfrac{1}{2}\Delta\Gamma_K\simeq |\Gamma_{12}|\,,
\end{equation} 
which lead us to 
\begin{equation}
\frac{1-r^2}{1+r^2}\simeq\dfrac{1}{4}\text{Im}\left(\dfrac{\Gamma_{12}}{M_{12}}\right)
\end{equation}
In order to relate $\bar{\epsilon}$ to measurable quantities  we need to look at decays in the kaon system. The best channels to look at are the decay to pion. The pions are pseudo-scalars, which tell us that under the discrete symmetries $C$, $P$ and $T$ they transform in the same way as the bilinear $\overline{\psi}\gamma_5\chi$ in Tab.~\ref{tab:Discrete}. Therefore, under $CP$ we have
\begin{align}
\begin{split}
\textbf{One pion state:}&\quad\mathcal{CP}|\pi^0\rangle=-|\pi^0\rangle\,,\\
\textbf{Two pion state:}&\quad\mathcal{CP}|\pi^0\pi^0\rangle=+|\pi^0\pi^0\rangle\,,\quad\mathcal{CP}|\pi^+\pi^-\rangle=+|\pi^+\pi^-\rangle\,,\\
\textbf{Three pion state:}&\quad\mathcal{CP}|\pi^0\pi^0\pi^0\rangle=-|\pi^0\pi^0\pi^0\rangle\,,\quad \mathcal{CP}|\pi^+\pi^-\pi^0\rangle=(-1)^l|\pi^+\pi^-\pi^0\rangle\,.
\end{split}
\end{align} 
For the state $|\pi^+\pi^-\pi^0\rangle$ the relative angular momentum $(l)$ between $\pi^0$ and $\pi^+\pi^-$ is relevant. We can then conclude, from the above properties, that a two pion final state is $CP$-even and a three pion final state (with zero angular momentum) $CP$-odd. The kaon decays to two or three pions can then be characterized as
\begin{equation}
\mathcal{CP}\textbf{ conserving:}
\left[
\begin{array}{l}
K_S\rightarrow 2\pi \quad (\text{via }K_1)\\\\
K_L\rightarrow 3\pi \quad (\text{via }K_2)
\end{array}\right.\quad
\mathcal{CP}\textbf{ violating:}
\left[
\begin{array}{l}
K_S\rightarrow 3\pi \quad (\text{via }K_2)\\\\
K_L\rightarrow 2\pi \quad (\text{via }K_1)
\end{array}\right.
\end{equation}
This type of $CP$ violation is called indirect since it comes from the presence of a small admixture of $CP$ eigenstates in the final mass eigenstates, and not from a explicit breaking in the decay. We define the decay amplitudes:
\begin{equation}\label{eq:cptdecay}
\textbf{Decays: }
\left[
\begin{array}{l}
\langle (\pi\pi)_{I=0}|\mathcal{H}|K^0\rangle=A_0e^{i\delta_0}\\\\
\langle (\pi\pi)_{I=2}|\mathcal{H}|K^0\rangle=A_2e^{i\delta_2} 
\end{array}\right.\,,
\quad
\begin{array}{c}
\textbf{CPT}\\
\textbf{decays}
\end{array}\textbf{: }
\left[
\begin{array}{l}
\langle (\pi\pi)_{I=0}|\mathcal{H}|\overline{K^0}\rangle=-A^\ast_0e^{i\delta_0}\\\\
\langle (\pi\pi)_{I=2}|\mathcal{H}|\overline{K^0}\rangle=-A^\ast_2e^{i\delta_2} 
\end{array}\right.\,.
\end{equation}
Here $\delta_0$ and $\delta_2$ are the phase shifts where isospin quantum number $I=0$ and $I=2$ in $\pi\pi$ scattering. These are strong phases,
%reason why 
and thus
they do not change sign under $CPT$ conjugation. These phases were factored out explicitly so that the phases of $A_{0,2}$ are all of weak nature
\begin{equation}
A_0=|A_0|\text{exp}[i\phi_0]\,,\quad A_2=|A_2|\text{exp}[i\phi_2]\,.
\end{equation}
From the combination of \Eref{eq:cptdecay} and \Eref{eq:masseigen} we get
\begin{align}
\begin{split}
A_0^{S,L}\equiv&\langle (\pi\pi)_{I=0}|\mathcal{H}_{EW}|K_{S,L}\rangle=\frac{pA_0\pm qA_0^\ast}{\sqrt{|p|^2+|q|^2)}}\text{exp}[i\delta_0]=\frac{(1+\bar{\epsilon})A_{0}\mp(1-\bar{\epsilon})A_0^\ast}{\sqrt{2(1+|\bar{\epsilon}|^2)}}\text{exp}[i\delta_0]\,,\\
A_2^{S,L}\equiv&\langle (\pi\pi)_{I=2}|\mathcal{H}_{EW}|K_{S,L}\rangle=\frac{pA_2\pm qA_2^\ast}{\sqrt{|p|^2+|q|^2)}}\text{exp}[i\delta_0]=\frac{(1+\bar{\epsilon})A_{2}\mp(1-\bar{\epsilon})A_2^\ast}{\sqrt{2(1+|\bar{\epsilon}|^2)}}\text{exp}[i\delta_2]
\end{split}
\end{align}
Using the isotopic spin decomposition for the two pion states
\begin{align}
\begin{split}
\langle \pi^0\pi^0|=\langle(\pi\pi)_{I=0}|\frac{1}{\sqrt{3}}-\langle(\pi\pi)_{I=2}|\sqrt{\frac{2}{3}}\,,\\
\frac{1}{\sqrt{2}}(\langle \pi^+\pi^-|+\langle \pi^-\pi^+|)=\langle(\pi\pi)_{I=0}|\sqrt{\frac{2}{3}}+\langle(\pi\pi)_{I=2}|\frac{1}{\sqrt{3}}\,,
\end{split}
\end{align}
%where the charged pion state is correctly normalized and is this one that should enter in the transition amplitudes. This allow us to define the following transition amplitudes
where the charged pion state is correctly normalized, the transition amplitudes are defined as follow:
\begin{align}
\begin{split}
A(K_{S,L}\rightarrow \pi^0\pi^0)\equiv&\langle\pi^0\pi^0|\mathcal{H}_{EW}|K_{S,L}\rangle=\frac{1}{\sqrt{3}}A_{0}^{S,L}-\sqrt{\frac{2}{3}}A_2^{S,L}\,,\\
A(K_{S,L}\rightarrow \pi^+\pi^-)\equiv&\langle\pi^+\pi^-|\mathcal{H}_{EW}|K_{S,L}\rangle=\sqrt{\frac{2}{3}}A_{0}^{S,L}+\frac{1}{\sqrt{3}}A_2^{S,L}\,,
\end{split}
\end{align}
and
\begin{align}
\begin{split}
A(K^0\rightarrow \pi^+\pi^-)\equiv&\langle\pi^+\pi^-|\mathcal{H}_{EW}|K^{0}\rangle=\frac{1}{\sqrt{3}}\left[\sqrt{2}A_0+e^{i(\delta_2-\delta_0)}A_2\right]\\
A(\overline{K^0}\rightarrow \pi^+\pi^-)\equiv&\langle\pi^+\pi^-|\mathcal{H}_{EW}|\overline{K^0}\rangle=-\frac{1}{\sqrt{3}}\left[\sqrt{2}A^\ast_0+e^{i(\delta_2-\delta_0)}A^\ast_2\right]\\
A(K^0\rightarrow \pi^0\pi^0)\equiv&\langle\pi^0\pi^0|\mathcal{H}_{EW}|K^0\rangle=\frac{1}{\sqrt{3}}\left[A_0-\sqrt{2}e^{i(\delta_2-\delta_0)}A_2\right]\\
A(\overline{K^0}\rightarrow \pi^0\pi^0)\equiv&\langle\pi^0\pi^0|\mathcal{H}_{EW}|\overline{K^0}\rangle=-\frac{1}{\sqrt{3}}\left[A_0^\ast-\sqrt{2}e^{i(\delta_2-\delta_0)}A^\ast_2\right].
\end{split}
\end{align}
Experimentally the decay of $K_L$ to two-pion final state is observed and one can define useful quantities that measure this CP violation, i.e.
\begin{align}
\begin{split}
\eta_{00}=&\frac{A(K_L\rightarrow \pi^0\pi^0)}{A(K_S\rightarrow \pi^0\pi^0)}=\frac{A_0^L-\sqrt{2}A_2^L}{A_0^S-\sqrt{2}A_2^S}=
\epsilon-\frac{2\epsilon^\prime}{1-\sqrt{2}\omega}
\,,\\
\eta_{+-}=&\frac{A(K_L\rightarrow \pi^+\pi^-)}{A(K_S\rightarrow \pi^+\pi^-)}=\frac{\sqrt{2}A_0^L+A_2^L}{\sqrt{2}A_0^S+A_2^S}=
\epsilon+\frac{\epsilon^\prime}{1+\omega/\sqrt{2}}\,,
\end{split}
\end{align}
where $\omega\equiv \text{Re}[A_2/A_0]e^{i(\delta_2-\delta_0)}$. 
The experimental values for these quantities are~\cite{Agashe:2014kda}
\begin{align}
\eta_{00}=&(2.221\pm0.011)\times 10^{-3}\,\text{exp}[i(43.52\pm 0.06)^\degree]\,,\\
\eta_{+-}=&(2.232\pm0.011)\times 10^{-3}\,,\text{exp}[i(43.51\pm0.05)^\degree]\,,
\end{align}
showing how close these two quantities are. However, the fact that $\eta_{00}\neq\eta_{+-}$ is the source of $CP$ violation in the kaon decay to two-pion final states. The parameter $\epsilon$ is the measure of indirect $CP$ violation, which can be parametrized by amplitude ratio
\begin{equation}\label{eq:epsilon}
\epsilon\equiv\frac{A_0^L}{A_0^S}\simeq\bar{\epsilon}+i\xi=\frac{e^{i\pi/4}}{\sqrt{2}\Delta m_K}\left(\text{Im}\, M_{12}+2\xi\text{Re}\,M_{12}\right)\,,\quad \xi=\frac{\text{Im}[A_0]}{\text{Re}[A_0]}\,.
\end{equation}
Both $\bar{\epsilon}$ and $\xi$ have phase dependent conventions; however, since $\eta_{+-}$ and $\eta_{00}$ are experimental quantities $\epsilon$ is convention independent (similar to $\epsilon^\prime$). For direct $CP$ violation parameter $\epsilon^\prime$, where we have a direct transition of a $CP$-odd (even) term to a $CP$-even (odd), it is convenient to parametrize it through the following relation 
\begin{equation}
\epsilon^\prime=\frac{1}{\sqrt{2}}\left(\frac{A_2^L}{A_0^S}-\frac{A_2^S}{A^S_0}\frac{A_0^L}{A_0^S}\right)\,.
\end{equation}
For small $\bar{\epsilon}$, i.e.$|\bar{\epsilon}|\ll 1$, we can then write 
\begin{equation}
 \epsilon\simeq \bar{\epsilon}+i\frac{\text{Im}[A_0]}{\text{Re}[A_0]}\,,\quad
\epsilon^\prime\simeq \frac{-ie^{i\Phi^\prime}}{\sqrt{2}}\dfrac{\text{Re}[A_2]}{\text{Re}[A_0]}\left[\dfrac{\text{Im}[A_2]}{\text{Re}[A_2]}-\dfrac{\text{Im}[A_0]}{\text{Re}[A_0]}\right]\,.
\end{equation}
It is possible by a choice of phase convention to set $\text{Im}[A_0]=0$, known as Wu and Yang phase convention. The expressions are then simplified to
\begin{equation}
\textbf{In Wu-Yang phase convention:}\quad\left\{
\begin{array}{rl}
 \epsilon&\simeq \frac{1}{3}(2\eta_{+-}+\eta_{00})\simeq \bar{\epsilon}\\\\
\epsilon^\prime&\simeq \frac{1}{3}(\eta_{+-}-\eta_{00})\simeq \dfrac{e^{i\Phi^\prime}}{\sqrt{2}}\dfrac{\text{Im}[A_2]}{\text{Im}[A_0]}
\end{array}\right.\,,
\end{equation}
where $\Phi^\prime=\pi/2+\delta_2-\delta_0\simeq \pi/4$. The parameter $\epsilon^\prime$, which is only non-zero if there is $CP$ violation in the decay amplitudes is proportional to the difference of $\eta_{+-}$ and $\eta_{00}$, which almost cancel. A more practical quantity to evaluate $\epsilon^\prime$ is the ratio given by 
\begin{equation}
\text{Re}(\epsilon^\prime/\epsilon)\simeq\frac{1}{6(1+\omega/\sqrt{2})}\left(1-\left|\frac{\eta_{00}}{\eta_{+-}}\right|^2\right)\,.
\end{equation}
The parameter $\omega$ is small, i.e. $|\omega|\sim 1/25$, and often ignored. This quantity can be accurately measured on the rations $\Gamma(K_L\rightarrow \pi^0\pi^0)/\Gamma(K_L\rightarrow \pi^+\pi^-)$ and $\Gamma(K_S\rightarrow \pi^0\pi^0)/\Gamma(K_S\rightarrow \pi^+\pi^-)$, in terms of which 
\begin{equation}
\left|\frac{\eta_{00}}{\eta_{+-}}\right|^2=\frac{\Gamma(K_L\rightarrow \pi^0\pi^0)/\Gamma(K_L\rightarrow \pi^+\pi^-)}{\Gamma(K_S\rightarrow \pi^0\pi^0)/\Gamma(K_S\rightarrow \pi^+\pi^-)}\,.
\end{equation}
From the fit to $K\rightarrow \pi\pi$ data we get~\cite{Agashe:2014kda}
\begin{equation}\label{eq:espilonexp}
|\epsilon|=(2.228\pm0.011)\times 10^{-3}\,,\quad \text{Re}[\epsilon^\prime/\epsilon]=(1/65\pm 0.26)\times 10^{-3}\,.
\end{equation}
Another important observable is the $CP$ asymmetry of time integrated semi-leptonic decay rates
\begin{equation}
\delta_L\equiv\frac{\Gamma(K_L\rightarrow \ell^+\nu_\ell\pi^-)-\Gamma(K_L\rightarrow \ell^-\bar{\nu}_\ell\pi^+)}{\Gamma(K_L\rightarrow \ell^+\nu_\ell\pi^-)+\Gamma(K_L\rightarrow \ell^-\bar{\nu}_\ell\pi^+)}=\frac{1-\left|\frac{q}{p}\right|^2}{1+\left|\frac{q}{p}\right|^2}=\frac{2\text{Re}[\bar{\epsilon}]}{1+|\bar{\epsilon}|^2}\longrightarrow\overbrace{\frac{2\text{Re}[\epsilon]}{1+|\epsilon|^2}}^{\textbf{Wu-Yang}}
\end{equation}
This observable measure the orthogonality between $K_L$ and $K_S$, see \Eref{eq:ortho}.

We can now shortly evaluate $\epsilon$ within the SM. The off-diagonal element $M_{12}$ in the kaon system is given by
\begin{equation}
2m_KM_{12}^\ast=\langle \overline{K^0}|\mathcal{H}_{eff}^{\Delta S=2}|K^0\rangle\,,
\end{equation}
where the factor $2m_K$ is due to the normalization of external states. $|\mathcal{H}_{\mbox{\scriptsize{eff}}}^{\Delta S=2}$ is the effective Hamiltonian for the $\Delta S=2$ transitions, this in lower order is given by the box diagrams in \Fref{fig:2}a. We can integrate out the heavy internal particles and run down to low energies with the renormalization group. By doing this we obtain the contact term
\begin{equation}
Q(\Delta S=2)=(\bar{s}d)_{V-A}(\bar{s}d)_{V-A}\,.
\end{equation}
The effective Hamiltonian, including leading and next-to-leading QCD corrections in the improved RGEs, for scales $\mu<\mu_c=\mathcal{O}(m_c)$ is given by
\begin{align}
\begin{split}
\mathcal{H}_{eff}^{\Delta S=2}=&\frac{G_F^2}{16\pi^2}M_W^2\left[(V_{cs}^\ast V_{cd})^2 \eta_1 S_{0}(x_c)+(V_{ts}^\ast V_{td})^2 \eta_2 S_{0}(x_t)+2(V_{cs}^\ast V_{cd})(V_{ts}^\ast V_{td}) \eta_3 S_{0}(x_c,x_t)\right]\\
&\times [\alpha_s^{(3)}(\mu)]^{-2/9}\left[1+\frac{\alpha_s^{(3)(\mu)}}{4\pi}J_3\right]Q(\Delta S=2)+\text{h.c.}
\end{split}
\end{align}
with $\alpha_s^{(3)}$ the strong coupling constant in an effective three flavour theory and $J_3=1.895$ in NDR scheme~\cite{Buras:1998raa}. The $S_0$ loop functions are given by ($x_i=m_i^2/M_W^2$)
\begin{align}
\begin{split}
S_0(x_t)=&2.39\left(\frac{m_t}{167\,\text{GeV}}\right)^{1.52}\,,\quad S_0(x_c)=x_c\,,\\
S_0(x_c,x_t)=&x_c\left[\ln \frac{x_t}{x_c}-\frac{3x_t}{4(1-x_t)}-\frac{3x_t^2\ln x_t}{4(1-x_t)^2}\right]\,.
\end{split}
\end{align}
The factors $\eta_{1,2,3}$ are correction factors describing short distance QCD effects and at NLO read~\cite{Buras:1998raa}: $\eta_1=1.38\pm 0.20$, $\eta_2=0.57\pm 0.01$, $\eta_3=0.47\pm 0.04$. We can now take the matrix element of our contact interaction, the non-perturbative part of the calculation, we get
\begin{equation}
\langle \overline{K^0}|Q(\Delta S=2)|K^0\rangle\equiv \frac{8}{3}B_K(\mu)F_K^2m_K^2\,,\quad \hat{B}_K=B_K(\mu)[\alpha_s^{(3)}(\mu)]^{-2/9}\left[1+\frac{\alpha_s^{(3)}(\mu)}{4\pi}J_3\right]\,,
\end{equation}
where $\hat{B}_K$ is a renormalization group invariant parameter and $F_K=160$ MeV is the kaon decay constant. We finally find the matrix element to be
\begin{align}\label{eq:sec3M12K}
\begin{split}
M_{12}=\frac{G_F^2}{12\pi^2}F_K^2\hat{B}_Km_KM_W^2&\left[(V_{cs}^\ast V_{cd})^2 \eta_1 S_{0}(x_c)+(V_{ts}^\ast V_{td})^2 \eta_2 S_{0}(x_t)\right.\\
&\left.+2(V_{cs}^\ast V_{cd})(V_{ts}^\ast V_{td}) \eta_3 S_{0}(x_c,x_t)\right]\,.
\end{split}
\end{align}
Inserting this last result into \Eref{eq:epsilon} we obtain, in the Wu-Yang phase convention,
\begin{equation}\label{eq:epsilon2}
\epsilon\simeq C_\epsilon \hat{B}_K\text{Im}[V_{ts}^\ast V_{td}]\left\{\text{Re}[V_{cs}^\ast V_{cd}]\left[\eta_1 S_0(x_c)-\eta_3S_0(x_c,x_t)\right]-\text{Re}[V_{ts}^\ast V_{td}] \eta_2 S_0(x_t)\right\}e^{i\pi/4}\,,
\end{equation}
with
\begin{equation}
C_\epsilon=\frac{G_F^2 F_K^2 m_K M_W^2}{6\sqrt{2}\pi^2 \Delta m_K}\simeq 3.837\times 10^4\,.
\end{equation}
Corrections of the order $\text{Re}[V_{ts}^\ast V_{td}]/\text{Re}[V_{cs}^\ast V_{cd}]=\mathcal{O}(\lambda^4)$ have been neglected and we have used the unitary relation $\text{Im}[(V_{cs}^\ast V_{cd})^\ast]=\text{Im}[V_{ts}^\ast V_{td}]$. Using the standard CKM parametrization, \Eref{eq:CKMPDG}, and comparing Eq.~\ref{eq:epsilon2} with the experimental value \Eref{eq:espilonexp} we can extract the CKM CP phase $\delta$, important for the unitary triangle analysis.   

The $K_L-K_S$ mass difference is now trivial to extract from Eqs.~(\ref{eq:sec3M12K}) and ~(\ref{eq:sec3MGrelations}). Using the fact that $|V_{ts}^\ast V_{td}|\ll |V_{cs}^\ast V_{cd}|$, the charm-quark contribution in the loop dominates and we get
\begin{equation}
\Delta m_K\simeq \frac{G_F^2}{12\pi^2}F_K^2\hat{B}_Km_KM_W^2|V_{cs}^\ast V_{cd}|^2S_0(x_c)\,.
\end{equation}

\subsubsection{The $B^0_{d,s}-\overline{B^0}_{d,s}$ system: $|B_d^0\rangle=|\bar{b}d\rangle\,,\,|\overline{B^0}_d\rangle=|b\bar{d}\rangle$, $|B_s^0\rangle=|\bar{b}s\rangle\,,\,|\overline{B^0}_s\rangle=|b\bar{s}\rangle$}\label{subsubsec:3.4.2}

Contrarily to the $K^0-\overline{K^0}$ system, in the $B^0_{d,s}-\overline{B^0}_{d,s}$ system the long distance effects are very small $ |\Gamma_{12}|\ll |M_{12}|$ (see discussion in~\cite{Silva:2004gz}). Therefore, to leading order in
$| \Gamma_{12}/M_{12}|$, we get
\begin{equation}\label{secB: relations}
\Delta m_{B_q}=2|M^{q}_{12}|\,,\quad\Delta\Gamma_{B_q}=2\text{Re}(M_{12}^{q\ast}\Gamma^q_{12})/|M^q_{12}|\,,\quad
\dfrac{q}{p}\simeq\dfrac{M_{12}^{q\ast}}{|M^q_{12}|}\left[1-\dfrac{1}{2}\text{Im}\left(\dfrac{\Gamma^q_{12}}{M^q_{12}}\right)\right]\,,
\end{equation}
with $q=d,s$ and the notation of $H$, $L$ states given in the general discussion is kept here. In the $B$-system we have $|V_{td}^\ast V_{tb}|\sim |V_{cd}^\ast V_{cb}|$, however due to the quarks spectrum, i.e. $m_{u,c}\ll m_t$, the top quark contribution is now the one dominating. 

\begin{figure}[h]
\centering\includegraphics[width=.7\linewidth]{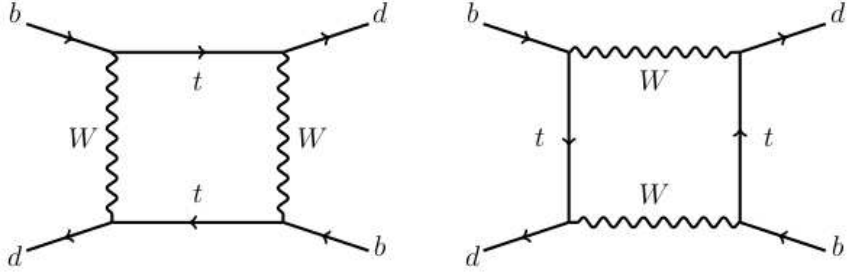}
\caption{Box diagram contributing to $B^0=\overline{B^0}$ mixing}
\label{fig:20}
\end{figure}

In a similar way as was done for the $K$-system, the off-diagonal element $M_{12}^{(q)}$ is given by
\begin{equation} \label{secB: M12}
2m_{B_q}|M_{12}^{(q)}|=|\langle \overline{B^0}_q|\mathcal{H}_{\mbox{\scriptsize{eff}}}^{\Delta B=2}|B_q^0\rangle|\,.
\end{equation}
The effective Hamiltonian , obtained from integrating out the top quark, is given by
\begin{equation}
\mathcal{H}_{\mbox{\scriptsize{eff}}}^{\Delta B=2}=\frac{G_F^2}{16\pi^2}M_W^2(V_{tb}^\ast V_{tq})^2
\eta_B S_0(x_t)[\alpha_s^{(5)}(\mu_q)]^{-6/23}\left[1+\frac{\alpha_s^{(5)}(\mu_q)}{4\pi}J_b\right]Q(\Delta B=2)+\text{h.c.}\,,
\end{equation}
with $\mu_q=\mathcal{O}(m_q)$ and $J_5=1.627$. The contact term is given by
\begin{equation}
Q(\Delta B=2)=(\bar{b}q)_{V-A}(\bar{b}q)_{V-A}\,.
\end{equation} 
Taking the matrix element we get, in an analogous ways as for the $K$ system,
\begin{equation}
\langle \overline{B^0}_q|Q(\Delta B=2)|B_q^0\rangle\equiv \frac{8}{3}B_{B_q}(\mu)F_{B_q}^2m_{B_q}^2\,,\quad \hat{B}_{B_q}=B_{B_q}(\mu)[\alpha_s^{(5)}(\mu_q)]^{-6/23}\left[1+\frac{\alpha_s^{(5)}(\mu_q)}{4\pi}J_b\right]\,,
\end{equation}
with $F_{B_q}$ the decay constant for $B_q$. Using Eq.~\eqref{secB: M12} and the first relation in Eq.~\eqref{secB: relations} one gets
\begin{equation}
\Delta m_{B_{q}}\simeq\frac{G_F^2}{6\pi^2}\eta_Bm_{B_q} \hat{B}_{B_q}F_{B_q}^2M_WS_0(x_t)|V_{tq}|^2\,.
\end{equation}
This relation for the mass difference in important in the standard analysis of the unitary triangle.

\section{Flavour Physics Beyond the SM}\label{sec:4}

$CP$ violation in the SM comes from the flavour sector. However, $CP$ violation observed so far is too small by a factor of $10^{-16}$ to explain the absence of anti-matter, which means that physics beyond the SM (BSM) must exist. Therefore, a right question wouldn't be whether BSM exist or not, but at which scale it will show up. For particle physicists, there are also two different reasons hinting us that surprises might be awaiting to be discovered by at around TeV scale.

The first reason is coming from so-called `the fine-tuning/hierarchy problem, which is related to the lightness of the Higgs particle compared to a arbitrarily high scale (below PLACK scale). The recently discovered Higgs particle, which is the only missing piece of the Standard Model (SM), may be the first fundamental scalar particles we have discovered. It is employed for the electroweak symmetry breaking (EWSB) and for generating masses for the fermions. While it explains why the weak force, unlike all other forces, is very short-ranged, it also provide us a problem. In order to obtain the observed $\sim 125$GeV, which is far much smaller than the size of quantum corrections from seemingly unrelated forces, a miraculous fine-tuning has to be invoked. However, this `naturalness' problem can be solved, if new physics exists beyond the Higgs particle. And the corresponding new physics and new particles are predicted to be observed in the scale of EWSB.

The other reason is coming from cosmology. According to the standard model of cosmology, which is now well established, some twenty percent of the energy of the universe comes from matter that does not shine (that is, electromagnetically neutral), but is much more massive than neutrinos. There are no candidates among particles in the SM for this type of matter, so called ``dark matter (DM)''. The cosmological and astrophysical observations suggest us that the mass of the DM particles is light enough to be produced and observed at the TeV scale.

In a general picture of physics beyond the SM one can see the amplitude of a given process being described in the form
\begin{equation}
A(\text{in}\rightarrow \text{out})\simeq A_0\left[\frac{C_{SM}}{M_W^2}+\frac{C_{NP}}{\Lambda_{NP}^2}\right]\,.
\end{equation}
The coefficients $C_{SM(NP)}$ will the depend of the process and SM extension. However, we can see that flavour physics can place strong constraints on new physics even beyond the LHC reach. In scenarios where new physics does not respect the SM symmetries or breaking pattern, the coefficients tend to be hierarchical $C_{SM}\ll C_{NP}$, allowing to probe large scales. 

For example, in the SM there are only two $|\Delta F|=2$ operators entering in $K^0-\bar{K}^0$ and $B^0-\bar{B}^0$ mixing, see Sec.~\ref{sec:3}. A common feature in NP flavour models is the presence of additional four-quark operators, which change the flavour number by two units. Those interactions can place a strong bounds on the NP scale. Without specifying its origin we can typically describe them through the effective Lagrangian 
\begin{align}
\begin{split}
\mathcal{L}^{|\Delta F|=2}_{NP}=&\frac{1}{\Lambda^2}\sum_{i=1}^5c^{q_\alpha q_\beta}_i\mathcal{Q}_i^{q_\alpha q_\beta}+\frac{1}{\Lambda^2}\sum_{i=1}^3\tilde{c}^{q_\alpha q_\beta}_i\tilde{\mathcal{Q}}_i^{q_\alpha q_\beta}
\end{split}
\end{align}
with the dimension six $|\Delta F|=2$ operators given by~\cite{{Grossman:2007bd}}
\begin{equation} \label{eq:dim6}
\begin{array}{rlrl}
\mathcal{Q}_1^{q_\alpha q_\beta }=&\left(\overline{q}_{\beta L}\gamma_\mu q_{\alpha L}\right)\left(\overline{q}_{\beta L}\gamma_\mu q_{\alpha L}\right)\,,&\tilde{\mathcal{Q}}_1^{q_\alpha q_\beta }=&\left(\overline{q}_{\beta R}\gamma_\mu q_{\alpha R}\right)\left(\overline{q}_{\beta R}\gamma_\mu q_{\alpha R}\right)\,,\\
\mathcal{Q}_2^{q_\alpha q_\beta }=&\left(\overline{q}_{\beta R} q_{\alpha L}\right)\left(\overline{q}_{\beta R} q_{\alpha L}\right)\,,&\tilde{\mathcal{Q}}_2^{q_\alpha q_\beta }=&\left(\overline{q}_{\beta L}  q_{\alpha R}\right)\left(\overline{q}_{\beta L} q_{\alpha R}\right)\,,\\
\mathcal{Q}_3^{q_\alpha q_\beta }=&\overline{q}_{\beta R}^a  q_{\alpha L}^b \overline{q}_{\beta R}^b  q_{\alpha L}^a \,,&\tilde{\mathcal{Q}}_3^{q_\alpha q_\beta }=&\overline{q}_{\beta L}^a  q_{\alpha R}^b \overline{q}_{\beta L}^b  q_{\alpha R}^a \,,\\
\mathcal{Q}_4^{q_\alpha q_\beta }=&\left(\overline{q}_{\beta R}q_{\alpha L}\right)\left(\overline{q}_{\beta L} q_{\alpha R}\right)\,,&&\\
\mathcal{Q}_5^{q_\alpha q_\beta }=&\overline{q}_{\beta R}^a  q_{\alpha L}^b  \overline{q}_{\beta L}^b  q_{\alpha R}^a \,.&&\\
\end{array}  
\end{equation}
\Table~\ref{tab:bounds} summarizes the bounds on the new physics scale or Wilson coefficient. 
\begin{table}[h]
\begin{center}
\caption{Summary of the most relevant bounds on $d=6$ four-quark flavour operators. Taken from~\cite{Isidori:2010kg}}
\label{tab:bounds}
\begin{tabular}{lccccl}
\hline\hline\\[-10pt]
\multirow{2}{*}{\textbf{Operator}}&\multicolumn{2}{c}{\textbf{Bounds on $\Lambda$ in TeV ($c_i^{\mbox{\scriptsize{NP}}}=1$)}}&\multicolumn{2}{c}{\textbf{Bounds on $c_i^{\mbox{\scriptsize{NP}}}$ ($\Lambda=1\,\text{TeV}$)}}&\multirow{2}{*}{\textbf{Observable}}\\
&Re&Im&Re&Im&\\
\hline
$(\overline{s_L}\gamma^\mu d_L)^2$&$9.8\times 10^2$&$1.6\times 10^4$&$9.0\times 10^{-7}$&$3.4\times 10^{-9}$&\multirow{2}{*}{$\Delta m_K;\, \epsilon_K$}\\
$(\overline{s_R} d_L)(\overline{s_L}d_R)$&$1.8\times 10^4$&$3.2\times 10^5$&$6.9\times 10^{-9}$&$2.6\times 10^{-11}$&\\
\hline
$(\overline{c_L}\gamma^\mu u_L)^2$&$1.2\times 10^3$&$2.9\times 10^3$&$5.6\times 10^{-7}$&$1.0\times 10^{-7}$&\multirow{2}{*}{$\Delta m_D;\, |q/p|,\, \phi_D$}\\
$(\overline{c_R} u_L)(\overline{c_L}u_R)$&$6.2\times 10^3$&$1.5\times 10^4$&$5.7\times 10^{-8}$&$1.1\times 10^{-8}$&\\
\hline
$(\overline{b_L}\gamma^\mu d_L)^2$&$6.6\times 10^2$&$9.3\times 10^2$&$2.3\times 10^{-6}$&$1.1\times 10^{-6}$&\multirow{2}{*}{$\Delta m_{B_d};\,S_{\psi K_S}$}\\
$(\overline{b_R} d_L)(\overline{b_L}d_R)$&$2.5\times 10^3$&$3.6\times 10^3$&$3.9\times 10^{-7}$&$1.9\times 10^{-7}$&\\
\hline
$(\overline{b_L}\gamma^\mu s_L)^2$&$1.4\times 10^2$&$2.5\times 10^2$&$5.0\times 10^{-5}$&$1.7\times 10^{-5}$&\multirow{2}{*}{$\Delta m_{B_s};\,S_{\psi \phi}$}\\
$(\overline{b_R} s_L)(\overline{b_L}s_R)$&$4.8\times 10^2$&$8.3\times 10^2$&$8.8\times 10^{-6}$&$2.9\times 10^{-6}$&\\
\hline\hline
\end{tabular}
\end{center}
\end{table}
As seen in \Tref{tab:bounds} new physics scale tends to be pushed to very high scales (several orders above the TeV scale) due to flavour constraints. Saying it in other way, in order to have new physics at the TeV scale we need it to have specific flavour structure not so different from that of the SM at low energies. The quest for viable new physics models is known as ``New Physics flavor problem". In this section we will look at some extensions and their confrontation with flavour observables.

\subsection{Minimal flavour Violation hypothesis}\label{subsec:4.1}
One popular solution to the flavour puzzle is the minimal flavour violation (MFV) hypothesis~\cite{D'Ambrosio:2002ex}. The MFV is not a model, but a simple framework for the flavour structure on new physics seen from and effective field theory point of view. The main assumptions are:
\begin{itemize}
\item No new operators beyond those present in the SM;
\item All flavour changing transitions are governed by $CKM$, i.e. no new complex phases beyond those present in the SM
\begin{equation}
A(\text{in}\rightarrow \text{out})\propto \lambda_{CKM}^i\underbrace{(F^i_{SM}+F_{NP}^i)}_{\text{real}}\,.
\end{equation}
\end{itemize}
In the SM the CKM is the only source of flavour violation and is approximately a unit matrix. The SM has no flavour changing neutral currents at tree level, and in this way CKM-induced flavour change interactions are guarantee to be small. If new physics is flavour-diagonal such that all the flavour-violation goes through the CKM, then we are guaranteed to have small effects. Therefore, just like in the SM, Yukawa couplings are the only sources of flavour symmetry breaking in physics beyond the SM. In MFV we then have a CKM and GIM suppression working in a similar way to the SM, allowing and EFT-like approach.

\begin{figure}[h]
\centering\includegraphics[width=.5\linewidth]{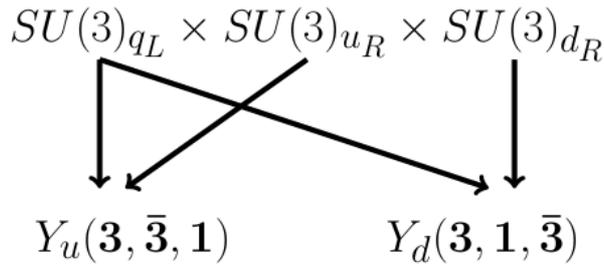}
\caption{Global flavour symmetry and spurious fields transformations}
\label{fig:21}
\end{figure}

The effective approach of MFV takes into account the larger flavour group in the SM when the Yukawa intersections are absent, see \Eref{eq:globalU3}. This symmetry is explicitly broken in the presence of the Yukawa terms, but we can formally restore it by promoting the Yukawa matrices to be spurions (appropriate dimensionless auxiliary fields), which transform under the flavour group in the appropriate way to make it invariant (see \Fref{fig:21}).

Using the $SU(3)_q^3\times SU(3)_l^2$ symmetry, we can rotate the background values of the auxiliary field $Y$, as we did in \Eref{eq:WBTIII},
\begin{equation}
Y_d=\lambda_d\,,\quad Y_u=V_{CKM}^\dagger \lambda_u\,,\quad Y_\ell=\lambda_{\ell}\,.
\end{equation}
MFV requires that the dynamics of flavour violation is completely determined by the structure of the ordinary Yukawa couplings. In particular, all $\mathcal{CP}$ violation effects originates from the $CKM$ phase. From the hierarchical structure of the Yukawa matrix, i.e. only top Yukawa is large, we can define the new physics flavour coupling
\begin{equation}
\left(\lambda_{FC}\right)_{ij}=
\left\{
\begin{array}{ll}
(Y_uY_u^\dagger)_{ij}\simeq y_t^2 V_{3i}^\ast V_{3j}\,,&i\neq j\\
0\,,&i=j
\end{array}
\right.
\end{equation}
\begin{table}[h]
\begin{center}
\caption{Relevant $d=6$ MFV flavour operators and their bounds on new physics. Taken from~\cite{Hurth:2008jc}. }
\begin{tabular}{ccc}
\hline\hline
\textbf{MFV $d=6$ operator}&\textbf{Observables}&\textbf{$\Lambda$ [TeV]}\\
\hline
$\frac{1}{2}(\overline{q_L}\lambda_{FC}\gamma_\mu q_L)^2$&$\epsilon_K$, $\Delta m_{D_d}$&5.9\\
$\phi^\dagger(\overline{d_R}\lambda_d\lambda_{FC}\sigma_{\mu\nu}q_L)(eF^{\mu\nu})$&$B\rightarrow X_s\gamma$, $B\rightarrow X_s\ell^+\ell^-$&6.1\\
$\phi^\dagger(\overline{d_R}\lambda_d\lambda_{FC}\sigma_{\mu\nu}T^aq_L)(eg_s G^{a\mu\nu})$&$B\rightarrow X_s\gamma$, $B\rightarrow X_s\ell^+\ell^-$&3.4\\
$(\overline{q_L}\lambda_FC\gamma_\mu q_L)(eD_\nu F^{\mu\nu} )$&$B\rightarrow X_s\ell^+\ell^-$&1.5\\
$i(\overline{q_L}\lambda_{FC}\gamma_\mu q_L)\phi^\dagger D^\mu\phi$&$B\rightarrow X_s\ell^+\ell^-$, $B_s\rightarrow \mu^+\mu^-$&1.1\\
$i(\overline{q_L}\lambda_{FC}\gamma_\mu \tau^aq_L)\phi^\dagger \tau^aD^\mu\phi$&$B\rightarrow X_s\ell^+\ell^-$, $B_s\rightarrow \mu^+\mu^-$&1.1\\
$(\overline{q_L}\lambda_{FC}\gamma_\mu q_L)(\overline{\ell_L}\gamma^\mu\ell_L)$&$B\rightarrow X_s\ell^+\ell^-$, $B_s\rightarrow \mu^+\mu^-$&1.7\\
$(\overline{q_L}\lambda_{FC}\gamma_\mu \tau^aq_L)(\overline{\ell_L}\gamma^\mu \tau^a\ell_L)$&$B\rightarrow X_s\ell^+\ell^-$, $B_s\rightarrow \mu^+\mu^-$&1.7\\
$(\overline{q_L}\lambda_{FC}\gamma_\mu q_L)(\overline{e_R}\gamma^\mu e_R)$&$B\rightarrow X_s\ell^+\ell^-$, $B_s\rightarrow \mu^+\mu^-$&2.7\\
\hline\hline
\end{tabular}
\end{center}
\end{table}
The basic building blocks of FCNC operators are
\begin{equation}
\overline{q_L}Y_uY_u^\dagger q_L\,,\quad \overline{d_R}Y_D^\dagger Y_uY_u^\dagger q_L\,,\quad \overline{d_R}Y_d^\dagger Y_uY_u^\dagger Y_d d_R
\end{equation} 
expanding in powers of the off-diagonal $CKM$ matrix elements and in powers of the small Yukawa couplings, such as
\begin{equation}
\overline{q_L}\lambda_{FC}q_{L}\quad \text{and}\quad \overline{d_R}\lambda_d\lambda_{FC}q_L
\end{equation}
The MFV framework is general and can be implemented in a given BSM scenario, e.g. SUSY and  composite Higgs models, resulting in reducing the cutoff scale (flavour bound) from $\mathcal{O}(1000)$ TeV to $\mathcal{O}(1)$ TeV, which in turn makes it a very predictive theory framework. Compared to SM, only the flavour-independent magnitude of the transition amplitudes can be modified. A fingerprint of this framework is the prediction $(\sin 2\beta)_{B\rightarrow \psi K_s}=(\sin 2\beta)_{K\rightarrow \pi\nu\bar{\nu}}$, which can be identified by experiments.

\subsection{Partial compositeness}\label{subsec:4.2}
Partial compositeness is a completely different way of flavour protection mechanism~\cite{Kaplan:1991dc}. The idea is to generate quark and lepton masses through linear couplings of the Standard Model fields to composite operators, i.e.
\begin{equation}
\Delta_L\overline{q_L}\mathcal{O}_R+\Delta^u_R\overline{u_R}\mathcal{O}^u_R+\Delta^d_R\overline{d_R}\mathcal{O}^d_R+\cdots\,,
\end{equation}
where $\Delta_{L,R}$ are known as pre-Yukawa couplings and $\mathcal{O}_{L,R}$ are fermionic operators arising from the strong sector. The nice aspect of this linear coupling is that no relevant operator can be built out of $\mathcal{O}_{L,R}$, since both have a classical mass dimension of $5/2$. Also, the quadratic operators $\mathcal{O}_L\mathcal{O}_L$, $\mathcal{O}_R\mathcal{O}_R$ vanish due to spinor identities and $\mathcal{O}_L\mathcal{O}_R$ is forbidden by gauge invariance. Therefore, the lowest-dimension operators on can build out of the composite operators are $\mathcal{O}_L\partial\!\!\!/\mathcal{O}_L$ and
$\mathcal{O}_R\partial\!\!\!/\mathcal{O}_R$, which have classical dimension six and therefore irrelevant. 

The physical light fermions will then be a mixture of both elementary and composite states, known as partial compositeness,
\begin{equation}
|\psi_{phys}\rangle=\cos\theta |\psi_{elem}\rangle+\sin\theta |\psi_{comp}\rangle\,.
\end{equation}
The flavour problem in theories with strong dynamics can be improved if partial compositeness is implemented.

\begin{equation}
m_\Psi\simeq g_\Psi f\quad \longrightarrow \quad y_{SM}\simeq \frac{\Delta_L\Delta_R}{m_\Psi}
\end{equation}

Partial compositeness provide partial solutions to both flavour and hierarchy puzzles. Still, this is a partial solution since from the kaon system $\epsilon_K$ and $\epsilon^\prime_K/\epsilon_K$ one still needs some sort of alignment, at least in the down sector. On the other hand, in this framework we can have a naturally sizable non-standard contribution to $\Delta a_{CP}$. This approach can be an alternative to MFV.

\subsection{B physics at the LHC}\label{subsec:4.3}
Rare decays based on the flavour transition $b\rightarrow s$ have for some time call the attention of the flavour community, as they can be sensitive probes of new physics~\cite{Altmannshofer:2014rta,Altmannshofer:2015sma}:
\begin{itemize}
\item[]\textbf{hadronic: }$B\rightarrow \phi K,\, B\rightarrow \eta^\prime K,\, B_s\rightarrow \phi\phi,\, B\rightarrow K\pi,\, B_s\rightarrow KK,\,\cdots$
\item[]\textbf{radiative: }$B\rightarrow X_s\gamma,\, B\rightarrow K^\ast \gamma,\, B_s\rightarrow \phi \gamma,\,\cdots$
\item[]\textbf{semi-leptonic: }$B\rightarrow X_s\ell\ell,\, B\rightarrow K\ell\ell,\, B\rightarrow K^\ast \ell\ell,\, B_s\rightarrow \phi \ell\ell,\,\cdots$
\item[]\textbf{leptonic: }$B_s\rightarrow \mu\mu$
\item[]\textbf{neutrino: }$B\rightarrow K\nu\bar{\nu},\, B\rightarrow K^\ast \nu\bar{\nu}$
\end{itemize}
The most relevant ones in order to constrain new physics in the LHC era are the leptonic, semi-leptonic and radiative exclusive decays. 

Recently, the LHCb collaboration observed an excess in $B\rightarrow K^\ast \mu^+\mu^-$ decay~\cite{Aaij:2013qta} by measuring the angular observables with a minimal sensitivity to the choice of form factors~\cite{Descotes-Genon:2013vna}. This tension can be soften by the presence of new physics. One useful way to search for  new physics that could induce these deviations is to look at the effective Hamiltonian relevant for this transition. From the complete list presented in Sec.~\ref{sec:2}, the current-current, QCD penguin and electroweak penguin operators ar typically dominated by the SM contribution at low energies and will only contribute to the considered observables though mixing with the dominant operators. This effect is therefore small. The chromomagnetic dipole operators, for leptonic and semi-leptonic decays enter only through mixing. Tensor operator do not appear in $d=6$ operator expansion the the SM. Having this information we can write the relevant effective Hamiltonian
\begin{equation}
\mathcal{H}_{eff}=-\frac{G_F}{\sqrt{2}}\frac{\alpha}{\pi}V_{tb}V_{ts}^\ast\sum_i\left(C_i^\ell\mathcal{O}_i^\ell+C_i^{\prime\ell}\mathcal{O}_i^{\prime\ell}\right)
\end{equation}
with $\alpha$ the fine structure constant and the operators considered are
\begin{equation}
\begin{array}{ll}
\mathcal{O}_7=\frac{m_b}{e}(\overline{s}\sigma_{\mu\nu}P_R b)F^{\mu\nu}\,,&\quad
\mathcal{O}_7^\prime=\frac{m_b}{e}(\overline{s}\sigma_{\mu\nu}P_L b)F^{\mu\nu}\,,\\
\mathcal{O}_9^\ell=(\overline{s}\gamma_\mu P_Lb)(\overline{\ell}\gamma^\mu\ell)\,,&\quad \mathcal{O}_9^{\prime\ell}=(\overline{s}\gamma_\mu P_Rb)(\overline{\ell}\gamma^\mu\ell)\,,\\ 
\mathcal{O}_{10}^\ell=(\overline{s}\gamma_\mu P_Lb)(\overline{\ell}\gamma^\mu\gamma_5\ell)\,,&\quad \mathcal{O}_{10}^{\prime\ell}=(\overline{s}\gamma_\mu P_Rb)(\overline{\ell}\gamma^\mu\gamma_5\ell)\,,\\ 
\mathcal{O}_S^\ell=(\overline{s} P_Rb)(\overline{\ell}\ell)\,,&\quad \mathcal{O}_S^{\prime\ell}=(\overline{s} P_Lb)(\overline{\ell}\ell)\,,\\
\mathcal{O}_P^\ell=(\overline{s} P_Rb)(\overline{\ell}\gamma_5\ell)\,,&\quad \mathcal{O}_P^{\prime\ell}=(\overline{s} P_Lb)(\overline{\ell}\gamma_5\ell)\,.
\end{array}
\end{equation}
The operators $\mathcal{O}_{7-10}$ have been listed before, they are just written in the $L,R$ notation instead of $V,A$ one. The scalar and pseudo-scalar operators were also added, even though their impact is small in the observables. The prime operators are not present in the SM expansion, they therefore correspond always to new physics effects.
 
The presence of new physics in the relevant observables can be tracked to the corresponding operators:
\begin{itemize}
\item[$\bullet$] $B\rightarrow K\mu^+\mu^-$:\quad $C_7^{(\prime)}\,,\quad C_{9}^{(\prime)}\,,\quad C_{10}^{(\prime)}$

\item[$\bullet$]  $B\rightarrow K^\ast \mu^+\mu^-$:\quad $C_7^{(\prime)}\,,\quad C_{9}^{(\prime)}\,,\quad C_{10}^{(\prime)}$

\item[$\bullet$]  $B\rightarrow K^\ast \gamma$:\quad $C_7^{(\prime)}$

\item[$\bullet$]  $B\rightarrow \phi \mu^+\mu^-$:\quad $C_{10}^{(\prime)}\,,\quad C_{S,P}^{(\prime)}$

\item[$\bullet$]  Lepton-nonuniversality: $C_{9}^{(\prime)},\quad C_{10}^{(\prime)}$

\item[$\bullet$] $B\rightarrow \mu^+\mu^-$: $C_{10}^{(\prime)},\quad C_{S,P}$
\end{itemize}

In $B\rightarrow K\mu^+\mu^-$, $B\rightarrow K^\ast \mu^+\mu^-$, and $B\rightarrow \phi \mu^+\mu^-$ the form factors and contributions of the hadronic weak Hamiltonian are the main theoretical challenges. Direct $CP$ asymmetries in $B$ decays can give a hint of new physics, specially in $B\rightarrow K^\ast \gamma$ since the $B$ factories measurements and LHCb are so precise. However, new physics in this observable is proportional to the strong phase that appears as a sub-leading effect and is also plagued with many uncertainties. 

Several global fits have been done~\cite{Altmannshofer:2014rta,Altmannshofer:2015sma}, under the assumption of new physics entering only through one operator or two real Wilson coefficients. These analysis tend to favour values $C_9^{NP}<0$ in order to accommodate the recent anomalies. 
New physics entering through $C_9$ can also contribute to the meson mixing. $B_s$-mixing is in general the most constraining observable. 

Lepton-nouniversility is also a power probe of new physics. In the SM the process $b\rightarrow s\ell\ell$ is lepton flavour universal. However, beyond the SM new flavour violating interactions can give substantial deviation form lepton-universality. Ratios of branching fractions, as well as double ratios can serve as a clean probe of new physics~\cite{Hiller:2014yaa,Hiller:2014ula}. A big advantage of considering ratios is the automatic cancelling of several uncertainties. Recently, the LHCb collaborations has reported~\cite{Aaij:2014ora}
\begin{equation}
R_K^{LHCb}=0.745^{+0.090}_{-0.074}\pm0.036
\end{equation}
which shows a $2.6\sigma$ deviation form the SM prediction $R_K^{SM}\simeq 1+\mathcal{O}(m_\mu^2/m_b^2)$~\cite{Hiller:2014ula}, in the dilepton invariant mass squared bin $1\,\text{GeV}^2\leq q^2< 6\,\text{GeV}^2$. The branching fractions rations of rare semi-leptonic $B$ decays of dimuons over dielectrons arge given by~\cite{Hiller:2014ula}
\begin{equation}
R_H=\frac{\mathcal{B}(\overline{B}\rightarrow \overline{H}\mu\mu)}{\mathcal{B}(\overline{B}\rightarrow \overline{H}ee)}\simeq
\left\{
\begin{array}{ll}
1+\Delta_++\Sigma_+\,,&H=K\\
1+\Delta_-+\Sigma_-\,,&H=K_{0}(1430)\\
1+p(\Delta_--\Delta_++\Sigma_--\Sigma_+)+\Delta_++\Sigma_+\,,&H=K^\ast\\
1+\frac{1}{2}(\Delta_-+\Delta_++\Sigma_-+\Sigma_+)\,,&H=X_s
\end{array}
\right.
\end{equation}
while the double ratios are defined as 
\begin{equation}
X_H\equiv\frac{R_H}{R_K}\simeq
\left\{
\begin{array}{ll}
1+(\Delta_--\Delta_++\Sigma_--\Sigma_+)\,,&H=K_{0}(1430)\\
1+p(\Delta_--\Delta_++\Sigma_--\Sigma_+)\,,&H=K^\ast\\
1+\frac{1}{2}(\Delta_--\Delta_++\Sigma_--\Sigma_+)\,,&H=X_s
\end{array}
\right.
\end{equation}
with 
\begin{equation}
\Delta_{\pm}=2\frac{\text{Re}\left(C_9^{SM}(C_9^{NP \mu}\pm C_9^{\prime \mu})^\ast\right)+\text{Re}\left(C_{10}^{SM}(C_{10}^{NP \mu}\pm C_{10}^{\prime \mu})^\ast\right)}{|C_{9}^{SM}|^2+|C_{10}^{SM}|^2}-(\mu\rightarrow e)
\end{equation}
the new physics contribution from the interference with the SM, and
\begin{equation}
\Sigma_{\pm}=\frac{|C_9^{NP\mu}\pm C_{9}^{\prime \mu}|^2+|C_{10}^{NP\mu}\pm C_{10}^{\prime \mu}|^2}{|C_{9}^{SM}|^2+|C_{10}^{SM}|^2}-(\mu\rightarrow e)
\end{equation}
the pure new physics contribution. At the $m_b$ scale we have for the SM Wilson coefficients $C_9^{SM}=-C_{10}^{SM}\simeq 4.2$. The factor $p$ is the polarization faction and is close to 1 (it is exactly 1 at zero recoil). These expression are valid to a very good accuracy given the current experimental uncertainties. The double ratios are very useful tool for precision tests new physics. They are only sensitive to new physics coupled to right-handed quarks, and therefore can be seen as complementary to $R_H$.

Another clean probe of new physics in the $B$ sector are the leptonic decays $B\rightarrow \ell\ell$.
The model independent average time-integrated branching ratio for $\overline{B}_s\rightarrow \ell\ell$ decays is~\cite{Buras:2013uqa}
\begin{equation}
\frac{\mathcal{B}(\overline{B}_s\rightarrow \ell\ell)}{\mathcal{B}(\overline{B}_s\rightarrow \ell\ell)^{SM}}=\left|1-0.24(C_{10}^{NP \ell}-C_{10}^{\prime \ell})-y_\ell C_{P-}^\ell\right|^2+|y_\ell C_{S-}^\ell|^2\,,
\end{equation}
with $y_u=7.7$, $y_e=(m_\mu/m_e)y_\mu=1.6\times 10^3$ and $C_{P,S-}=C_{P,S}-C_{P,S}^\prime$. The current reported experimental value for $\overline{B}_s\rightarrow \mu^+\mu^-$ decays is~\cite{CMSandLHCbCollaborations:2013pla}
\begin{equation}
\frac{\mathcal{B}(\overline{B}_s\rightarrow \mu^+\mu^-)^{exp}}{\mathcal{B}(\overline{B}_s\rightarrow \mu^+\mu^-)^{SM}}=0.79\pm0.20\,.
\end{equation}
Being a purely leptonic final state, the theoretical prediction of these processes is very clean and serves as a good probe for NP.

\section{Brief conclusions}
We have presented a short overview in the topics of flavour and $CP$ violation in and beyond the SM. The  most relevant aspects can be summarized as:
\begin{itemize} 
\item[$\bullet$] Often we have seen the indirect evidence of New particles in flavour physics before directly discovering them;

\item[$\bullet$] The SM flavour sector has been tested with impressive and increasing precision;

\item[$\bullet$] In the SM, fermions come in 3 generations of quarks and leptons; flavour physics is all about them;

\item[$\bullet$] All flavour violation in the SM is from the CKM matrix;

\item[$\bullet$] CPV in SM is small, and comes from flavour;

\item[$\bullet$] We have developed non relativistic QM tools for meson mixing;

\item[$\bullet$] We have schematically shown how to calculate hadronic observables;

\item[$\bullet$] Theoretical tools to understand the underlying physics is important. For example, effective field theory allows separation of different scales (separation of calculable parts and nonperturbative parts);

\item[$\bullet$] Any sensitivity to high scales (including to physics beyond the Standard Model) can be treated using perturbative methods;

\item[$\bullet$]Flavour structure of New Physics has to be special in order to be compatible with TeV scale New Physics. A popular example is MFV, but other possibilities exist such a partial compositeness, etc;

\item[$\bullet$] If new particles discovered, their flavour properties can teach us about the underlying structure of New Physics:  masses (degeneracies), decay rates (flavour decomposition), cross sections;

\item[$\bullet$] Flavour physics provide important clues to model building in the LHC era;

\item[$\bullet$] LHC era is also a Flavour Precision era, and a lot of interesting measurements are coming, as we have already seen some tensions with SM.
\end{itemize}

\section*{Acknowledgements}

We wish to thank Monika Blanke for useful suggestions while preparing the lectures.
We would like to also thank all the organisers of the second AEPSHEP School for the very warm hospitality extended to us and for the very nice atmosphere in the School.
This work was supported by the National Research Foundation of Korea grant MEST No. 2012R1A2A2A01045722.

\end{document}